\documentclass[reqno]{amsart}
\usepackage{amsmath,amssymb,euscript,graphicx}
\usepackage[arrow,matrix,curve]{xy}
\numberwithin{equation}{section}

\newcommand\bZ{\mathbb{Z}}

\newcommand\bR{\mathbb{R}}
\newcommand\bC{\mathbb{C}}
\newcommand\B{{\boldsymbol{B}}}
\newcommand\be{\mathfrak{e}}
\newcommand\bm{\mathfrak{m}}
\newcommand\fg{\mathfrak{g}}

\newcommand\ad{{\operatorname{ad}}}
\newcommand\Ad{{\operatorname{Ad}}}

\newcommand\Hom{\operatorname{Hom}}
\newcommand\Ext{\operatorname{Ext}}
\newcommand\coker{\operatorname{coker}}
\newcommand\tr{\operatorname{tr}}
\newcommand\vol{\operatorname{vol}}
\newcommand\id{\mathrm{id}}
\newcommand\cc{\mathrm{cc}}
\newcommand\sus{\operatorname{s}}
\newcommand\im{\operatorname{im}}
\renewcommand\Re{\operatorname{Re}}
\renewcommand\Im{\operatorname{Im}}

\newcommand\al\alpha
\newcommand\gam\gamma
\newcommand\del\delta
\newcommand\veps\varepsilon
\newcommand\lam\lambda
\newcommand\sig\sigma
\newcommand\tht\theta
\newcommand\om\omega
\newcommand\vph\varphi
\newcommand\vro\varrho
\newcommand\Gam\varGamma
\newcommand\Sig\varSigma
\newcommand\Om\varOmega
\newcommand\PSI\varPsi

\newcommand\cA{\EuScript{A}}
\newcommand\cB{\EuScript{B}}
\newcommand\cG{\EuScript{G}}
\newcommand\cH{\EuScript{H}}
\newcommand\cL{\EuScript{L}}
\newcommand\cM{\EuScript{M}}
\newcommand\cN{\EuScript{N}}

\newcommand\cU{\EuScript{U}}
\newcommand\cF{\mathcal{F}}
\newcommand\cJ{\mathcal{J}}
\newcommand\AH{\cA_{\mathrm{H}}}
\newcommand\MH{\cM_{\mathrm{H}}}
\newcommand\UH{\cU_{\mathrm{H}}}
\newcommand\Mf{\cM_\mathrm{flat}}

\newcommand\ta{{\tilde a}}
\newcommand\tb{{\tilde b}}
\newcommand\tSig{{\tilde\Sig}}
\newcommand\bI{{\partial I}}
\newcommand\bSig{{\partial\Sig}}
\newcommand\LG{{}^L\!G}
\newcommand\LT{{}^L\!T}
\newcommand\LW{{}^L\!W}
\newcommand\Lh{{}^L\!h}
\newcommand\Li{{}^L\!i}
\newcommand\Lj{{}^L\!j}
\newcommand\Lq{{}^L\!q}

\newcommand\Ldel{{}^L\!\del}
\newcommand\ZG{{Z(G)}}
\newcommand\pG{\pi_1(G)}
\newcommand\pGad{\pi_1(G_\ad)}
\newcommand\tG{{\widetilde G}}
\newcommand\tP{{\widetilde P}}
\newcommand\tLG{{\widetilde\LG}}
\newcommand\tq{\tilde q}
\newcommand\tM{{\widetilde\cM}}
\newcommand\tU{{\widetilde\cU}}
\newcommand\tB{{\widetilde\cB}}
\newcommand\mm{{\overline m}}
\newcommand\ee{{\overline e}}
\newcommand\CP{\bC P}
\newcommand\sfrac[2]{\raisebox{-.1ex}{\scalebox{1.1}[1.2]{$\frac{#1}{#2}$}}}
\newcommand\mfrac[2]{\raisebox{-.1ex}{\scalebox{1.2}[1.3]{$\frac{#1}{#2}$}}}
\newcommand\msum[2]{\raisebox{.1ex}{\scalebox{0.9}[0.9]
                   {$\displaystyle\sum_{#1}^{#2}$}}}
\newcommand\moplus[2]{\raisebox{.15ex}{\scalebox{0.9}[0.9]
                   {$\displaystyle\bigoplus_{#1}^{#2}$}}}

\newcommand\msqcup[2]{\raisebox{.15ex}{\scalebox{0.9}[0.9]
                   {$\displaystyle\bigsqcup_{#1}^{#2}$}}}
\newcommand\mprod[2]{\raisebox{.1ex}{\scalebox{0.9}[0.9]
                   {$\displaystyle\prod_{#1}^{#2}\,$}}}
\newcommand\mint[2]{\raisebox{.1ex}{\scalebox{0.9}[0.9]
                   {$\displaystyle\int_{#1}^{#2}$}}}
\newcommand\map[1]{\stackrel{{#1}}\longrightarrow}
\newcommand\ui{\underline i{}}
\newcommand\uj{\underline j{}}
\newcommand\udel{\underline\del{}}
\newcommand\uA{\underline A}
\newcommand\upG{\underline{\hspace{5ex}}\hspace{-5ex}
                \pi_1\raisebox{0.2ex}{\small(}G\raisebox{0.2ex}{\small)}}
\newcommand\uZG{\underline{\hspace{5ex}}\hspace{-5ex}
                Z\raisebox{0.2ex}{\small(}G\raisebox{0.2ex}{\small)}}
\newcommand\uZoG{\underline{\hspace{5ex}}\hspace{-5ex}
                Z\raisebox{0.2ex}{\small(}\tG\raisebox{0.2ex}{\small)}}
\newcommand\ii{\sqrt{-1}}
\newcommand\bra\langle
\newcommand\ket\rangle
\newcommand\dcirc{{\displaystyle\circ}}
\newcommand\scirc{{\scriptstyle\circ}}

\begin{document}

\hfill {\tt arXiv:1804.11343\,[hep-th]}

\vspace{2em}

\title{Non-orientable surfaces and electric-magnetic duality}
\maketitle

\vspace{-1em}

\begin{center}
{\large Siye Wu}\\
\bigskip
{\small Department of Mathematics, National Tsing Hua University,
Hsinchu 30013, Taiwan\\
E-mail address: {\tt swu@math.nthu.edu.tw}}
\end{center}

\begin{abstract}
We consider the reduction along two compact directions of a twisted $N=4$
gauge theory on a $4$-dimensional orientable manifold which is not a global
product of two surfaces but contains a non-orientable surface.
The low energy theory is a sigma-model on a $2$-dimensional worldsheet with
a boundary which lives on branes constructed from the Hitchin moduli space of
the non-orientable surface.
We modify 't~Hooft's notion of discrete electric and magnetic fluxes in gauge
theory due to the breaking of discrete symmetry and we match these fluxes with
the homotopy classes of maps in the sigma-model.
We verify the mirror symmetry of branes as predicted by $S$-duality in gauge
theory.\\

\end{abstract}

\vspace{.25cm}

\section{Introduction}
In a celebrated work \cite{KW} (see \cite{GW08,W08,FW08,GW10,W18} for
subsequent developments), Kapustin and Witten showed that a twisted version
of the $N=4$ supersymmetric gauge theory in four dimensions compactifies to
a $2$-dimensional sigma-model whose target space is the Hitchin moduli space.
Electric-magnetic duality or $S$-duality in four dimensions reduces to mirror
symmetry in two dimensions that explains the geometric Langlands programme.
In this paper, we consider the reduction along two compact directions of the
gauge theory on a $4$-dimensional orientable spacetime manifold which is not a
global product of two surfaces but contains embedded non-orientable surfaces.
The low energy theory is a sigma-model on a $2$-dimensional worldsheet with
a boundary which lives on branes constructed from the Hitchin moduli space of
the non-orientable surface.
We  show that the discrete symmetry from the centre of the gauge group is
broken by fixing the topology types of the gauge bundles and we modify
't~Hooft's notion of discrete electric and magnetic fluxes accordingly.
We match these rectified discrete fluxes with the homotopy classes of
relative maps in the sigma-model.
We verify the mirror symmetry of branes as predicted by $S$-duality in
gauge theory.

Throughout this paper, the gauge group $G$ is a connected compact semisimple
but not necessarily simply connected Lie group.
When concerned with the coupling constant or the action of the theory, we
further assume for simplicity that $G$ is simple, but this restriction is
not essential as the theory always decouples to ones whose gauge groups
are the simple factors of $G$.
The rest of the paper is organised as follows.

In \S\ref{sec:4d}, we revisit a few global aspects of quantum gauge theories
in four dimensions.
We find that the usual concept of discrete electric and magnetic fluxes of
't Hooft \cite{tH78,tH79} requires a subtle modification.
As is well known, the discrete magnetic fluxes in $H^2(Y,\pG)$ classify the
topology of the gauge bundles over a time slice $Y$, which is a $3$-manifold,
whereas the discrete electric fluxes are the momenta of discrete symmetries
in $H^1(Y,\ZG)$ that are present when the centre $\ZG$ of the gauge group $G$
is non-trivial.
In \S\ref{sec:em4d}, we show that a discrete symmetry may change the topology
of the bundles and is therefore broken in a fixed topological sector.
This leads naturally to the notion of rectified discrete electric and magnetic 
fluxes, which form the sets $\be(Y,G)$ and $\bm(Y,G)$, respectively.
The two sets are defined by the connecting homomorphism $\del_Y^1$ in the long
exact sequence \eqref{eqn:long} of cohomology groups of $Y$.
In \S\ref{sec:Gad}, we revisit quantum gauge theory in light of this adjustment
in both canonical quantisation and the path integral formalism.
In \S\ref{sec:sdual}, we establish $S$-duality between the rectified discrete
electric and magnetic fluxes, given by \eqref{eqn:Srec}.

In \S\ref{sec:2d}, we recall the reduction along a closed orientable surface
$C$ from a twisted $N=4$ gauge theory in four dimensions to a sigma-model in
two dimensions \cite{KW}, with an emphasis on the gauge group $G$ being an
arbitrary compact semisimple Lie group.
Though this is a straightforward generalisation of \cite{KW}, where $G$ is
often taken as the universal covering group $\tG$ or the adjoint group
$G_\ad$, we introduce a few techniques in topology that will be useful in
\S\ref{sec:non-or}.
In \S\ref{sec:4d2d}, we review the dimensional reduction of gauge theories
on the $4$-manifold $X=\Sig\times C$ to sigma-models whose worldsheet $\Sig$
is an orientable surface and whose target space is the Hitchin moduli space
$\MH(C,G)$ and we verify their anomaly-free conditions.
In \S\ref{sec:e2d}, we interpret the discrete fluxes in gauge theory from the
point of view of two dimensions.
Curiously, there are less topological types of gauge bundles over a
$4$-dimensional spacetime than the homotopy classes of maps from the
worldsheet to the target.
This mismatch is resolved by summing over some flat $B$-fields on the target
space.
In \S\ref{sec:S2d-o}, we recall the relation between $S$-duality in gauge
theory and mirror symmetry in sigma-model and explain its consistency when
the gauge group $G$ is changed to $G_\ad$ or $\tG$.

In \S\ref{sec:non-or}, we consider gauge theory on an orientable $4$-manifold
$X$ which is not a global product of two surfaces but contains embedded
non-orientable surfaces $C'$ whose orientation double cover is $C$.
When the size of $C'$ (and hence that of $C$) becomes small, the theory
reduces to a sigma-model whose target space is $\MH(C,G)$ on a worldsheet
$\Sig$ with a boundary which lives on branes constructed from the Hitchin
moduli space $\MH(C',G)$ for $C'$.
In \S\ref{sec:4dn2d}, we explain this reduction and show that the sigma-model
obtained is anomaly-free from the $2$-dimensional point of view.
When the worldsheet $\Sig$ is a cylinder, the $4$-dimensional spacetime $X$
splits as a product of time and a spacelike $3$-manifold $Y$.
In \S\ref{sec:em4dn}, we compute the sets of rectified discrete electric and
magnetic fluxes, $\be(Y,G)$ and $\bm(Y,G)$, in the $4$-dimensional gauge
theory.
They are given, respectively, in \eqref{eqn:beC} and \eqref{eqn:bmC} using
the cohomology groups of $C$, $C'$ as well as the sets $\be(\bZ_2,G)$ and
$\bm(\bZ_2,G)$; the latter are defined in \eqref{eqn:Z2G} using the connecting
homomorphism $\del_{\bZ_2}^1$ in the exact sequence \eqref{eqn:del2}.
In \S\ref{sec:em2dn}, we match the discrete parameters in the $4$- and
$2$-dimensional theories.
In particular, the open string configurations are classified by the homotopy
classes of relative maps whereas the branes $\cB^{\ee_2,\mm_2}$ are labelled
by the discrete data $\ee_2\in\be(\bZ_2,G)$, $\mm_2\in\bm(\bZ_2,G)$.
Remarkably, while in two dimensions the $G$-bundles on $C'$ always pull back
to topologically trivial bundles on $C$, and the discrete $B$-field on
$\MH(C,G)$ must be trivial to have an anomaly-free theory, these discrete
parameters are also absent in the discrete fluxes in $\bm(Y,G)$ and $\be(Y,G)$
from the $4$-dimensional calculation.
In \S\ref{sec:mir-n}, we verify the exchange of relative winding and momenta,
and mirror symmetry of the branes as predicted by $S$-duality;
this provides yet another non-trivial test of $S$-duality.

In \S\ref{sec:qt}, we apply the results in \S\ref{sec:non-or} to the
quantisation of $\MH(C',G)$ via branes and mirror symmetry \cite{GW09,G}.

We collect and work out various results in topology in the Appendix.

\section{Global aspects of four dimensional gauge theory}\label{sec:4d}
\subsection{Discrete electric and magnetic fluxes}\label{sec:em4d}
If the spacetime $X$ is an orientable $4$-manifold, for any compact Lie group
$G$, characteristic classes of a $G$-bundle $P$ over $X$ are in $H^2(X,\pG)$
and $H^4(X,\pi_3(G))$ \cite{St}.
If $G$ is simple, then $H^4(X,\pi_3(G))\cong\bZ$ contains the instanton
number $k(P)$ which is summed over in path integral, and each term is weighted
by a phase determined by a theta angle in the dual group $\mathrm U(1)$ of
$\bZ$ \cite{BPST,JR,CDG}.
More generally, $k(P)$ is a class in a higher rank free Abelian group
$H^4(X,\pi_3(G))$ and the theta angle is in its dual torus.
On the other hand, the class $\xi(P)$ in the torsion group $H^2(X,\pG)$ is the
obstruction to lifting its structure group $G$ to the universal cover $\tG$.
It is called the discrete flux because it gives rise to the discrete electric
and magnetic fluxes \cite{tH78,tH79}.

Suppose the spacetime manifold is $X=T^1\times Y$, where $T^1$ is a circle in
the time direction and $Y$ is a spacelike closed orientable $3$-manifold.
As in \cite{KW}, we write $\xi(P)=a+m$, $a=\xi^{1,1}(P)$, $m=\xi^{0,2}(P)$
according to the K\"unneth decomposition
$H^2(X,\pG)\cong H^1(Y,\pG)\oplus H^2(Y,\pG)$.
Since $m$ describes the topology of the $G$-bundle over a time slice $Y$, it
depends only on the field configuration at a particular time.
So quantisation can be performed with a fixed value of $m$.
We call $m$ a discrete magnetic flux until we meet the need for modification
below.

In contrast, $a\in H^1(Y,\pG)$ contains information from not only the initial
data but also the entire time interval.
At the quantum level, fixing $a$ is compatible with neither the cluster
decomposition property nor a Hilbert space interpretation.
Instead, we define discrete electric fluxes as the momenta of a discrete
translation group $H^1(Y,\ZG)$.
Any $g\in H^1(Y,\ZG)\cong\Hom(\pi_1(Y),\ZG)$ acts on a connection by modifying
its holonomy along a loop $\gam$ representing $[\gam]\in\pi_1(Y)$ by
$g([\gam])\in\ZG$.
Geometrically, $g\in H^1(Y,\ZG)$ determines a flat $\ZG$-bundle $Q^g\to Y$
and thus a new $G$-bundle $P^g:=Q^g\times_\ZG P$; a connection on $P$ and
the flat connection on $Q^g$ defines a new connection on $P^g$.
If the matter fields, such as the adjoint matter, are in representations of
$G$ in which the centre $\ZG$ acts trivially, the action of $g$ on them is
trivial.
Since the curvature and hence the action functional is invariant under
$H^1(Y,\ZG)$, the group acts as classical symmetries.
The symmetry survives at the quantum level because it preserves the path
integral measure.
The quantum Hilbert space therefore decomposes according to representation
types of $H^1(Y,\ZG)$, each labelled by a character $e\in H^1(Y,\ZG)^\vee$.
We tentatively call $e$ a discrete electric flux.

In the most general situation, however, we can not fix $e\in H^1(Y,\ZG)^\vee$
and $m\in H^2(Y,\pG)$ simultaneously because the symmetry group $H^1(Y,\ZG)$
changes the topology of a bundle.
In fact \cite{Wu15}, $g\in H^1(Y,\ZG)$ maps $\xi(P)=m$ to
$\xi(P^g)=m+\del^1_Y(g)$, where $\del^1_Y$ is the connecting homomorphism
in the long exact sequence
\begin{equation}\label{eqn:long}
\scalebox{.95}[1]{$0\to H^1(Y,\pG)\map{i^1_Y}H^1(Y,Z(\tG))\map{j^1_Y}H^1(Y,\ZG)
\map{\del^1_Y}H^2(Y,\pG)\map{i^2_Y}H^2(Y,Z(\tG))\map{j^2_Y}H^2(Y,\ZG)\to0$}
\end{equation}
of cohomology groups of $Y$ induced by the short exact sequence
\begin{equation}\label{eqn:piG}
0\to\pG\map{i}Z(\tG)\map{j}\ZG\to0
\end{equation}
of coefficient groups.
In \eqref{eqn:long}, $i^1_Y$ is injective and $j^2_Y$ is surjective because
$Y$ is  a closed orientable $3$-manifold.
Actually, $\del^1_Y$ maps only to the subgroup $\Ext(H_1(Y,\bZ),\pG)$ of
$H^2(Y,\pG)$ (see \cite{Wu15} for related results on flat $G$-bundles).
So $\del^1_Y$ is zero if $H_1(Y,\bZ)$ is torsion-free.
This torsion-free assumption was made in \cite{VW,Wu08} and is satisfied when
$Y$ is a product of $S^1$ and a closed orientable surface \cite{KW}.
The map $\del_Y^1$ is also zero if the orders of torsion elements in
$H_1(Y,\bZ)$ are coprime to those of $\pG$ or if the exact sequence
\eqref{eqn:piG} splits.
In all cases when $\del_Y^1=0$, the symmetry $H^1(Y,\ZG)$ preserves the
discrete magnetic flux $m\in H^2(Y,\pG)$, and the quantum Hilbert space
decomposes into sectors labelled by
$(e,m)\in H^1(Y,\ZG)^\vee\times H^2(Y,\pG)$.

For a general orientable $3$-manifold $Y$, such as the one constructed in
\S\ref{sec:non-or} (or \S\ref{sec:HYX}) from a non-orientable surface,
$H_1(Y,\bZ)$ does contain torsion elements and $\del^1_Y$ can be non-zero.
A simpler example is $Y=\bR P^3$, which is orientable, and $G=\mathrm{SO}(6)$,
whose universal cover is $\tG=\mathrm{SU}(4)$.
Then \eqref{eqn:piG} is $0\to\bZ_2\to\bZ_4\to\bZ_2\to0$ and all cohomology
groups in \eqref{eqn:long} are $\bZ_2$.
As a result, $\del_Y^1$ is an isomorphism.
So in general, a fixed $m$ breaks the discrete symmetry to the subgroup
$\ker(\del^1_Y)\subset H^1(Y,\ZG)$, and thus the quantum Hilbert space
decomposes according to the characters in $\be(Y,G):=\ker(\del_Y^1)^\vee$.
At first sight, this suggests an apparent asymmetry between the discrete
electric and magnetic fluxes.
But a gauge theory sector with $m\in H^2(Y,\pG)$ is isomorphic to the one
with $m+\del^1_Y(g)$ because of the symmetry $g\in H^1(Y,\ZG)$ itself.
So the actual discrete parameters labelling non-isomorphic sectors are in
$\bm(Y,G):=H^2(Y,\pG)/\im(\del^1_Y)=\coker(\del^1_Y)$.
The sets $\be(Y,G)$ and $\bm(Y,G)$ form an exact sequence
\begin{equation}\label{eqn:exactem}
0\to\be(Y,G)^\vee\to H^1(Y,\ZG)\map{\del_Y^1}H^2(Y,\pG)\to\bm(Y,G)\to0.
\end{equation}
Henceforth we call elements in $\be(Y,G)$ and $\bm(Y,G)$ discrete electric
and magnetic fluxes, respectively, or the rectified discrete fluxes to
emphasise the adjustment.

We can describe these discrete fluxes using the universal cover $\tG$, which
has a trivial $\pi_1(\tG)$ but a maximal centre $Z(\tG)$, or the adjoint group
$G_\ad:=G/\ZG$, which has a trivial centre but has $\pGad\cong Z(\tG)$.
{}From \eqref{eqn:long} we get
\begin{equation}\label{eqn:bm}
\bm(Y,G)\cong\im(i^2_Y)=\ker(j^2_Y)\subset H^2(Y,\pGad)=\bm(Y,G_\ad).
\end{equation}
So a rectified discrete magnetic flux $m$ in a gauge theory with group $G$ can
be naturally regarded as a discrete magnetic flux of a gauge theory with group
$G_\ad$.
The $G$-bundles $P$ with $\xi(P)\in m$, of possibly different topological
types, produces the same $G_\ad$-bundle $P_\ad:=P/\ZG$ with $\xi(P_\ad)=m$,
and $\bm(Y,G)$ is precisely the set of discrete fluxes of $G_\ad$-bundles
on $Y$ whose structure group can be lifted from $G_\ad$ to $G$.

On the other hand, we have $\be(Y,G)=\coker((\del^1_Y)^\vee)$, where
$(\del^1_Y)^\vee\colon H^2(Y,\pG)^\vee\to H^1(Y,\ZG)^\vee$ is the dual map of
$\del^1_Y$ with
$\im((\del^1_Y)^\vee)\cong\{e\in H^1(Y,\ZG)^\vee:e(\ker(\del^1_Y))=1\}$.
Using the dual sequence of \eqref{eqn:long}, we obtain
\begin{equation}\label{eqn:be}
\be(Y,G)\cong\im((j^1_Y)^\vee)=\ker((i^1_Y)^\vee)\subset H^1(Y,Z(\tG))^\vee
=\be(Y,\tG).
\end{equation}
So a rectified discrete electric flux in a theory of gauge group $G$ can be
identified as a discrete electric flux in a theory with gauge group $\tG$
which is trivial on $\ker(j^1_Y)\cong H^1(Y,\pG)$.
By \eqref{eqn:dis}, we obtain a useful identity
\begin{equation}\label{eqn:disF}
\msum{e\in\be(Y,G)}{}e(a)=\left\{\!\begin{array}{ll}
      |\ker(\del^1_Y)| & \mbox{if } a\in\ker(j^1_Y),\\
      0 & \mbox{if otherwise}.
      \end{array}\right.
\end{equation}

Geometrically, each $g\in H^1(Y,\ZG)$ acts on the configuration space
$\cB(Y,G):=\bigsqcup_{[P]}\cB(P)$, where the union is over the topological
types $[P]$ of $G$-bundles $P$ over $Y$ and $\cB(P)=\cA(P)/\cG(P)$ is the
space of gauge equivalence classes of connections on $P$
(cf.~\S\ref{sec:pi01}).
If $\del_Y^1(g)=0$, then for any connection $A$ on $P$, $g\cdot A$ can be
identified as a connection on the same bundle $P$.
Since $\cA(P)$ is an affine space, the two connections $A$ and $g\cdot A$
are joined by a line segment.
So the gauge equivalent classes $[A]$ and $[g\cdot A]$ are in the same
connected component of $\cB(Y,G)$.
However if $\del_Y^1(g)\ne0$, then $g$ sends one connected component of
$\cB(Y,G)$ to another, but the subgroup $\ker(\del_Y^1)$ preserves each
component.
A simple analogy is the quantum mechanics of a particle on a disconnected
configuration space with a discrete symmetry acting on the total space and
hence on the set of its connected components.

For a closed $4$-manifold $X$, we have a similar long exact sequence
\begin{equation}\label{eqn:longX}
\scalebox{.9}[1]{$0\to H^1(X,\pG)\map{i_X^1}H^1(X,Z(\tG))\map{j_X^1}
H^1(X,\ZG)\map{\del_X^1}H^2(X,\pG)\map{i^2_X}H^2(X,Z(\tG))\map{j^2_X}
H^2(X,\ZG)\to\cdots.$}
\end{equation}
We obtain the set $\ker(j_X^2)\subset H^2(X,\pGad)$ of discrete fluxes of
$G_\ad$-bundles over $X$ that can be lifted to $G$-bundles.
In the (relativistic invariant) path integral formulation, we sum over both
the instaton number and the discrete flux in $\ker(j_X^2)$.
If $X=T^1\times Y$, then \eqref{eqn:longX} reduces to two copies of
\eqref{eqn:long}.
Thus in this case we have
\[ \ker(\del_X^1)\cong\ZG\oplus\ker(\del^1_Y)\subset H^1(X,\ZG)
   \cong\ZG\oplus H^1(Y,\ZG),\quad
   \im(\del_X^1)\cong\im(\del_Y^1)\subset H^2(Y,\pG),   \]
\[ \ker(j_X^2)\cong\ker(j_Y^1)\oplus\ker(j_Y^2)\subset H^2(X,\pGad)
   \cong H^1(Y,\pGad)\oplus H^2(Y,\pGad).     \]

\subsection{Quantum gauge theories revisited}\label{sec:Gad}
We revisit quantum gauge theories, first in the path integral formalism and
then in canonical quantisation, taking into account the role of the rectified
discrete fluxes introduced in \S\ref{sec:em4d}.
We pay particular attention to the relation among theories with various gauge
groups of the same Lie algebra.

As before, the spacetime is a closed orientable $4$-manifold $X$ and the
gauge group $G$ is a compact semisimple Lie group.
The partition function of the gauge theory is
\begin{equation}\label{eqn:ZXG}
Z_{X,G}=\msum{[P]}{}\mfrac1{\vol(\cG(P))}\mint{}{}DA\cdots\;e^{-S(A,\cdots)},
\end{equation}
where the sum is over the topological types $[P]$ of principal $G$-bundles over
$X$ and the integral is over the gauge and matter fields.
The classical action (when $G$ is simple) is
\begin{equation}\label{eqn:4dS}
S(A,\dots)=-\mfrac1{e^2}\!\mint{X}{}\tr F_A\wedge*F_A
+\mfrac{\ii\,\tht}{8\pi^2}\!\mint{X}{}\tr F_A\wedge F_A+\cdots,
\end{equation}
where $e>0$ is the coupling constant, $\tht$ is the theta angle, ``$-\tr$'' is
the inner product on the Lie algebra $\fg$ of $G$ such that the long roots are
of length $\sqrt2$, and the matter fields are omitted.

If the matter fields are in representations of $G$ in which the centre $\ZG$
acts trivially (such as the adjoint representation), the gauge group can be
changed to the adjoint group $G_\ad:=G/\ZG$.
A $G$-bundle $P$ defines a $G_\ad$-bundle $P_\ad:=P/\ZG$.
In the gauge theories with gauge groups $G$ and $G_\ad$, the gauge and matter
fields are the same, but the groups of gauge transformations $\cG(P)$ and
$\cG(P_\ad)$ are related by the exact sequence \eqref{eqn:GGad} with $X$ as
the base manifold.
Formally, the volumes of the two groups are related by
\begin{equation}\label{eqn:vol}
\vol(\cG(P_\ad))=\mfrac{|\ker(\del^1_X)|}{|\ZG|}\vol(\cG(P)).
\end{equation}
When $H_1(X,\bZ)$ has no torsion elements, we have $\del_X^1=0$ and hence
$\ker(\del_X^1)=H^1(X,\ZG)\cong\ZG^{\oplus b_1(X)}$, where $b_1(X)$ is the
first Betti number of $X$.
Therefore we recover from \eqref{eqn:vol} the formula \cite{VW}
\begin{equation}\label{eqn:volVW}
\vol(\cG(P_\ad))=|Z(G)|^{b_1(X)-1}\vol(\cG(P)).
\end{equation}

To relate the partition functions of the two gauge theories, we note that
$\xi(P_\ad)=i_X^2(\xi(P))$, but the map
$i_X^2\colon H^2(X,\pG)\to H^2(X,\pGad)$ is many to one with multiplicity
$|\ker(i_X^2)|=|\im(\del_X^1)|$.
On the other hand, $k(P_\ad)=k(P)$ under the isomorphism
$\pi_3(G_\ad)\cong\pi_3(G)$.
Using these facts and \eqref{eqn:vol}, we obtain
\begin{align}\label{eqn:ZGGad}
Z_{X,G}&=|\im(\del_X^1)|\msum{\xi(P_\ad)\in\im(i^2_X)}{}\;\;
\msum{k(P)\in H^4(X,\pi_3(G))}{}\mfrac{|\ker(\del^1_X)|}{|\ZG|}
\mfrac1{\vol(\cG(P_\ad))}\mint{}{}DA\cdots\;e^{-S(A,\cdots)}  \nonumber  \\
&=\mfrac{|H^1(X,\ZG)|}{|\ZG|}\!\!\msum{\xi(P_\ad)\in\ker(j^2_X)}{}\;\;
\msum{k(P_\ad)\in H^4(X,\pi_3(G_\ad))}{}\mfrac1{\vol(\cG(P_\ad))}
\mint{}{}DA\cdots\;e^{-S(A,\cdots)}.
\end{align}
Here the sum is over topological types of $G_\ad$-bundles $P_\ad$ over $X$
that can be lifted to $G$-bundles.
In the case $\del_X^1=0$, $H^2(X,\pG)$ is a subset of $H^2(X,\pGad)$ and
the derivation of \eqref{eqn:ZGGad} can be simpler \cite{Wu08}.

In canonical quantisation, the spacetime is $X=T^1\times Y$, where $Y$ is a
closed orientable $3$-manifold.
We can fix a discrete magnetic flux $m\in\bm(Y,G)$.
Suppose $P^m$ is a $G$-bundle over $Y$ such that $\xi(P^m)$ is in the coset
$m$; other elements in the same coset give rise to isomorphic quantum theories.
Let $\cH_{Y,G}^m$ be the quantum Hilbert space.
The partition function is
$Z_{Y,G}^m(\beta)=\tr_{\cH_{Y,G}^m}e^{-\beta\hat H}$, where $\beta>0$ is the
length of $T^1$ (or the inverse temperature) and $\hat H$ is the Hamiltonian
operator.
Let $\hat g$ be the action of $g\in\ker(\del_Y^1)$ on $\cH_{Y,G}^m$.
Under the action of the unbroken symmetry group $\ker(\del^1_Y)$, the Hilbert
space $\cH_{Y,G}^m$ and the partition function $Z_{Y,G}^m(\beta)$ decompose as
\[ \cH_{Y,G}^m=\moplus{e\in\be(Y,G)}{}\cH_{Y,G}^{e,m},\qquad
Z_{Y,G}^m(\beta)=\msum{e\in\be(Y,G)}{}Z_{Y,G}^{e,m}(\beta).   \]
Here $\cH_{Y,G}^{e,m}$ is the sector that transforms according to the
character $e\in\be(Y,G)$ of the group $\ker(\del_Y^1)$ whereas
$Z_{Y,G}^{e,m}(\beta)$ is obtained by inserting the corresponding projection
operator in the trace, i.e.,
\begin{equation}\label{eqn:Zemg}
Z_{Y,G}^{e,m}(\beta)=\mfrac1{|\ker(\del^1_Y)|}\msum{g\in\ker(\del^1_Y)}{}
    e(g)^{-1}\,\tr_{\cH_{Y,G}^m}\!\big(\hat g\;e^{-\beta\hat H}\big).
\end{equation}

The above trace with $g\in\ker(\del_Y^1)$ contains a sum over the homotopy
types of paths $[A(t)]$ in the space $\cB(P^m):=\cA(P^m)/\cG(P^m)$ such that
$[A(\beta)]=g\cdot[A(0)]$.
Here, $[A]$ means the gauge equivalence class of a connection $A$ on $P^m$.
If $g=0$ (and $\hat g=\id_{\cH_{Y,G}^m}$), these paths in $\cB(P^m)$ are loops
representing elements of $\pi_1(\cB(P^m))\cong\pi_0(\cG(P^m))$.
Homotopy groups of the group $\cG(P^m)$ of gauge transformations are studied
in \S\ref{sec:pi01}.
By \eqref{eqn:pi0GY3}, an element of $\pi_0(\cG(P^m))$ is a pair $(k,a)$,
where $a\in H^1(Y,\pG)$ and $k\in\eta_*^{-1}(a)$.
Here, $\eta_*\colon\pi_0(\cG(P))\to H^1(Y,\pi_1(G))$ maps (the homotopy class
of) a gauge transformation to the primary obstruction to its topological
triviality.
The set $\eta_*^{-1}(a)$ is a torsor over $H^3(Y,\pi_3(G))$; it can be
identified, though not naturally, with $H^3(Y,\pi_3(G))$, and we can roughly
write $k\in H^3(Y,\pi_3(G))$.
Using a gauge transformation on $P^m$ in the homotopy class $(k,a)$, we can
construct a $G$-bundle $P^{k,a,m}$ over $X=T^1\times Y$.
In the path integral for the trace with $g=0$, we integrate over the gauge
field $A\in\cA(P^{k,a,m})$ and the matter fields that couple to it.
If $g\ne0$, the homotopy types of such paths $[A(t)]$ in $\cB(P^m)$ form a set
$\pi_1(\cB(P^m),g)$ which is naturally identified with $\zeta_*^{-1}(g)$ and
is a torsor over $\pi_1(\cB(P^m))$.
Here the map $\zeta_*\colon\pi_1(\cB(P^m_\ad))\to\ker(\del_Y^1)$ is explained
in the short exact sequence \eqref{eqn:pi0GGad} or \eqref{eqn:pi1BBad}.
An element of $\pi_1(\cB(P^m),g)$ is a pair $(k,a)$, where
$k\in H^3(Y,\pi_3(G))$ and $a\in(j_Y^1)^{-1}(g)\subset H^1(Y,\pGad)$, and
it determines a twisted $G$-bundle $P^{k,a,m}$ over $X$ in the presence of a
discrete $B$-field $g\in H^1(Y,\ZG)\subset H^2(X,\ZG)$ (see \S\ref{sec:GGad}).
Note that $P^{k,a,m}$ is not determined by $g$ alone but requires the more
refined information $a$.
To summarise, we get
\begin{equation}\label{eqn:Zkam}
\tr_{\cH_{Y,G}^m}\big(\hat g\,e^{-\beta\hat H}\big)=\msum{a\in(j_Y^1)^{-1}(g)}
\;\;\msum{k\in H^3(Y,\pi_3(G))}{}Z_{Y,G}^{k,a,m}(\beta),
\end{equation}
where $Z_{Y,G}^{k,a,m}(\beta)$ is an integral over the gauge and matter fields
associated to a possibly twisted $G$-bundle $P^{k,a,m}$.

Combining the sums over $g$ in \eqref{eqn:Zemg} and $a$ in \eqref{eqn:Zkam},
we get
\begin{equation}\label{eqn:Zem}
Z_{Y,G}^{e,m}(\beta)=\mfrac1{|\ker(\del^1_Y)|}\msum{a\in H^1(Y,\pGad)}{}
\,e(a)^{-1}\!\msum{k\in H^3(Y,\pi_3(G))}{}Z_{Y,G}^{k,a,m}(\beta).
\end{equation}
Here we identified $e\in\be(Y,G)$ with an element in
$\be(Y,\tG)=(H^1(Y,\pGad))^\vee$ by the inclusion \eqref{eqn:be}.
The bundle $P^{k,a,m}$ is an honest $G$-bundle if
$a\in\ker(j_Y^1)\cong H^1(Y,\pG)$ but is twisted if otherwise.
But $P^{k,a,m}_\ad:=P^{k,a,m}/\ZG$ is always an honest $G_\ad$-bundle over $X$
with $\xi(P^{k,a,m}_\ad)=a+m$, where $a=\xi^{1,1}(P^{k,a,m}_\ad)$,
$m=\xi^{0,2}(P^{k,a,m}_\ad)$.
So if $e\in\be(Y,G)$ and $g=j_Y^1(a)\in H^1(Y,\ZG)$ as before, we get a
geometric formula
\begin{equation}\label{eqn:geomph}
e(g)=e(\xi^{1,1}(P^{k,a,m}_\ad))
\end{equation}
for computing the phase $e(g)$ in \eqref{eqn:Zemg}.
Using \eqref{eqn:disF} for summing over $e\in\be(Y,G)$, we get
\[  Z_{Y,G}^m(\beta)=\msum{a\in H^1(Y,\pG)}{}\;\msum{k\in H^3(Y,\pi_3(G))}{}
    Z_{Y,G}^{k,a,m}(\beta),   \]
in which only honest $G$-bundles remain.
The partition function of the entire gauge theory is therefore
\[  Z_{X,G}=|\im(\del^1_Y)|\msum{m\in\bm(Y,G)}{}Z_{Y,G}^m(\beta),  \]
where $|\im(\del^1_Y)|$ is the size of the coset $m\in\bm(Y,G)$.
So canonical quantisation, after summing over various sectors labelled by
$(e,m)$, agrees with the manifestly relativistic invariant path integral
expression \eqref{eqn:ZXG}.

Finally, the relation \eqref{eqn:ZGGad} to the $G_\ad$-theory is compatible
with canonical quantisation. 
The $G_\ad$-bundle $P^m_\ad:=P^m/\ZG$ over $Y$ has
$\xi(P_\ad^m)=i_Y^2(\xi(P^m))=m$.
As in the derivation of \eqref{eqn:ZGGad}, we have
\[ Z_{Y,G}^{k,a,m}(\beta)
   =\mfrac{|\ker(\del^1_X)|}{|\ZG|}\,Z_{Y,G_\ad}^{k,a,m}(\beta)
   =|\ker(\del^1_Y)|\,Z_{Y,G_\ad}^{k,a,m}(\beta).   \]
In agreement with \eqref{eqn:ZGGad}, if we sum over
$a\in H^1(Y,\pG)\cong\ker(j^1_Y)$ and $m\in\bm(Y,G)\cong\ker(j^2_Y)$, we have
\[ Z_{X,G}=|H^1(Y,\ZG)|\msum{(a,m)\in\ker(j^1_Y)\times\ker(j^2_Y)}{}\;
   \msum{k\in H^3(Y,\pi_3(G_\ad))}{}Z_{Y,G_\ad}^{k,a,m}(\beta).      \]

\subsection{$S$-duality of the rectified discrete fluxes}\label{sec:sdual}
Electric charges in a gauge theory with gauge group $G$ are in one-to-one
correspondence with irreducible representations of $G$, or characters of $G$.
If the spacetime is $4$-dimensional, magnetic charges in the theory form the
set $\Hom(\mathrm U(1),G)$ of homomorphisms from $\mathrm U(1)$ to $G$ modulo
conjugations by $G$; they are the cocharacters of $G$.
Given a homomorphism $\gam\colon\mathrm U(1)\to G$, we can construct, from
the Hopf fibration $S^3\to S^2$ or from the Dirac monopole of magnetic charge
$1$, a principal $G$-bundle over $S^2$ or a non-Abelian monopole on
$\bR^3\backslash\{0\}$ with a Yang-Mills connection \cite{AB,FL}.
Conjugations of $\gam$ by $G$ yield gauge equivalent bundles and connections.
This classification of solutions to the Yang-Mills equation on $S^2$ or on
$\bR^3\backslash\{0\}$ by magnetic charges is more refined than the
topological classification of $G$-bundles \cite{Ka06}.

In \cite{GNO}, Goddard, Nyuts and Olive introduced a dual, magnetic group of 
$G$, which is, by an observation of Atiyah (cf.~\cite{KW}), the Langlands
dual group $\LG$.
The electric charges in a gauge theory with gauge group $G$ are in
one-to-one correspondence with the magnetic charges in another gauge theory
with gauge group $\LG$, and vice versa.
A simple way to see this correspondence is by the natural identifications
\[ \Hom(T,\mathrm U(1))/W\simeq\Hom(\mathrm U(1),\LT)/\LW,\qquad
   \Hom(\mathrm U(1),T)/W\simeq\Hom(\LT,\mathrm U(1))/\LW,   \]
where $T,\LT$ are the respective maximal tori of $G,\LG$ and $W,\LW$ are the
respective Weyl groups.

A further step is the electric-magnetic duality (also called $S$-duality)
conjecture of Montonen and Olive \cite{MO}.
It states that a gauge theory with gauge group $G$ is, at the quantum level,
isomorphic to another gauge theory with gauge group $\LG$, in the sense that
we can match the states in the quantum Hilbert spaces and the operators acting
on them, exchanging electricity and magnetism in the two theories.
If $G$ is simple, under the $S$-duality which exchanges $G$ and $\LG$, the
complex coupling constant $\tau:=\frac\tht{2\pi}+\frac{4\pi\ii}{e^2}$ (in the
upper half plane) goes to $^L\!\tau=-\frac1{n_\fg\tau}$, where $n_\fg=1$ if
$\fg$ is simply-laced and $n_\fg=2,3$ is the ratio of the square lengths of
the long and short roots if $\fg$ is not \cite{DFHK}.
The duality transformation $S\colon\tau\mapsto-\frac1{n_\fg\tau}$ and the
transformation $T\colon\tau\mapsto\tau+1$ (periodicity in the theta angle)
generate an action on $\tau$ of the modular group $\mathrm{SL}(2,\bZ)$ if
$n_\fg=1$ \cite{Sen} and the Hecke group $G(\sqrt{n_\fg})$ if $n_\fg>1$
\cite{AKS}.
$S$-duality is more plausible in supersymmetric gauge theories \cite{WO} and
is believed to be exact in the pure $N=4$ gauge theory \cite{Os}.
That $S$-duality in various twisted $N=4$ gauge theories makes remarkable
predictions that are compatible with highly non-trivial mathematical results
\cite{VW,KW} is a strong indication of its validity.

We now establish the matching under $S$-duality of the rectified discrete
electric and magnetic fluxes in \S\ref{sec:em4d}.
Consider a gauge theory on $X=T^1\times Y$, where $Y$ is a closed orientable
$3$-manifold.
If the gauge group $G$ is compact and semisimple, there are isomorphisms of
finite Abelian groups
\begin{equation}\label{eqn:LZG}
\pi_1(\LG)\cong\ZG^\vee,\quad Z(\LG)\cong\pG^\vee,
\end{equation}
and the dual of the exact sequence \eqref{eqn:piG} is the same sequence but
for $\LG$.
{}From \eqref{eqn:Poin}, we have natural isomorphisms
\begin{equation}\label{eqn:Ydual}
H^2(Y,\pi_1(\LG))\cong H^1(Y,\ZG)^\vee,\qquad
H^1(Y,Z(\LG))^\vee\cong H^2(Y,\pG).
\end{equation}
If $\del^1_Y=0$ in \eqref{eqn:long}, then $H^1(Y,\ZG)^\vee$ and $H^2(Y,\pG)$
are respectively the sets of discrete electric and magnetic fluxes, and they
are exchanged under $S$-duality by the isomorphisms \eqref{eqn:Ydual}.
More generally, the rectified discrete electric and magnetic fluxes are
elements of the sets $\be(Y,G)$ and $\bm(Y,G)$ defined in \S\ref{sec:em4d}.
For a gauge theory with the dual gauge group $\LG$, the rectified discrete
fluxes are in $\be(Y,\LG)=H^1(Y,Z(\LG))^\vee/\im((\Ldel^1_Y)^\vee)$ and
$\bm(Y,\LG)=H^2(Y,\pi_1(\LG))/\im(\Ldel^1_Y)$, where $\Ldel^1_Y$ is the
connecting homomorphism in the long exact sequence
\begin{equation}\label{eqn:Llong}
\scalebox{.9}[1]{$0\to H^1(Y,\pG)\map{\Li^1_Y}H^1(Y,Z(\tLG))
\map{\Lj^1_Y}H^1(Y,Z(\LG))\map{\Ldel^1_Y}H^2(Y,\pi_1(\LG))\map{\Li^2_Y}
H^2(Y,Z(\tLG))\map{\Lj^2_Y}H^2(Y,Z(\LG))\to0$}
\end{equation}
for the Langlands dual group $\LG$.
Under the isomorphisms \eqref{eqn:Ydual}, the sequence \eqref{eqn:Llong} is
the dual of \eqref{eqn:long} and $\Ldel^1_Y$ can be identified with
$(\del^1_Y)^\vee$ (see \S\ref{sec:Poin}).
Thus, taking the dual of \eqref{eqn:exactem}, we obtain two natural
isomorphisms
\begin{equation}\label{eqn:Srec}
\bm(Y,\LG)\cong\be(Y,G),\qquad\be(Y,\LG)\cong\bm(Y,G).
\end{equation}
Another way to see the above isomorphisms is through the counterparts of
\eqref{eqn:bm}, \eqref{eqn:be} for $\LG$, which are
\[ \bm(Y,\LG)\cong\im(\Li^2_Y)=\ker(\Lj^2_Y),\qquad
   \be(Y,\LG)\cong\im((\Lj^1_Y)^\vee)=\ker((\Li^1_Y)^\vee),  \]
and the identification of maps $\Li^k_Y=(j^{3-k}_Y)^\vee$,
$\Lj^k_Y=(i^{3-k}_Y)^\vee$, $k=1,2$ under \eqref{eqn:Ydual}
(see again \S\ref{sec:Poin}).

Now suppose the spacetime $4$-manifold $X$ is not necessarily of the product
form $T^1\times Y$.
As in \eqref{eqn:ZGGad}, the partition function of the theory with gauge group
$G$ contains a sum over $\ker(j_X^2)\subset H^2(X,\pGad)$, the set of discrete
fluxes of $G_\ad$-bundles on $X$ whose structure group can be lifted to $G$.
However, $\ker(j_X^2)$ is not respected by $S$-duality.
Instead, when $G$ is exchanged with $\LG$, $\ker(^L\!j_X^2)$ becomes
complimentary to $\ker(j_X^2)$ in the sense that $\ker(^L\!j_X^2)$ can be
identified with the set of homomorphisms from $H^2(X,\pGad)$ to $\mathrm U(1)$
that are trivial on $\ker(j_X^2)$.
That is,
\begin{equation}\label{eqn:SdualX}
\ker(^L\!j_X^2)\cong\ker(H^2(X,\pGad)^\vee\to\ker(j_X^2)^\vee)
\cong\Big(\!\sfrac{H^2(X,\pGad)}{\ker(j_X^2)}\!\Big)^{\!\vee}.
\end{equation}
So the smaller $\ker(j_X^2)$ is in the original theory, the larger
$\ker(^L\!j_X^2)$ becomes in the dual theory, and vice versa.
For example, if $G=\tG$ is simply connected, then $\ker(j_X^2)=0$.
But since $\LG=(\LG)_\ad$ and $^L\!j_X^2=0$, the set 
$\ker(^L\!j_X^2)=H^2(X,\pi_1((\LG)_\ad))$ is maximally possible.
On the contrary, if $G=G_\ad$, then $\ker(j_X^2)=H^2(X,\pGad)$ but
$\ker(^L\!j_X^2)=0$.
To verify \eqref{eqn:SdualX}, we note that $\ker(j_X^2)=\im(i_X^2)$ and that
the Pontryagin dual of the maps $i_X^2$, $j_X^2$ are $^L\!j_X^2$, $^L\!i_X^2$,
respectively (cf.~\S\ref{sec:Poin}).
Therefore the right hand side of \eqref{eqn:SdualX} equals
$(\coker(i_X^2))^\vee\cong\ker((i_X^2)^\vee)=\ker(^L\!j_X^2)$.

\section{Reduction to two dimensions along an orientable surface}\label{sec:2d}
\subsection{Reduction from four to two dimensions}\label{sec:4d2d}
The $N=4$ supersymmetric gauge theory in four dimensions admits three
inequivalent twists \cite{Ya,Ma}.
In the one that is related to the geometric Langlands programme there are two
independent supersymmetric transformations, $\del_l$ and $\del_r$, that can be
defined on an arbitrary curved spacetime $4$-manifold $X$ \cite{KW}.
So there is a family of topological field theories parametrised by
$t\in\bC\cup\{\infty\}=\CP^1$, each with a BRST operator
$\del_t=\del_l+t\del_r$.
Among the bosonic fields are a gauge field $A$ (connection on a $G$-bundle
$P$), a `Higgs' field $\phi\in\Om^1(X,\ad P)$ and a complex scalar field
$\sig\in\Om^0(X,\ad\,P^\bC)$.
The transformation $\del_t$ satisfies $\del_t^2=-\ii(1+t^2)\pounds_\sig$,
where $\pounds_\sig$ the infinitesimal gauge transformation on fields
generated by $\sig$ \cite{KW}.
If $G$ is simple, the action is
\begin{align}\label{eqn:action}
S(A,\phi,\sig,\dots)
&=\del_tV+\mfrac{\ii\,\PSI}{4\pi}\mint{X}{}\tr F_A\wedge F_A    \nonumber\\
&=-\mfrac1{e^2}\mint{X}{}\tr\big(\cF\wedge*\overline{\cF}
   +d_A^*\phi\wedge*\,d_A^*\phi\big)
  +\mfrac{\ii\,\tht}{8\pi^2}\mint{X}{}\tr F_A\wedge F_A+\cdots,
\end{align}
where $\cF:=(F_A-\frac12[\phi,\phi])+\ii\,d_A\phi$ is the curvature of the
connection $A+\ii\phi$ on the complexified bundle $P^\bC:=P\times_GG^\bC$,
and only the terms with $A,\phi$ are displayed in \eqref{eqn:action}.
The canonical parameter \cite{KW}
\begin{equation}\label{eqn:Ps}
\PSI=\mfrac\tht{2\pi}+\mfrac{4\pi\ii}{e^2}\cdot\mfrac{t-t^{-1}}{t+t^{-1}}
=\mfrac{\tau t+\bar\tau t^{-1}}{t+t^{-1}}
\end{equation}
takes values in the entire $\CP^1$ (not just the upper half plane).
It is real (i.e., in the $\bR P^1$ inside $\CP^1$) if and only if $|t|=1$;
in particular, $\PSI=\infty$ if $t=\pm\ii$.
The theory depends only on $\PSI$ because the $t$-dependence of $\del_t$ can
be eliminated by a redefinition of $\del_t$, $\del'_t:=(1+t^2)^{-1/2}\del_t$
when $t\ne\pm\ii$, satisfying $(\del'_t)^2=-\ii\pounds_\sig$ \cite{KW}.

Suppose now that the spacetime manifold is $X=\Sig\times C$, where $C$ is a
closed orientable surface of small size while the surface $\Sig$ is also
orientable and is either open or closed but of large size.
At low energies, fields on $X$ have to achieve (or nearly achieve) minimal
energy along $C$ but can be slowly varying along $\Sig$.
So a gauge theory on $X$ reduces to a sigma-model on the worldsheet $\Sig$
\cite{BJSV,HMS}.
For the twisted $N=4$ gauge theory with the action \eqref{eqn:action}, the
equations for minimal energy configuration along $C$ are, for all $t\in\CP^1$,
Hitchin's equations \cite{Hi87}
\begin{equation}\label{eqn:H}
F_A=\tfrac12[\phi,\phi],\quad d_A*\phi=0,\quad d_A\,\phi=0,
\end{equation}
where $A$ is a connection on a principal $G$-bundle $P$ over $C$ and
$\phi\in\Om^1(C,\ad\,P)$.
The Hitchin moduli space $\MH(C,G)$ is the space of pairs $(A,\phi)$
satisfying \eqref{eqn:H} modulo gauge transformations.
Among the low energy degrees of freedom is a map $u\colon\Sig\to\MH(C,G)$.
The fundamental relation between theories in four and two dimensions is
\cite{KW}
\begin{equation}\label{eqn:pull}
P_\ad\cong(u\times\id_C)^*\cU,
\end{equation}
where $\cU:=\UH(C,G)\to\MH(C,G)\times C$ is the universal bundle with
structure group $G_\ad$.
The connection $A$, curvature $F_A$ and the `Higgs' field $\phi$ in the gauge
theory on $X$ are the pull-backs via the map
$u\times\id_C\colon X\to\MH(C,G)\times C$ of the corresponding universal
objects $A^\cU$, $F^\cU$, $\phi^\cU$ of $\cU$ (see \S\ref{sec:univ}).

We assume that $C$ is of genus $g(C)>1$.
Then a generic flat $G$-connection or Hitchin pair on $C$ is irreducible and
its stabiliser is $\ZG$.
So it represents a smooth point on the moduli space.
We shall focus on the smooth part of $\MH(C,G)$ (which we denote by the same
notation) as near singular points, there are extra light degrees of freedom
that requires a more intricate study.
The space $\MH(C,G)$ is hyper-K\"ahler \cite{Hi87} and is of real dimension
$4(g(C)-1)\dim G$.
Following \cite{KW}, we let $I$ be the complex structure on $\MH(C,G)$ induced
by that on $C$, $J$ be the rotation from $\del A$ to $\del\phi$ (both are in
$\Om^1(C,\ad P)$), and $K=IJ$.
Let $\om_I$, $\om_J$, $\om_K$ be the corresponding K\"ahler forms.
The sigma-model metric on the target space $\MH(C,G)$ is
$\frac{4\pi}{e^2}=\Im\tau$ times the standard hyper-K\"ahler metric \cite{KW}.

The theta term in the action \eqref{eqn:action} reduces to a $B$-field given
by a globally defined $2$-form
$B_\tht=-\frac\tht{2\pi}\om_I=-(\Re\tau)\,\om_I$ in the sigma-model \cite{KW}.
A conceptual construction of the $B$-field is to use the connection $A^\cU$
and curvature $F^\cU$ of the universal bundle $\cU$.
Let $B_\tht$ be the $2$-form on $\MH(C,G)$ obtained by integrating the
$4$-form $\frac\tht{8\pi^2}\tr F^\cU\wedge F^\cU$ along $C$.
Using the $(1,1)$-part of $F^\cU$ in \eqref{eqn:univ-F}, we can show that the
$2$-form $B_\tht$ has the same expression as above.
By \eqref{eqn:pull}, the theta term in gauge theory becomes
\begin{equation}\label{Bfield}
\mfrac{\ii\tht}{8\pi^2}\mint{X}{}\tr F\wedge F
=\mfrac{\ii\tht}{8\pi^2}\mint{\Sig}{}\mint{C}{}u^*\tr F^\cU\wedge F^\cU
=\ii\mint{\Sig}{}u^*B_\tht,
\end{equation}
justifying that $B_\tht$ is the $B$-field to which the sigma-model couples.
If $G$ is simply connected, then $\frac1{4\pi}\tr F^\cU\wedge F^\cU$
has integral periods and so does its integration along $C$.
Hence the holonomy $\exp(\ii\int_\Sig u^*B_\tht)$ is invariant under
$\tht\mapsto\tht+2\pi$ or $\tau\mapsto\tau+1$.
In fact the resulting change in the $B$-field
$B_\tht\mapsto B_\tht+\frac{\om_I}{2\pi}$ can be interpreted as a change of
trivialisations \cite{Mu} of a topologically trivial $B$-field on $\MH(C,G)$.
If $G$ is not simply connected, then the instanton number contribution can be
fractional \cite{VW,Wu08} and the class $[\om_I/2\pi]$ need not be integral
\cite{KW}.
So $\tau\mapsto\tau+1$ brings a non-trivial phase in both $4$- and
$2$-dimensional theories.

Generally, whether $B$ is closed or not, an $N=(2,2)$ supersymmetry requires a
pair (for the left and right movers on the worldsheet) of complex structures
$J_+,J_-$ on the target space.
They are parallel under connections preserving an Hermitian metric $g$ (with
respect to both $J_\pm$) but whose torsions are proportional to $\pm H$,
where $H:=dB$ \cite{GHR}.
These conditions are equivalent to having a pair of commuting (twisted)
generalised complex structures
\[  \cJ_\pm:=\mfrac12{J_+\pm J_-\hspace{2ex}-(\om_+^{-1}\mp\om_-^{-1})\choose
    \hspace*{0ex}\om_+\mp\om_-\hspace{3.5ex}-(J_+^t\pm J_-^t)},  \]
where $\om_\pm=gJ_\pm$, that form a generalised K\"ahler metric on the target
\cite{Gu}.
The theory can then be twisted in two ways to make topological sigma-models,
each of which depends only on one of $\cJ_\pm$ (cf.~\cite{W92,KL07,Pu}).
We choose one, say $\cJ:=\cJ_+$.
If $J_+=J_-$, the twisted theory is a $B$-model in the complex structure
$J_+$, whereas if $J_+=-J_-$, it is an $A$-model in a symplectic form 
proportional to $\om_+$.

If the target space is a hyper-K\"ahler manifold such as $\MH(C,G)$, there is
a family of complex structures \cite{KW}
\[  J_w:=\mfrac{1-\bar ww}{1+\bar ww}I+\ii\mfrac{w-\bar w}{1+\bar ww}J
         +\mfrac{w+\bar w}{1+\bar ww}K                                   \]
parametrised by $w\in\bC\cup\{\infty\}=\CP^1$.
Then for any pair $(J_+,J_-)$ of complex structures given by
$(w_+,w_-)\in\CP^1\times\CP^1$ and for any closed $B$-field, the conditions
for $N=(2,2)$ supersymmetry are automatically satisfied.
The theory is a $B$-model if $w_+=w_-$.
If $w_+\ne w_-$, the target space has a K\"ahler structure $(g,\om',J')$
given by
\begin{equation}\label{eqn:om-J}
\om':=\mfrac{(1+|w_+|^2)(1+|w_-|^2)}{2|w_+-w_-|^2}(\om_+-\om_-),\qquad
J':=\mfrac{(1+|w_+|^2)^{1/2}(1+|w_-|^2)^{1/2}}{2|w_+-w_-|}(J_+-J_-)
\end{equation}
and the theory is an $A$-model with the symplectic form $\om'$ by a $B$-field
transform of
\[  B':=(\om_++\om_-)(J_+-J_-)^{-1}=\mfrac1{2|w_+-w_-|^2}
    \scalebox{.9}[.9]{$\begin{vmatrix}
    \om_I & \om_J & \om_K \\ 1-|w_+|^2 & \ii(w_+-\bar w_+) & w_++\bar w_+ \\
    1-|w_-|^2 & \ii(w_--\bar w_-) & w_-+\bar w_- \end{vmatrix}$}.    \]

While the above is sufficient for the existence of a classical theory with
$N=(2,2)$ supersymmetry, the quantum theory is anomaly-free only if
$c_1(T_+^{1,0})+c_1(T_-^{1,0})=0$, where $T_\pm^{1,0}$ are the holomorphic
tangent bundles of the target space in the complex structures $J_\pm$,
respectively \cite{KL07}.
Equivalently, the $c_1$ of the $\ii$-eigenbundle of $\cJ$ should vanish.
If $J_+=J_-$ ($B$-model), this condition is satisfied if $c_1(T_+^{1,0})=0$
whereas if $J_+=-J_-$ ($A$-model), it is always satisfied \cite{W92}.
Moreover, there should be a nowhere zero pure spinor of the generalised
tangent bundle which is closed under the (twisted) de Rham operator
\cite{KL07}.
When the $B$-field is closed, the condition reduces to the definition of
generalised Calabi-Yau manifolds \cite{Hi03}, of which the usual Calabi-Yau
and symplectic manifolds are examples.

If the target is hyper-K\"ahler, the twisted theory is anomaly-free for all
pairs $(w_+,w_-)\in\CP^1\times\CP^1$.
If $w_+\ne w_-$, the target is, before the $B$-field transform, generalised
Calabi-Yau with a pure spinor $\exp(B'+\ii\,\om')$, where
\[ B'+\ii\,\om'=-\ii\Big(\mfrac{w_++w_-}{w_+-w_-}\om_I+
   \ii\mfrac{w_+w_-+1}{w_+-w_-}\om_J+\mfrac{w_+w_--1}{w_+-w_-}\om_K\Big). \]
As $w_\pm\to w\in\CP^1$, $B'+\ii\,\om'$ becomes proportional to a $(2,0)$-form
(with respect to $J_w$)
\[ \Om'_w:=\left\{\!\begin{array}{ll}
      2w\,\om_I+\ii(w^2+1)\om_J+(w^2-1)\om_K & \mbox{if } w\in\bC,\\
      \ii\,\om_J+\om_K & \mbox{if } w=\infty.
      \end{array}\right. \]
When $w_+=w_-=w$, the $B$-model target space is Calabi-Yau with a holomorphic
pure spinor $\exp(\ii\Om'_w)$.

The twisted $N=4$ gauge theory on $X=\Sig\times C$ with the BRST
transformation $\del_t$ ($t\in\CP^1$) reduces at low energies to the
topological sigma-model on $\Sig$ with $(w_+,w_-)=(-t,t^{-1})$ \cite{KW}.
The generalised complex structure is
\begin{equation}\label{eqn:Jt}
\cJ_t=\mfrac1{1+\bar tt}{\hspace*{6ex}-\ii(t-\bar t)J\hspace{8ex}
-(\Im\tau)^{-1}((1-\bar tt)\om_I^{-1}-(t+\bar t)\om_K^{-1})
\choose\hspace*{-12ex}
\Im\tau((1-\bar tt)\om_I-(t+\bar t)\om_K)\hspace{12ex}\ii(t-\bar t)J^t},
\end{equation}
taking into account the kinetic factor $\Im\tau$ from gauge theory.
If $t=\pm\ii$, $w_+=w_-=\mp\ii$, the $2$-dimensional theory is a $B$-model
in the complex structures $\pm J$.
If $t\in\bR\cup\{\infty\}$, the theory is an $A$-model; for example, the
symplectic structure is $\pm(\Im\tau)\,\om_K$ if $t=\mp1$ and
$\pm(\Im\tau)\,\om_I$ if $t=0,\infty$.
For other values of $t$, the theory is equivalent to an $A$-model with a
symplectic form $\om_t$ upon a $B$-field transform by $B_t$, where \cite{KW}
\begin{equation}\label{eqn:omt}
\om_t:=(\Im\tau)\mfrac{1-\bar t^2t^2}{(1+t^2)(1+\bar t^2)}
       \Big(\om_I-\mfrac{t+\bar t}{1-\bar tt}\,\om_K\Big),\quad
B_t:=-(\Im\tau)\mfrac{\ii(t^2-\bar t^2)}{(1+t^2)(1+\bar t^2)}
	  \Big(\om_I+\mfrac{1-\bar tt}{t+\bar t}\,\om_K\Big).
\end{equation}
As expected, these $2$-dimensional theories are anomaly-free.
Since $\om_K$ is exact, the complexified K\"ahker class is
\[  [\,B_t+\ii\,\om_t\,]=-\ii(\Im\tau)\mfrac{t-t^{-1}}{t+t^{-1}}\,[\om_I].  \]
Together with the $B$-field $B_\tht$ from the theta term, the cohomology class
is $[B_\tht+B_t+\ii\,\om_t]=-\PSI[\om_I]$, showing that $\PSI$ is the relevant
parameter in the $2$-dimensional theory as well \cite{KW}.
These $2$-dimensional topological field theories depend only on $J$, $\om_I$,
$\om_K$ which can be defined without choosing a complex structure on $C$.
This reflects the metric independence of the $4$-dimensional topological
theories.

\subsection{Two-dimensional interpretation of the discrete fluxes}
\label{sec:e2d}
Suppose the worldsheet of the sigma-model has a splitting
$\Sig=T^1\times S^1$, where the circle $T^1$ is in the time direction and
$S^1$ is in the spatial direction.
Then so does the $4$-dimensional spacetime $X=T^1\times Y$ as in
\S\ref{sec:em4d}, where $Y=S^1\times C$ is a closed orientable $3$-manifold.
Recall that $H^2(X,\pG)=H^1(Y,\pG)\oplus H^2(Y,\pG)$.
We write $a=a_0+a_1$ according to $H^1(Y,\pG)=H^0(C,\pG)\oplus H^1(C,\pG)$
and $m=m_0+m_1\in H^2(Y,\pG)$ according to
$H^2(Y,\pG)=H^2(C,\pG)\oplus H^1(C,\pG)$ \cite{KW}.
Since $H_1(Y,\bZ)$ is torsion-free in this case, the action of a discrete
symmetry group element $g\in H^1(Y,\ZG)$ does not change the topology of
$G$-bundles over $Y$.
If we write $g=g_0+g_1$ according to $H^1(Y,\ZG)=H^0(C,\ZG)\oplus H^1(C,\ZG)$,
then $g_0$ and $g_1$ modify the holonomies of a $G$-bundle $P\to Y$ along
$S^1$ and $C$, respectively.
They both preserve the topology of the bundles or equivalently, the connected
components of $\cB(Y,G)$ and hence of $\MH(C,G)$.
The discrete electric and magnetic fluxes are thus in
$\be(Y,G)=H^1(Y,\ZG)^\vee$ and $\bm(Y,G)=H^2(Y,\pG)$ without
the adjustment in \S\ref{sec:em4d}.
We write $e=e_0+e_1$ according to
$H^1(Y,\ZG)^\vee=H^0(C,\ZG)^\vee\oplus H^1(C,\ZG)^\vee$.

In the sigma-model, $m_0\in H^2(C,\pG)$ labels the connected component
$\cM:=\MH^{m_0}(C,G)$ in which the map $u$ takes values and
$m_1\in H^1(C,\pG)\cong\Hom(\pi_1(S^1),\pi_1(\cM))$ labels the homotopy type
of $u$ at a fixed time \cite{KW}.
The symmetry group $H^1(C,\ZG)$ acts as discrete translations on $\cM$.
So the quantum Hilbert space further decomposes into sectors labelled by
discrete momenta $e_1\in H^1(C,\ZG)^\vee$.
In each sector of the sigma-model labelled by $e_1$, we integrate over maps
$u\colon[0,\beta]\times S^1\to\cM$ such that
$u(\beta,\cdot)=g_1\cdot u(0,\cdot)$ for some $g_1\in H^1(C,\ZG)$, weighted
by the phase $e_1(g_1)^{-1}\in\mathrm U(1)$.
If $g_1=0$, the homotopy classes of such maps along the time direction are in
the fundamental group $\pi_1(\cM)$.
In general, they form a set $\pi_1(\cM,g_1)$ which is a torsor over
$\pi_1(\cM)$.
The sums over $\pi_1(\cM,g_1)$ and $g_1\in H^1(C,\ZG)$ combine to a sum over
$a_1\in H^1(C,\pGad)$.
This is reminiscent of a similar pattern in gauge theory (see \S\ref{sec:Gad});
in fact, such a map $u$ yields a $G_\ad$-bundle $P_\ad$ via \eqref{eqn:pull}
with $\xi^{1,1}(P_\ad)=m_1+a_1$ and is weighted by the same of phase $e_1(a_1)^{-1}$
in gauge theory by \eqref{eqn:geomph}.
If we sum over $e_1$, then $a_1$ is restricted to $H^1(C,\pG)$ by
\eqref{eqn:dis}.
This means $g_1=0$ and $u$ becomes a map from the closed worldsheet $\Sig$
to $\cM$.
If we also sum over $m_1$, then we obtain a partition function
$Z_{\Sig,\cM}^{a_0}$ which is relativistic invariant in two dimensions.

Similarly, $g_0\in\ZG$ gives rise to $G$-bundles that are twisted along
$\Sig$.
But since it does not act on the target space $\cM$, it is not a symmetry
of the sigma-model in the usual sense.
Instead, each character $e_0\in\ZG^\vee$ determines a flat $B$-field
$e_0(\bar\xi(\cU))\in H^2(\cM,\mathrm U(1))$ on $\cM$ that the sigma-model
couples to, producing a phase that matches the one from the gauge theory
\cite{KW}.
Here $\bar\xi(\cU)\in H^2(\cM,\ZG)$ is the only non-zero component of
$\xi(\cU)\in H^2(\cM\times C,\pGad)$ after changing the coefficient group to
$\ZG$ (see \S\ref{sec:univ-c}).
For $e_0$ to be a true discrete parameter in two dimensions, we need to
verify that the $B$-fields $e_0(\bar\xi(\cU))$ on $\cM$ are different for
different $e_0$, that is, the map
$e_0\in\ZG^\vee\mapsto e_0(\bar\xi(\cU))\in H^2(\cM,\mathrm U(1))$ is injective.
In fact, the composition of this map with
$\vro^2_{\cM,\mathrm U(1)}\colon H^2(\cM,\mathrm U(1))\to\pi_2(\cM)^\vee$
in \eqref{eqn:vrho} is still an injective map $\ZG^\vee\to\pi_2(\cM)^\vee$
because it is Pontryagin dual to the surjective map
$j\circ\zeta_*\colon\pi_2(\cM)\to\ZG$ (see also \S\ref{sec:univ-c}).
This establishes $e_0\in\ZG^\vee$ as faithful parameter in the
$2$-dimensional theory.

Although the $4$-dimensional gauge theory requires
$a_0\in H^2(\Sig,\pG)\cong\pG$, we will show below that from the
$2$-dimensional point of view alone, $a_0$ can actually take values in a
larger set $H^2(\Sig,\pGad)\cong\pGad$, and each $a_0$ comes with a phase
$e_0(a_0)^{-1}$ in the path integral in the presence of the flat $B$-field
$e_0(\bar\xi(\cU))$ on $\cM$.
For each $e_0$,
\begin{equation}\label{eqn:e0a0}
Z_{\Sig,\cM}^{e_0}=\mfrac1{|\ZG|}
\msum{a_0\in\pGad}{}e_0(a_0)^{-1}Z_{\Sig,\cM}^{a_0}
\end{equation}
is already the partition function of a relativistic invariant theory on the
closed worldsheet $\Sig$.
The constraint $a_0\in\pG$ is enforced by \eqref{eqn:dis} only after summing
over $e_0\in\ZG^\vee$, which is not required in two dimensions but is
necessary to achieve relativistic invariance in four dimensions.
The resulting partition function is
\begin{equation}\label{eqn:e0circ}
Z^\dcirc_{\Sig,\cM}:=\msum{e_0\in\ZG^\vee}{}Z^{e_0}_{\Sig,\cM}
=\msum{a_0\in\pG}{}Z_{\Sig,\cM}^{a_0}.
\end{equation}

That $a_0\in\pG$ comes from summing over $e_0\in\ZG^\vee$ is a relativistic
invariant statement in two dimensions and is valid for worldsheets not
necessarily of the product form $T^1\times S^1$.
Let $\Sig$ be any closed orientable worldsheet.
The map $u\colon\Sig\to\MH(C,G)$ in sigma-model is related to the $G$-bundle
$P$ over $X=\Sig\times C$ in gauge theory by \eqref{eqn:pull}.
We write $\xi(P)=\xi^{2,0}(P)+\xi^{1,1}(P)+\xi^{0,2}(P)$ according to
$H^2(X,\pG)=H^2(\Sig,\pG)\oplus H^1(\Sig,H^1(C,\pG))\oplus H^2(C,\pG)$.
In the special case $\Sig=T^1\times S^1$, we have $\xi^{2,0}(P)=a_0$,
$\xi^{1,1}(P)=a_1+m_1$ and $\xi^{0,2}(P)=m_0$.
In general, let $[u]^{0,2}$ denote the element $m_0\in H^2(C,\pG)$ such
that the image of $u$ is contained in $\MH^{m_0}(C,G)$.
Then it follows easily from \eqref{eqn:pull} that $[u]^{0,2}=\xi^{0,2}(P)$,
agreeing with the above interpretation of $m_0$ when $\Sig$ is a product.
With $m_0$ fixed, we proceed to find the range of the parameter $a_0$ in the
sigma-model defined on any closed worldsheet $\Sig$.

To achieve this goal, we compute the set $[\Sig,\cM]$ of homotopy classes of
maps from $\Sig$ to $\cM:=\MH^{m_0}(C,G)$.
Since $\pi_1(\cM)=H^1(C,\pG)$ is Abelian and it acts trivially on $\pi_2(\cM)$
(see \S\ref{sec:MH}), the Puppe sequence \eqref{eqn:[S,Z]} is
\begin{equation}\label{eqn:[S,M]}
0\to\pi_2(\cM)\to[\Sig,\cM]\to H^1(\Sig,H^1(C,\pG))\to0.
\end{equation}
Denote the image of $[u]\in[\Sig,\cM]$ in $H^1(\Sig,H^1(C,\pG))$ by
$[u]^{1,1}$.
We first show the identity $[u]^{1,1}=\xi^{1,1}(P)$.
The isomorphism $\vro^1_{\Sig,\pi_1(\cM)}$ in \eqref{eqn:vrho} sends
$[u]^{1,1}$ to the induced homomorphism $u_*\colon\pi_1(\Sig)\to\pi_1(\cM)$
of $u$ on fundamental groups.
For the universal bundle $\cU\to\cM\times C$, we have
$\xi^{1,1}(\cU)\in H^1(\cM,H^1(C,\pGad)$ and
$\vro^1_{\cM^1,H^1(C,\pGad)}(\xi^{1,1}(\cU))$ is the map
$i_C^1\colon\pi_1(\cM)=H^1(C,\pG)\to H^1(C,\pGad)$ (see \S\ref{sec:univ-c}).
Therefore
\[ i_C^1\circ\vro^1_{\Sig,\pi_1(\cM)}([u]^{1,1})
   =\vro^1_{\Sig,H^1(C,\pGad)}(u^*\xi^{1,1}(\cU))
   =\vro^1_{\Sig,H^1(C,\pGad)}(\xi^{1,1}(P_\ad))
   =i_C^1\circ\vro^1_{\Sig,\pi_1(\cM)}(\xi^{1,1}(P)),  \]
where we used \eqref{eqn:pull} for the middle equality.
The result follows since $i_C^1$ is injective and $\vro^1_{\Sig,\pi_1(\cM)}$
is an isomorphism.

Next, we note the exact sequence \eqref{eqn:pi2M} for $\pi_2(\cM)$.
The sequence \eqref{eqn:[S,M]} splits partially because there is a natural
map $[\Sig,\cM]\to H^2(\Sig,\pGad)\cong\pGad$ sending the homotopy class
$[u]$ of a map $u\colon\Sig\to\cM$ to
$[u]^{2,0}:=u^*(\xi^{2,0}(\cU))\in H^2(\Sig,\pGad)\cong\pGad$.
The fact that $\vro^2_{\cM,\pGad}(\xi^{2,0}(\cU))\in\Hom(\pi_2(\cM),\pGad)$
is the map $\zeta_*$ in \eqref{eqn:pi2M} (see \S\ref{sec:univ-c}) shows that
$[u]\mapsto[u]^{2,0}$ is indeed a partial splitting of \eqref{eqn:[S,M]}.
Thus we can rewrite \eqref{eqn:[S,M]} as
\[  0\to H^2(C,\pi_3(G))\to[\Sig,\cM]\to H^2(\Sig,\pGad)\oplus
    H^1(\Sig,H^1(C,\pG))\to0,  \]
where $[u]\in[\Sig,\cM]$ maps to $([u]^{2,0},[u]^{1,1})$.
Using \eqref{eqn:pull} again, we obtain $[u]^{2,0}=\xi^{2,0}(P_\ad)$, which
is $a_0$ above, but now it takes values in $H^2(\Sig,\pGad)$.
The $4$-dimensional gauge theory imposes a constraint
$[u]^{2,0}\in H^2(\Sig,\pG)$.
The set $[\Sig,\cM]^\dcirc$ of such restricted homotopy classes fits in the
exact sequence
\[ 0\to H^2(C,\pi_3(G))\to[\Sig,\cM]^\dcirc\to H^2(\Sig,\pG)\oplus
    H^1(\Sig,H^1(C,\pG))\to0.  \]

We can identify the counterpart of the instanton number $k(P)$ in two
dimensions.
If $[u]^{1,1}=0$, then $u$ is homotopically trivial on the $1$-skeleton
$\Sig^{(1)}$ of $\Sig$.
So $u$ descends (up to homotopy) to a map $\bar u\colon S^2\to\cM$ and the
bundle $\bar P:=(\bar u\times\id_C)^*\cU$ over $S^2\times C$ pulls back to
$P$ over $\Sig\times C$.
If furthermore $[u]^{2,0}=0$, then $[u]$ reduces to an element $k(u)$ in
$H^2(C,\pi_3(G))$, which is a subgroup of $\pi_2(\cM)$ where $[\bar u]$
belongs.
In this case, $\bar P$ is topologically trivial along $S^2$ and can be
constructed by a loop in $\cG(P^{m_0})$ whose homotopy class is identified
with both $[\bar u]$ and the (appropriately normalised) instanton number
$k(\bar P)\in H^2(S^2\times C,\pi_3(G))\cong H^2(C,\pi_3(G))$
(see \S\ref{sec:univ-c}).
Pulling back from $S^2\times C$ to $X=\Sig\times C$, we obtain the matching
of $k(u)$ with $k(P)$.

With the restriction $a_0=[u]^{2,0}\in\pG$, the partition function of the
$2$-dimensional sigma-model is
\begin{equation}\label{eqn:Zsig}
Z^\dcirc_{\Sig,\cM}=
\msum{[u]\in[\Sig,\cM]^\scirc}{}\,\mint{}{}Du\cdots\,e^{-S(u,\dots)}.
\end{equation}
A map $u$ does not uniquely determine a $G$-bundle $P$ and its connection via
\eqref{eqn:pull}; there is a $|H^1(\Sig,\ZG)|$-fold ambiguity to reconstruct
$P$ from $P_\ad$ along $\Sig$.
In addition, in the partition function of the gauge theory, we divide by the
order of the centre $\ZG$ that acts trivially on the fields.
So the relation of the partition functions of the gauge theory (in the limit
of small $C$) and the sigma-model is
\begin{equation}\label{eqn:ZZ}
Z_{X,G}^{m_0}=\mfrac{|H^1(\Sig,\ZG)|}{|\ZG|}Z^\dcirc_{\Sig,\cM}
=|\ZG|^{b_1(\Sig)-1}Z^\dcirc_{\Sig,\cM},
\end{equation}
Because both theories are topological, \eqref{eqn:ZZ} is valid for any size
of $C$.

For each $e_0\in\ZG^\vee$, the partition function of the sigma-model coupled
to the flat $B$-field $e_0(\bar\xi(\cU))$ is
\begin{equation}\label{eqn:e0Sig}
Z^{e_0}_{\Sig,\cM}:=\msum{[u]\in[\Sig,\cM]}{}\,
\mint{}{}Du\cdots\,e^{-S(u,\dots)}\bra u^*(e_0(\bar\xi(\cU))),[\Sig]\ket^{-1},
\end{equation}
where $\bra u^*(e_0(\bar\xi(\cU))),[\Sig]\ket\in\mathrm U(1)$ is the holonomy
of the $B$-field on $\Sig$.
The sum is now over all homotopy classes in $[\Sig,\cM]$ as it should be
from the $2$-dimensional point of view.
This means that we allow $G$-bundles in gauge theory on $X=\Sig\times C$ that
are twisted along $\Sig$ by a discrete $B$-field
$u^*(\bar\xi(\cU))\in H^2(\Sig,\ZG)$; such a twisted $G$-bundle defines an
honest $G_\ad$-bundle $P_\ad$ over $X$ with $\xi^{2,0}(P_\ad)=a_0$
contributing the same phase $e_0(a_0)^{-1}$. 
So we have
\[ Z_{X,G}^{e_0,m_0}=\mfrac{|H^1(\Sig,\ZG)|}{|\ZG|}\,Z^{e_0}_{\Sig,\cM}
=|\ZG|^{b_1(\Sig)-1}Z^{e_0}_{\Sig,\cM}   \]
for all $e_0\in\ZG^\vee$ as a refinement of \eqref{eqn:ZZ}.
On the sigma-model side, we have \eqref{eqn:e0a0}, where $Z_{\Sig,\cM}^{e_0}$
comes from all $u\colon\Sig\to\cM$ with $[u]^{2,0}=a_0\in\pGad$.
Summing over $e_0$ and using \eqref{eqn:e0circ}, we obtain \eqref{eqn:ZZ}.
In this way, the gauge theory has honest $G$-bundles and the
sigma-model has maps with restricted homotopy classes $[\Sig,\cM]^\dcirc$.

We have thus completed the matching of homotopy classes of maps $u$ in the
$2$-dimensional sigma-model and topological types of $G$-bundles $P$ in the
$4$-dimensional gauge theory when $X=\Sig\times C$.

If the gauge group is $G_\ad$, consider the sigma-model whose target is
$\cM_\ad:=\MH^{m_0}(C,G_\ad)$.
If $m_0\in H^2(C,\pG)$, there is a regular $H^1(C,\ZG)$-covering
$q\colon\cM\to\cM_\ad$ of the smooth parts of the moduli spaces
\cite{HWW,Wu15} (see also \S\ref{sec:univ-c}).
A map $u_\ad\colon\Sig\to\cM_\ad$ lifts to $u\colon\Sig\to\cM$ if and only if
$(u_\ad)_*(\pi_1(\Sig))\subset q_*(\pi_1(\cM))$, and different lifts of a
given $u_\ad$ are related by the deck transformations of $q$.
Thus the partition function $Z^\dcirc_{\Sig,\cM}$ can be written as
\begin{equation}\label{eqn:ZZMad}
Z^\dcirc_{\Sig,\cM}=|H^1(C,\ZG)|\msum{[u_\ad]\in q_*([\Sig,\cM]^\scirc)}{}
\,\mint{}{}Du_\ad\cdots\,e^{-S(u_\ad,\dots)}.
\end{equation}
The above sum is over homotopy classes $[u_\ad]\in[\Sig,\cM_\ad]$ of maps
$u_\ad\colon\Sig\to\cM_\ad$ such that $[u_\ad]^{0,2}=m_0$,
$[u_\ad]^{1,1}\in H^1(\Sig,H^1(C,\pG))$ and $[u_\ad]^{2,0}\in H^2(\Sig,\pG)$.
The identity \eqref{eqn:ZZMad} is consistent with its gauge theoretic 
counterpart \eqref{eqn:ZGGad} via the relation \eqref{eqn:ZZ} and its
analogue for $G_\ad$.

\subsection{$S$-duality and mirror symmetry}\label{sec:S2d-o}
We recall the reduction along a closed orientable surface of $S$-duality in
four dimensions to mirror symmetry in two dimensions \cite{BJSV,HMS,KW}.
In $4$-dimensional gauge theories, the duality transformation on the complex
coupling $\tau$ is $S\colon\tau\mapsto-1/n_\fg\tau$.
Duality also acts non-trivially on supersymmetry generators because it acts
on the central charges of the extended supersymmetry algebra.
The transformation on the parameter $t\in\CP^1$ of topological gauge theories
is $S\colon t\mapsto-\frac\tau{|\tau|}t$ \cite{KW}.
Together with $T\colon\tau\mapsto\tau+1$ which acts trivially on $t$,
the transformations of the canonical parameter $\PSI$ are
$S\colon\PSI\mapsto-1/n_\fg\PSI$ and $T\colon\PSI\mapsto\PSI+1$.
For example, if $\PSI=\infty$, $t=\ii$ and $\tau$ is purely imaginary, the
dual theory has $^L\PSI=0$, $^Lt=1$ and $^L\tau$ is also purely imaginary.

For the sigma-model with a hyper-K\"ahler target space and parametrised by
$(w_+,w_-)\in\CP^1\times\CP^1$, the mirror theory has
$(^L\!w_+,{}^L\!w_-)=(-\frac\tau{|\tau|}w_+,-\frac{|\tau|}\tau w_-)$ \cite{KW}.
This is compatible with the above transformation on $t\in\CP^1$ when the
$2$-dimensional theory comes from the reduction of a $4$-dimensional theory,
i.e., when $(w_+,w_-)=(-t,t^{-1})$.
For example, if $t=\ii$ and $\tau$ is purely imaginary, the sigma-model is a
$B$-model in the complex structure $J$.
Its mirror, with $^Lt=1$, is an $A$-model with the symplectic form
$(\Im\tau)\om_K$.
This is the important special case in which $S$-duality in four dimensions
gives rise to the geometric Langlands programme when reduced to two dimensions
\cite{KW}.
Outside the family of dimensional reductions, the sigma-model with
$(w_+,w_-)=(0,0)$ is a $B$-model in the complex structure $I$.
Its mirror, with the same $(^L\!w_+,{}^L\!w_-)=(0,0)$, is again a $B$-model
in the complex structure $I$ \cite{KW}.

In four dimensions, $S$-duality interchanges the discrete fluxes $e_0$ and
$m_0$, as well as $e_1$ and $m_1$ \cite{KW}.
While they are invariant under $T$, the transformations under $S$ are
$(e_0,m_0)\mapsto(-m_0,e_0)$ and $(e_1,m_1)\mapsto(-m_1,e_1)$.
This is the discrete analogue of the Hodge star operation on the curvature
$2$-form.
For $2$-dimensional sigma-models, mirror symmetry exchanges the connected
components of Hitchin's moduli space with the non-trivial flat $B$-fields
and the windings in the moduli space with the discrete momenta.
Indeed, from \eqref{eqn:Poin}, there are natural isomorphisms
\begin{align}\begin{split}\label{eqn:Cdual}
H^0(C,\ZG)^\vee\cong H^2(C,\pi_1(\LG)),\quad&
H^2(C,\pG)^\vee\cong H^0(C,Z(\LG)), \\
H^1(C,\ZG)^\vee\cong H^1(C,\pi_1(\LG)),\quad&
H^1(C,\pG)^\vee\cong H^1(C,Z(\LG)),
\end{split}\end{align}
making the duality of the discrete parameters possible.
The mirror symmetry between $e_0$ and $m_0$ and between $e_1$ and $m_1$ can be
explained concretely from the geometry of the target space.

To see the interchange of $m_0$ and $e_0$, recall the Hitchin fibration
$h\colon\MH(C,G)\to\B$, where $\B$ is a complex vector space of dimension
$(g(C)-1)\dim G$, defined by evaluating the invariant polynomials of the Lie
algebra $\fg$ on the holomorphic part of the `Higgs' field $\phi$ \cite{Hi87}.
The map $h$ is holomorphic in the complex structure $I$.
If $b\in\B$ is regular or outside a discriminant divisor, the fibre
$\MH(C,G)_b:=h^{-1}(b)$ is smooth, of real dimension $2(g(C)-1)\dim G$,
and is special Lagrangian in the symplectic structures $\om_J,\om_K$ but
holomorphic in the complex structure $I$.
Each $\MH(C,G)_b$ is a disjoint union of affine complex tori
$\MH^{m_0}(C,G)_b$ labelled by $m_0\in\pG$.
For a general simple gauge group $G$, these affine tori are torsors over
an Abelian variety which is a generalised Prym variety \cite{DG}.
The torus $\MH^{m_0=0}(C,G)_b$ has a distinguished point once a Hitchin
section \cite{Hi92} of $\MH^{m_0=0}(C,G)\to\B$ is chosen.

Now consider the dual fibration $\Lh\colon\MH(C,\LG)\to\B$ with gauge
group $\LG$, whose discriminant divisor in $\B$ coincides with that of $h$.
If $b\in\B$ is regular, the fibre $\MH(C,\LG)_b:=\Lh^{-1}(b)$ is a disjoint
union of affine tori $\MH^{e_0}(C,\LG)_b$ labelled by $e_0\in\ZG^\vee$.
Each torus $\MH^{e_0}(C,\LG)_b$ is naturally identified with the set of
trivialisations of the flat $B$-field $e_0(\bar\xi(\cU))$ on $\MH(C,G)_b$
(see \cite{HT} for $G=\mathrm{SU}(n),\mathrm{PU}(n)$, $n=2,3$ and \cite{DP12}
for the general case).
Note that $e_0(\bar\xi(\cU))$ is trivial on $\MH(C,G)_b$, though not
naturally trivial.
But any two trivialisations of the $B$-field differ by a flat line bundle
\cite{Mu}.
Thus mirror symmetry between $\MH(C,G)$ and $\MH(C,\LG)$ is $T$-duality in the
sense of Strominger, Yau and Zaslow \cite{SYZ}, extended by \cite{Hi01,DP08}
in the presence of non-zero $B$-fields.
The exchange of $B$-fields with topology under $T$-duality is a general
pattern that also appears in \cite{BEM}.

We can rephrase the above in the language of sigma-models and branes.
The mirror symmetry between the two $B$-models in complex structure $I$
identifies the rings of holomorphic functions on $\B$ as observables of
both theories.
Consequently, if a brane in the original theory is supported on a fibre
$\MH(C,G)_b$, its mirror is supported on the dual fibre $\MH(C,\LG)_b$.
For example, a twisted flat line bundle $\ell$ over $\MH^{m_0=0}(C,G)_b$
trivialising the $B$-field $e_0(\bar\xi(\cU))$ defines a brane of type
$(B,A,A)$ in the terminology of \cite{KW}.
Its mirror is a brane of type $(B,B,B)$ supported on the point $\ell^\vee$ on
$\MH^{e_0}(C,\LG)_b$ representing $\ell$.
Generally, if $\cB$ is a $B$-brane in the complex structure $I$, the dual
brane $\cB^\vee$ can be obtained by a (fibrewise) Fourier-Mukai transform
\cite{Muk,DP08,BBH} that is holomorphic in $I$ \cite{KW,DP12}.

The moduli space $\Mf^{m_0}(C,G)$ of flat $G$-connections on $C$ is in the
most singular fibre $h^{-1}(0)\supset\Mf(C,G)$.
Suppose the $B$-field is given by $e_0=0$.
The trivial line bundle on $\Mf^{m_0}(C,G)$ defines a brane on $\MH(C,G)$ of
type $(B,A,A)$.
Its mirror is a brane of type $(B,B,B)$ supported on the most singular point
of $\MH^{e_0=0}(C,\LG)$ represented by the trivial $\LG$-connection on $C$
with ${}^L\!\phi=0$, though its internal structure at that point is not
geometrical \cite{G}.
More generally, a character of $\pi_1(\Mf^{m_0}(C,G))\cong H^1(C,\pG)$
defines a flat line bundle over $\Mf^{m_0}(C,G)$ and hence another brane of
type $(B,A,A)$ on $\Mf^{m_0}(C,G)$.
Since $\pi_1(\Mf^{m_0}(C,G))^\vee\cong H^1(C,Z(\LG))$, the character can be
identified with a homomorphism from $\pi_1(C)$ to $Z(\LG)$ or a point in
$\MH^{e_0=0}(C,\LG)$, represented by a flat $\LG$-connection whose holonomy
is in $Z(\LG)$ together with ${}^L\!\phi=0$.
This point is the support of the mirror brane of type $(B,B,B)$.

We study the Hitchin fibration when the gauge group changes.
Since the discrete symmetry group $H^1(C,\ZG)$ acts trivially on the `Higgs'
fields, it commutes with $h$ and preserves its fibres.
Each torus component $\MH^{m_0}(C,G)_b$, if $b\in\B$ is regular, is also
invariant under $H^1(C,\ZG)$.
Changing the gauge group to $G_\ad$, the Hitchin fibration
$h_\ad\colon\MH(C,G_\ad)\to\B$ satisfies $h=h_\ad\circ q$, where
$q\colon\MH(C,G)\to\MH^\dcirc(C,G_\ad)$ is the regular $H^1(C,\ZG)$-covering
onto the subset $\MH^\dcirc(C,G_\ad)$ of $\MH(C,G_\ad)$ from bundles whose
gauge group can be lifted to $G$, as in \eqref{eqn:union-ad}.
On each torus component $\MH^{m_0}(C,G)_b$, $q$ restricts to an (affine)
isogeny onto the torus $\MH^{m_0}(C,G_\ad)_b$ of
$\MH(C,G_\ad)_b:=h_\ad^{-1}(b)$ with kernel $H^1(C,\ZG)$.
The $\tG$-theory involves moduli spaces $\MH^{m_0}(C,\tG)$ of Hitchin pairs
from $\tG$-bundles twisted by the discrete $B$-field $m_0$
(cf.~\S\ref{sec:MH}); we have $\MH^{m_0=0}(C,\tG)=\MH(C,\tG)$.
Let $\MH^\dcirc(C,\tG)$ be the union of $\MH^{m_0}(C,\tG)$ over
$m_0\in H^2(C,\pG)$ as in \eqref{eqn:union-cov}.
The Hitchin fibration $\tilde h\colon\MH^\dcirc(C,\tG)\to\B$ is defined and
is  preserved by the action of $H^1(C,Z(\tG))$.
The quotient map $\tq\colon\MH^\dcirc(C,\tG)\to\MH(C,G)$ by $H^1(C,\pG)$
is a regular covering, and $\tilde h=h\circ\tq$.
The fibre $\MH(C,\tG)_b:=\tilde h^{-1}(b)$ is a disjoint union of the tori
$\MH^{m_0}(C,\tG)_b$.
The restriction of $\tq$ to $\MH^{m_0}(C,\tG)_b$ is an (affine) isogeny onto
$\MH^{m_0}(C,\tG)_b$ with kernel $H^1(C,\pG)$.

On the dual side, we have Hitchin maps $\Lh\colon\MH(C,\LG)\to\B$,
$\Lh_\ad\colon\MH(C,(\LG)_\ad)\to\B$ and
${}^L\!\tilde h\colon\MH(C,\tLG)\to\B$.
The regular $H^1(C,Z(\LG))$-covering
$\Lq\colon\MH(C,\LG)\to\MH^\dcirc(C,(\LG)_\ad)$ and
$H^1(C,\pi_1(\LG))$-covering
${}^L\!\tq\colon\MH^\dcirc(C,\tLG)\to\MH(C,\LG)$ satisfy
$\Lh=\Lh_\ad\circ\Lq$ and ${}^L\!\tilde h=\Lh\circ{}^L\!\tq$.
For a regular $b\in\B$, let $\MH^{e_0}(C,\LG)_b$, $\MH^{e_0}(C,(\LG)_\ad)_b$
and $\MH^{e_0}(C,\tLG)_b$ be the torus components in the fibres
$\MH(C,\LG)_b:=\Lh^{-1}(b)$, $\MH(C,(\LG)_\ad)_b:=\Lh_\ad^{-1}(b)$ and
$\MH(C,\tLG)_b:={}^L\!\tilde h^{-1}(b)$, respectively, labelled by
$e_0\in H^2(C,\pi_1(\LG))\cong H^0(C,\ZG)^\vee$.
The restrictions to the tori,
${}^L\!q\colon\MH^{e_0}(C,\LG)_b\to\MH^{e_0}(C,(\LG)_\ad)_b$ and
${}^L\!\tq\colon\MH^{e_0}(C,\tLG)_b\to\MH^{e_0}(C,\LG)_b$ are (affine)
isogenies dual to $\tq\colon\MH^{m_0}(C,\tG)_b\to\MH^{m_0}(C,G)_b$ and
$q\colon\MH^{m_0}(C,G)_b\to\MH^{m_0}(C,G_\ad)_b$, respectively;
their kernels $H^1(C,Z(\LG))$ and $H^1(C,\pi_1(\LG))$ are dual to the kernels
$H^1(C,\pG)$ and $H^1(C,\ZG)$ of $\tq$ and $q$ by \eqref{eqn:Cdual}.

The coverings $q$, $\tq$ also determine the mappings of branes when
the gauge group $G$ becomes $G_\ad$ or $\tG$.
If $\cB$ is a B-brane on $\MH(C,G)$ in the complex structure $I$, then the
brane in the $G_\ad$-theory is $q_*\cB$ and the brane in the $\tG$-theory is
$\tq^*\cB$.
Suppose $\cB$ is supported on $\MH^{m_0}(C,G)_b$, where $b$ is regular, and
the Chan-Paton bundle is the flat line bundle $\ell^e$ defined by
$e\in(\pi_1(\MH^{m_0}(C,G)_b))^\vee$.
By the exact sequence
\[ 0\to\pi_1(\MH^{m_0}(C,G)_b)\to\pi_1(\MH^{m_0}(C,G_\ad)_b)\to
   H^1(C,\ZG)\to0,      \]
the element $e$ can be identified as a coset in
$\pi_1(\MH^{m_0}(C,G_\ad)_b)^\vee/H^1(C,\ZG)^\vee$ and as such, it is a
torsor over $H^1(C,\ZG)^\vee$.
Then $q_*\ell^e=\bigoplus_{e_\ad\in e}\ell^{e_\ad}$, where $\ell^{e_\ad}$ is
the flat line bundle over $\MH^{m_0}(C,G_\ad)_b$ determined by
$e_\ad\in\pi_1(\MH^{m_0}(C,G_\ad)_b)^\vee$.
In the special case $e=0$, $q_*\ell^{e=0}$ is the restriction to
$\MH^{m_0}(C,G_\ad)_b$ of $\bigoplus_{e_1\in H^1(C,\ZG)^\vee}\ell_\ad^{e_1}$,
where $\ell_\ad^{e_1}:=\MH(C,G)\times_{e_1}\bC$ is a flat line bundle over
$\MH^\dcirc(C,G_\ad)$.
On the other hand, $\tq^*\cB$ is the brane whose Chan-Paton bundle is
$\tq^*\ell^e$ on $\MH^{m_0}(C,\tG)_b$.
If instead $\cB$ is the brane $\bC_a$ of rank $1$ supported at a single point
$a\in\MH^{m_0}(C,G)_b$, then $q_*\cB$ is the brane $\bC_{a_\ad}$ supported at
$a_\ad:=q(a)\in\MH^{m_0}(C,G_\ad)_b$ whereas $\tq^*\cB$ is supported on the
discrete set $\tq^{-1}(a)\subset\MH^{m_0}(C,\tG)_b$, which a torsor over
$H^1(C,\pG)$.

The above operations on branes under the changes of the gauge group are
compatible with duality.
Suppose $\cB$ is given by a flat line bundle $\ell$ over over a regular torus
component $\MH^{m_0}(C,G)_b$ twisted by a $B$-field $e_0(\bar\xi(\cU))$, then
the mirror brane is of rank $1$ and is supported at $\ell^\vee$ in the torus
$\MH^{e_0}(C,\LG)_b$ with a $B$-field $m_0(\bar\xi({}^L\!\cU))$.
Changing the gauge group of the original theory to $G_\ad$, the brane $\cB$
becomes $q_*\cB$ given by $q_*\ell$ as above.
In the dual theory, the gauge group is changed to $\tLG$ and the brane
$({}^L\!\tq)^*(\cB^\vee)$ in the $\tLG$-theory is supported on
$({}^L\!\tq)^{-1}(\ell^\vee)\subset\MH^{e_0=0}(C,\tLG)_b$.
The latter is clearly the mirror of $q_*\cB$.
If instead the gauge group is changed to $\tG$ in the original theory, then
the group of the dual theory is $(\LG)_\ad$.
The brane $\tq^*\cB$ is given by the pull-back bundle $\tq^*\ell$;
it is mirror to the brane $\Lq_*(\cB^\vee)$ supported at $\Lq(\ell^\vee)$.
In general, by fibrewise Fourier-Mukai transform \cite{Muk}, the identities
\begin{equation}\label{eqn:qLB}
(q_*\cB)^\vee=({}^L\!\tq)^*(\cB^\vee),\qquad
(\tq^*\cB)^\vee=(\Lq)_*(\cB^\vee)
\end{equation}
hold for any $B$-brane $\cB$ in the complex structure $I$.
This verifies the desired compatibility with duality.

The interchange of winding numbers $m_1$ and conserved momenta $e_1$ under
duality is standard.
Here we explain a related phenomenon which will appear more pertinently
in \S\ref{sec:mir-n}.
Suppose the two ends of an open string are on the same brane defined by a
twisted flat line bundle $\ell^e$ over $\MH^{m_0}(C,G)_b$.
Then as explained above, $q_*\ell^e=\bigoplus_{e_\ad\in e}\ell^{e_\ad}$.
The $e_1$-sector of the theory is described by an open string in the
$G_\ad$-theory whose two ends are on the branes $\ell^{e_\ad}$ and
$\ell^{e'_\ad}$ such that $e'_\ad-e_\ad=e_1$.
In the mirror $\tLG$-theory, the dual brane of $\ell^{e_\ad}$ is supported
on the point $(\ell^{e_\ad})^\vee\in\MH^{e_0}(C,\tLG)_b$.
So the two ends of the open string in the $\tLG$-theory are related by the
action of $e_1$ on $\MH^{e_0}(C,\tLG)_b$, and in the $\LG$-theory, the
string is closed but has the winding number $e_1$.
Likewise, a string with a winding number $m_1$ in the original $G$-theory
lifts to an open string in the $\tG$-theory whose two ends are related by
a discrete translation $m_1$.
This corresponds to the tensoring of a flat line bundle determined by $m_1$
in the dual $(\LG)_\ad$-theory, and therefore the $\LG$-theory is in the
$m_1$-sector.
The matching of translations on one side and tensoring of line bundles on
the dual side, and vice versa, is another hallmark of the Fourier-Mukai
transform \cite{Muk}.

\section{Non-orientable surfaces, branes and duality}\label{sec:non-or}
In this section, $C'$ is a compact non-orientable surface with a conformal
structure.
Given a conformal structure on $C'$, the Hodge star $*$ on the $1$-forms on
$C'$ is defined up to a sign, but this is harmless to Hitchin's equations in
\eqref{eqn:H}.
So we do have a moduli space $\MH(C',G)$ of solutions to \eqref{eqn:H} on $C'$
up to gauge equivalence \cite{HWW} (see \S\ref{sec:MHn} for a summary and
further developments).
We will however avoid gauge theories on non-orientable $4$-manifolds, which do
not allow the full range of the theta angle or the full $\mathrm{SL}(2,\bZ)$
duality, although $S$-duality for Abelian gauge theories on a non-orientable
spacetime was recently studied in \cite{Me}.
Instead, we consider the twisted $N=4$ gauge theory on a suitably constructed
orientable $4$-manifold so that its reduction to two dimensions is a
sigma-model in which the moduli space $\MH(C',G)$, where $C'$ is
non-orientable, plays a role.

\subsection{Dimensional reduction from an orientable $4$-manifold}
\label{sec:4dn2d}
An orientable $4$-manifold that contains a non-orientable surface $C'$ can not
be a product space $\Sig\times C'$ as it would be non-orientable whether the
surface $\Sig$ is orientable or not.
Instead, we use the orientation double cover $\pi\colon C\to C'$ whose
non-trivial deck transformation $\iota$ is an orientation reversing
involution on $C$.
If $C'$ is a connected sum of $g(C')$ copies of $\bR P^2$, then $C$ is a
compact surface of genus $g(C)=g(C')-1$.
For example, $C=S^2$, $g(C)=0$ if $C'=\bR P^2$, $g(C')=1$ and $C=T^2$,
$g(C)=1$ if $C'$ is the Klein bottle, $g(C')=2$.
On the other hand, let $\tSig$ be a closed orientable surface with an
orientation reversing involution, also denoted by $\iota$, and let
$\Sig=\tSig/\iota$.
We assume that the fixed point set $\tSig^\iota$ of $\iota$ on $\tSig$ is
non-empty and $\Sig$ is an orientable surface whose boundary is identified
with $\tSig^\iota$.
For example, $\tSig=S^2$ if $\Sig$ is a disc (with one boundary circle) and
$\tSig=T^2$ if $\Sig$ is a cylinder (with two boundary circles).
We choose our spacetime manifold as $X=\tSig\times_\iota C$.
Since the diagonal action of $\iota$ on $\tSig\times C$ is free and orientation
preserving, the quotient space $X$ is a closed orientable smooth $4$-manifold.
Globally, $X$ is not a product of two surfaces, but there is a projection map
$\pi_X\colon X\to\Sig$ (by forgetting $C$).
If $\sig\in\Sig_0$, the interior of $\Sig$, then $\pi_X^{-1}(\sig)$ is a copy
of $C$.
But if $\sig\in\bSig$, then $\pi_X^{-1}(\sig)$ is a copy of $C'$.

We shall consider the dimensional reduction of the twisted $N=4$ gauge theory
on $X$ with a compact, semisimple gauge group $G$.
We can find a metric on $X$ so that $\Sig$ is large but $C$ and $C'$ are
small.
In fact, choosing $\iota$-invariant metrics on $C$ and on $\tSig$ such that
$\tSig$ is large but $C$ is small, the product metric on $\tSig\times C$
descends to such a metric on $X$.
If $C'$ is the connected sum of $g(C')>2$ copies of $\bR P^2$ or if $g(C)>1$,
a generic flat $G$-connection on $C'$ or on $C$ is irreducible.
At low energies, for the same reason as in \cite{BJSV,HMS,KW}, the fields
$A,\phi$ have to satisfy Hitchin's equations along the fibres $C$ or $C'$ of
$\pi_X$.
So the gauge theory on $X$ reduces to a sigma-model on $\Sig$: the interior
$\Sig_0$ is mapped to $\MH(C,G)$ (we denote the map by $u$) while the
boundary $\bSig$ is mapped to $\MH(C',G)$ (we denote the map by $u'$).
To stay at low energies, as a point $\sig\in\Sig_0$ goes to $\sig'\in\bSig$,
$u(\sig)$ must approach the pull-back to $C$ of the Hitchin pair on $C'$
given by $u'(\sig')$.
Therefore $u$ extends to a map from $\Sig=\Sig_0\cup\bSig$ to $\MH(C,G)$
such that $u|_{\bSig}=p\circ u'$, where
$p\colon\MH(C',G)\to\MH(C,G)$ pulls back bundles, connections and sections
from $C'$ to $C$.
The image of $p$, denoted by $\cN(C,G)$, is part of the $\iota$-invariant
subspace $\MH(C,G)^\iota$.
The low energy theory is a sigma-model on the worldsheet $\Sig$ with
boundary, with a pair of maps
$(u,u')\colon(\Sig,\bSig)\to(\MH(C,G),\MH(C',G))$ in the sense of
\S\ref{sec:rel}.
The map $p\colon\MH(C',G)\to\cN(C,G)$ is a regular $Z(G)_{[2]}$-covering
\cite{HWW} (cf.~\S\ref{sec:MHn}) and defines a brane which wraps around its
world-volume $\cN(C,G)$ non-trivially.
In this way, the sigma-model gains extra degrees of freedom at the boundary.

Conversely, given a pair of maps $(u,u')$ as above, we can reconstruct the
adjoint bundle $P_\ad$ over $X$ in gauge theory.
Note that $X=X_0\cup X_1$, where $X_0:=\pi_X^{-1}(\Sig_0)$, $\Sig_0$ is the
interior of $\Sig$ as above and $X_1:=\pi_X^{-1}(\Sig_1)$, $\Sig_1$ is a
tubular neighbourhood of $\bSig$ in $\Sig$.
Then $X_0$ is homeomorphic to $\Sig_0\times C$, $X_1$ is homotopic to
$\bSig\times C'$ via a deformation retract from $\Sig_1$ to $\bSig$, and
$X_0\cap X_1$ is homotopic to $\bSig\times C$.
Recall from \S\ref{sec:univ-c} the universal bundles
$\cU\to\MH(C,G)\times C$ and $\cU'\to\MH(C',G)\times C'$.
Consider the $G_\ad$-bundles $(u|_{\Sig_0}\times\id_C)^*\cU$ over $X_0$ and
$(u'\times\id_{C'})^*\cU'$ over $\bSig\times C'$ and hence over $X_1$ via the
retract.
By \eqref{eqn:ppull}, we have a bundle isomorphism
\[ (\id_\bSig\times\pi)^*(u'\times\id_{C'})^*\cU'=(u'\times\id_C)^*
 (\id_{\MH(C',G)}\times\pi)^*\cU'\cong(u'\times\id_C)^*(p\times\id_C)^*\cU
 =(u|_\bSig\times\id_C)^*\cU,   \]
showing that the above two $G_\ad$-bundles over $X_0$ and $X_1$ agree on the
overlap $X_0\cap X_1$.
Thus we obtain a $G_\ad$-bundle $P_\ad$ over $X$ from $(u,u')$, generalising
\eqref{eqn:pull} when the worldsheet $\Sig$ has a boundary.

The presence of branes whose world-volume is $\cN(C,G)$ is compatible with
the supersymmetry $\del_t$ that is made a BRST transformation in the
topological field theory.
Recall that the $4$-dimensional theories are parametrised by $t\in\CP^1$ or
by the canonical parameter $\PSI$ in \eqref{eqn:Ps}.
When $t=\pm\ii$ or equivalently when $\PSI=\infty$, the reduction to two
dimensions is a $B$-model in the complex structure $\pm J$.
For other values of $t$ or $\PSI$, the reduction is an $A$-model
(possibly after a $B$-field transform) with the symplectic form $\om_t$ in
\eqref{eqn:omt}, which is a linear combination of $\om_I$ and $\om_K$.
Happily, $\cN(C,G)$ is both a complex submanifold in $J$ and a Lagrangian
submanifold in $\om_I$ and $\om_K$ (cf.~\S\ref{sec:MHn}) and hence also in
$\om_t$.
This can also be phrased as a statement in generalised geometry.
Since the involution $\iota$ preserves $J$ but reverses $\om_I,\om_K$, the
generalised tangent bundle of $\cN(C,G)\subset\MH(C,G)^\iota$, which is
$\ker(\iota_*-1)\oplus\ker(\iota^*+1)\subset(T\oplus T^*)\MH(C,G)$, is
preserved by the generalised complex structures $\cJ_t$ in \eqref{eqn:Jt}.
Therefore $\MH(C,G)^\iota$, and hence $\cN(C,G)$, is a generalised complex
submanifold with respect to $\cJ_t$ for all $t\in\CP^1$.

At the quantum level, when the worldsheet has a non-empty boundary, the
theory requires more conditions to be anomaly-free than the bulk theory
(discussed in \S\ref{sec:4d2d}) does.
For the $A$-model with a Lagrangian $A$-brane, the extra condition
\cite{Hor,KL03} is that the Lagrangian submanifold that supports the brane
is gradable in the sense of \cite{Kon}.
(Mathematically, this is related to the orientability of the moduli space of
holomorphic curves \cite{dS,Liu,So,FO3,Ge}.)
In particular, a special Lagrangian submanifold in a Calabi-Yau manifold is
always gradable \cite{Hor,Sei}.
If the target space is hyper-K\"ahler such as the Hitchin moduli space
$\MH(C,G)$, then for any $(w_+,w_-)$ with $w_+\ne w_-$, the K\"ahler
structure in \eqref{eqn:om-J} is Calabi-Yau; it is generalised Calabi-Yau
before the $B$-field transform.
We restrict our attention to the subfamily $(w_+,w_-)=(-t,t^{-1})$ from
$4$-dimensional gauge theories.
The world-volume $\cN(C,G)$ is in fact special Langrangian for all
$t\ne\pm\ii$ and is calibrated by the top exterior power of the holomorphic
$2$-form
\[  \Om_t:=\om_J+\mfrac{\ii(1-\bar t^2t^2)}{(1+t^2)(1+\bar t^2)}
       \Big(\om_K+\mfrac{t+\bar t}{1-\bar tt}\,\om_I\Big).  \]
Consequently, the $2$-dimensional theory with $t\ne\pm\ii$, which is
equivalent to an $A$-model with an $A$-brane supported on $\cN(C,G)$, is
anomaly-free.
Since being anomaly-free is a closed condition, the $B$-model with $t=\pm\ii$
and with the same brane is also anomaly-free.

The above analysis does not take into account $B$-fields that a sigma-model
can couple to.
In general, a $B$-field $B$ on a target space $M$ defines a line bundle
$\cL_B$ over its loop space $LM$.
If the worldsheet $\Sig$ is mapped to $M$ by $u$ and if, for simplicity, the
boundary $\bSig$ is a single circle, the holonomy $\exp(\ii\int_\Sig u^*B)$
is a unit vector in the fibre of $\cL_B$ over $u|_\bSig$.
To obtain a phase in $\mathrm U(1)$, we often assume that $u(\bSig)$ is
contained in the support (or world-volume) $N$ of a brane on which there is
a Chan-Paton bundle twisted by $B$.
If the latter is a (twisted) line bundle, it trivialise the $B$-field on $N$
and the line bundle $\cL_B$ on $LN$.
The holonomy $\exp(\ii\int_{\bSig}(u|_\bSig)^*A')$ along $\bSig$ of a
connection $A'$ on the Chan-Paton bundle is a unit vector in the same line
over $u|_\bSig$ and thus the phase
$\exp(\ii\int_\Sig u^*B-\ii\int_{\bSig}(u|_\bSig)^*A')$
in the path integral is well defined \cite{Ka00}.
When $\bSig$ contains several boundary circles, $\exp(\ii\int_\Sig u^*B)$ is
an element of a tensor product of lines, each for one boundary circle.
Its phase ambiguity can then be canceled by the holonomies of several
(twisted) Chan-Paton bundles.

When the $B$-field is flat, i.e., $dB=0$, we can sometimes define the holonomy
$\exp(\ii\int_\Sig u^*B)$ when $\bSig\ne\emptyset$ as a $\mathrm U(1)$ phase
without twisting the Chan-Paton bundles.
But this is subject to an additional anomaly, though of a classical origin.
Suppose we have defined $\exp(\ii\int_\Sig u^*B)$ for a map
$u\colon(\Sig,\bSig)\to(M,N)$.
We deform $u$ smoothly until it returns to the same map.
In this way, we extend $u$ to $\hat u\colon I\times(\Sig,\bSig)\to(M,N)$
so that $\hat u(0,\cdot)=\hat u(1,\cdot)$ is the original map $u$.
By the flatness of the $B$-field, we have an identity
\[ \exp\Big(\ii\mint{\{1\}\times\Sig}{}\hat u^*B\Big)
   =\exp\Big(\ii\mint{\{0\}\times\Sig}{}\hat u^*B\Big)
   \,\exp\Big(\ii\mint{I\times\bSig}{}\hat u^*B\Big).   \]
So the phase $\exp(\ii\int_\Sig u^*B)$ can be defined if $B$ is trivial on
any torus in $N$.
While any class in $H_2(N,\bZ)$ can be realised by a smoothly embedded closed
surface \cite{Th}, not all of them can be realised by tori.
To circumvent this difficulty, we recall that a flat $B$-field on $M$ is an 
element of $H^2(M,\mathrm U(1))$.
Using the long exact sequence
\begin{equation}\label{eqn:u1-seq}
\cdots\to H^1(N,\mathrm U(1))\to H^2((M,N),\mathrm U(1))\to
H^2(M,\mathrm U(1))\to H^2(N,\mathrm U(1))\to\cdots,
\end{equation}
we find that if and only if the restriction of $B$ to $N$ is trivial can the
flat $B$-field in $H^2(M,\mathrm U(1))$ be lifted to $H^2((M,N),\mathrm U(1))$,
defining a phase $\exp(\ii\int_\Sig u^*B)\in\mathrm U(1)$ for every relative
map $u\colon(\Sig,\bSig)\to(M,N)$.

So if a theory on a worldsheet with boundary is already anomaly-free in the
absence of $B$-fields, its coupling to a flat $B$-field $B$ is consistent
if $B$ is trivial on $N$.
If the original theory without a $B$-field is anomalous, then it is possible
to cancel the anomaly by a non-trivial flat $B$-field on the world-volume of
the brane.
The topology of the normal bundle does not play a role here.
This can be compared with the Freed-Witten anomaly-free condition \cite{FW}
for untwisted open strings, which involves the $w_2$ of the normal bundle
of the world-volume.

In our case, the $B$-field $B_\tht$ on $\MH(C,G)$ induced by the theta term
is a globally defined closed $2$-form and it is topologically trivial.
We claim that $B_\tht=0$ on $\cN(C,G)$ and hence it is also geometrically
trivial on $\cN(C,G)$.
In fact, we have a stronger statement $\iota^*B_\tht=-B_\tht$ on $\MH(C,G)$
because $B_\tht=-\frac\tht{2\pi}\om_I$ \cite{KW} and because $\om_I$ satisfies
the same property $\iota^*\om_I=-\om_I$.
Alternatively, suppose as in \S\ref{sec:4d2d} that $B_\tht$ is the integration
along $C$ of $-\frac\tht{8\pi^2}\tr F^\cU\wedge F^\cU$, where $F^\cU$ is the
curvature of the universal bundle $\cU\to\MH(C,G)\times C$.
The diagonal action of $\iota$ on $\MH(C,G)\times C$ lifts to $\cU$
preserving the universal connection.
So the instanton density $\tr F^\cU\wedge F^\cU$ is a $\iota$-invariant
$4$-form on $\MH(C,G)\times C$.
Since $\iota$ reverses the orientation of $C$ on which the fibre integration
is performed, we obtain $\iota^*B_\tht=-B_\tht$ again, completing our
explanation that $B_\tht$ is trivial both topologically and geometrically on
$\cN(C,G)$.

The reduction of the theta term to two dimensions can be calculated
explicitly.
We note that the map $u\colon(\Sig,\bSig)\to(\MH(C,G),\cN(C,G))$ lifts to a
$\iota$-equivariant map $\tilde u\colon\tSig\to\MH(C,G)$, and
$(\tilde u\times\id_C)^*\cU\cong\tilde\pi^*P_\ad$, where
$\tilde\pi\colon\tSig\times C\to X$ is the projection.
Since $\iota$ is orientation preserving on $\tSig\times C$, we have
\[ \mint{X}{}\tr F\wedge F
   =\mfrac12\mint{\tSig\times C}{}\tilde\pi^*\tr F\wedge F
   =\mfrac12\mint{\tSig}{}\mint{C}{}\tilde u^*\tr F^\cU\wedge F^\cU
   =\mfrac12\mint{\tSig}{}\tilde u^*B_\tht=\mint{\Sig}{}u^*B_\tht    \]
showing explicitly that the holonomy $\exp(-\ii\int_\Sig u^*B_\tht)$ on $\Sig$
is well defined despite $\bSig\ne\emptyset$.

In \S\ref{sec:em2dn}, we will see that the rectified discrete fluxes from four
dimensions do produce non-trivial (but still untwisted) flat Chan-Paton bundles
over $\cN(C,G)$ through the covering map $p\colon\MH(C',G)\to\cN(C,G)$ in
\cite{HWW} or its refinement in \cite{Wu15} (cf.~\S\ref{sec:MHn}).
In view of the exact sequence \eqref{eqn:u1-seq}, these flat line bundles
reflect the ambiguity in $H^1(\cN(C,G),\mathrm U(1))$ when lifting a flat
$B$-field on $\MH(C,G)$ to a class in
$H^2((\MH(C,G),\cN(C,G)),\mathrm U(1))$.
All these branes define theories which are anomaly-free from the
$2$-dimensional point of view.

\subsection{Discrete electric and magnetic fluxes}\label{sec:em4dn}
We consider the case when the $4$-manifold $X=\tSig\times_\iota C$ has a
space-time splitting and determine the sets of rectified discrete electric
and magnetic fluxes.
To achieve this, we take $\tSig=T^1\times S^1$, and let $\iota$ be the
reflection in the spatial direction $S^1$ with two fixed points, so that the
quotient $S^1/\iota=I$ is an interval and $\Sig=\tSig/\iota$ is a cylinder
with two timelike boundary circles.
Then $X=T^1\times Y$, where $Y=S^1\times_\iota C$ is a smooth orientable
closed $3$-manifold, but $X$ is not a global product of two surfaces.

Though $H_1(Y,\bZ)$ contains $2$-torsion elements, the map $\del^1_Y$ in
\eqref{eqn:long} can still vanish.
For example, if all elements of $\pG$ are of odd order or if the sequence
\eqref{eqn:piG} splits, then $\del^1_Y=0$.
In such cases, the sets of discrete electric and magnetic fluxes remain
$\be(Y,G)=H^1(Y,\ZG)^\vee$ and $\bm(Y,G)=H^2(Y,\pG)$, respectively.
These cohomology groups of $Y$ are calculated in \S\ref{sec:HYX}.
But generally, the sets of discrete fluxes need to be adjusted as in
\S\ref{sec:em4d}.
We wish to calculate the sets $\bm(Y,G)$ and $\be(Y,G)$ of rectified discrete
fluxes for the above $3$-manifold $Y$ in the general case.

For subsequent applications, we start with the standard long exact sequence
\begin{equation}\label{eqn:longZ}
0\to\pG_{[2]}\map{i_{\bZ_2}^1}Z(\tG)_{[2]}\map{j_{\bZ_2}^1}\ZG_{[2]}
\map{\del_{\bZ_2}^1}\pG/2\pG\map{i_{\bZ_2}^2}Z(\tG)/2Z(\tG)\map{j_{\bZ_2}^2}
\ZG/2\ZG\to0.
\end{equation}
derived from the short exact sequence \eqref{eqn:piG}.
It is in fact \eqref{eqn:longA} for the classifying space $B\bZ_2$.
The map
\begin{equation}\label{eqn:del2}
\del_{\bZ_2}^1\colon\ZG_{[2]}\cong\Hom(\bZ_2,\ZG)\to
\pG/2\pG\cong\Ext(\bZ_2,\pG)
\end{equation}
is defined abstractly but has the following explicit description.
Given $z\in\ZG$, lift it to $\tilde z\in Z(\tG)$.
If $2z=0$ in $\ZG$, then $2\tilde z\in Z(\tG)$ is in the subgroup $\pG$.
With a different lift of $z$ to $Z(\tG)$, $2\tilde z$ changes by an element
in $2\pG$.
Thus the coset in $\pG/2\pG$ containing $2\tilde z\in\pG$, which is precisely
the image $\del_{\bZ_2}^1(z)$, depends only on $z$.
Let
\begin{equation}\label{eqn:Z2G}
\be(\bZ_2,G):=\ker(\del_{\bZ_2}^1)^\vee,\qquad
\bm(\bZ_2,G):=\coker(\del_{\bZ_2}^1).
\end{equation}
Similar to \eqref{eqn:exactem}, we have a short exact sequence
\[ 0\to\be(\bZ_2,G)^\vee\to\ZG_{[2]}\to\pG/2\pG\to\bm(\bZ_2,G)\to0.  \]
Furthermore, we have $\be(\bZ_2,G_\ad)=0$, $\be(\bZ_2,\tG)=Z(\tG)_{[2]}^\vee$,
$\bm(\bZ_2,G_\ad)=Z(\tG)/2Z(\tG)$, $\bm(\bZ_2,\tG)=0$ and
\[ \bm(\bZ_2,G)=\ker(j_{\bZ_2}^2)\subset\bm(\bZ_2,G_\ad),\qquad
   \be(\bZ_2,G)=\ker((i_{\bZ_2}^1)^\vee)\subset\be(\bZ_2,\tG).   \]

We consider a similar problem for the non-orientable surface $C'$.
There is a commutative diagramme
\[ \scalebox{.95}[1]{$
\xymatrix@=.75pc{&0\ar[d]&0\ar[d]&0\ar[d]&0\ar[d]&0\ar[d]&0\ar[d]&        \\
0\ar[r]&\pG_{[2]}\ar[d]\ar[r]^{i_{\bZ_2}^1}&Z(\tG)_{[2]}\ar[d]
\ar[r]^{j_{\bZ_2}^1}&\ZG_{[2]}\ar[d]\ar[r]^{\hspace*{-2em}\del_{\bZ_2}^1}&
\pG/2\pG\ar[r]^{i_{\bZ_2}^2}\ar[d]^\cong&Z(\tG)/2Z(\tG)\ar[r]^{j_{\bZ_2}^2}
\ar[d]^\cong&\ZG/2\ZG\ar[r]\ar[d]^\cong&0                                 \\
0\ar[r]&H^1(C',\pG)\ar[d]^{\pi^*}\ar[r]^{i_{C'}^1}&H^1(C',Z(\tG))
\ar[d]^{\pi^*}\ar[r]^{j_{C'}^1}&H^1(C',\ZG)\ar[d]^{\pi^*}\ar[r]^{\del_{C'}^1}&
H^2(C',\pG)\ar[d]\ar[r]^{i_{C'}^2}&H^2(C',Z(\tG))\ar[d]\ar[r]^{j_{C'}^2}&
H^2(C',\ZG)\ar[d]\ar[r]&0                                                 \\
0\ar[r]&\pi^*H^1(C',\pG)\ar[d]\ar[r]^{i_C^1}&\pi^*H^1(C',Z(\tG))\ar[d]
\ar[r]^{j_C^1}&\pi^*H^1(C',\ZG)\ar[r]\ar[d]&0&0&0&                        \\
&0&0&0&&&&}$}                                                             \]
containing exact columns \eqref{eqn:C1}.
The first two rows are the long exact sequences \eqref{eqn:longZ} and
\eqref{eqn:longA} for $C'$.
A routine diagramme chasing shows that the third row is also exact and that
there is a short exact sequence
\begin{equation}\label{eqn:kerdel}
0\to\ker(\del_{\bZ_2}^1)\to\ker(\del^1_{C'})\to\pi^*H^1(C',\ZG)\to0
\end{equation}
in which $\ker(\del_{\bZ_2}^1)=\im(j_{\bZ_2}^1)\cong\coker(i_{\bZ_2}^1)$.
In addition, there are isomorphisms
\begin{equation}\label{eqn:cokerdel}
\coker(\del^1_{C'})\cong\coker(\del_{\bZ_2}^1)\cong\im(i_{\bZ_2}^2)
=\ker(j_{\bZ_2}^2).
\end{equation}
Taking the dual of \eqref{eqn:kerdel}, we obtain the exact sequence
\begin{equation}\label{eqn:kerdel^}
0\to(\pi^*H^1(C',\ZG))^\vee\to\ker(\del^1_{C'})^\vee\to\be(\bZ_2,G)\to0.
\end{equation}

There is also a commutative diagramme of homology groups whose Poincar\'e
dual (cf.~\S\ref{sec:HC}) is
\[ \scalebox{.95}[1]{$\xymatrix@=.75pc{&&&&0\ar[d]&0\ar[d]&0\ar[d]        \\
&0\ar[d]&0\ar[d]&0\ar[r]\ar[d]&\sfrac{H^1(C,\pG)}{\pi^*H^1(C',\pG)}
\ar[r]^{i_C^1}\ar[d]&\sfrac{H^1(C,Z(\tG))}{\pi^*H^1(C',Z(\tG))}\ar[r]^{j_C^1}
\ar[d]&\sfrac{H^1(C,\ZG)}{\pi^*H^1(C',\ZG)}\ar[r]\ar[d]&0                 \\
0\ar[r]&H^0(C',\upG)\ar[d]^\cong\ar[r]^{\ui_{C'}^0}&H^0(C',\uZoG)
\ar[d]^\cong\ar[r]^{\uj_{C'}^0}&H^0(C',\uZG)\ar[d]^\cong\ar[r]^{\udel_{C'}^0}&
H^1(C',\upG)\ar[d]\ar[r]^{\ui_{C'}^1}&H^1(C',\uZoG)\ar[d]\ar[r]^{\uj_{C'}^1}&
H^1(C',\uZG)\ar[d]\ar[r]&0                                                \\
0\ar[r]&\pG_{[2]}\ar[d]\ar[r]^{i_{\bZ_2}^1}&Z(\tG)_{[2]}\ar[d]
\ar[r]^{j_{\bZ_2}^1}&\ZG_{[2]}\ar[d]\ar[r]^{\hspace*{-2em}\del_{\bZ_2}^1}&
\pG/2\pG\ar[r]^{i_{\bZ_2}^2}\ar[d]&Z(\tG)/2Z(\tG)\ar[r]^{j_{\bZ_2}^2}
\ar[d]&\ZG/2\ZG\ar[r]\ar[d]&0                                             \\
&0&0&0&0&0&0&}$}                                                          \]
containing exact columns \eqref{eqn:C1'}.
The second row is the long exact sequence of cohomology groups from the exact
sequence of local systems $0\to\upG\to\uZoG\to\uZG\to0$ on $C'$
(cf.~\S\ref{sec:Poin}) whereas the last row is identical to the first in the
previous diagramme.
It follows that the first row is also exact; in fact, by \eqref{eqn:Ctau},
it is Poincar\'e dual to the last row of the previous diagramme.
Moreover, there is an exact sequence
\begin{equation}\label{eqn:cokerdel0}
0\to\sfrac{H^1(C,\pG)}{\pi^*H^1(C',\pG)}\to\coker(\udel{}^0_{C'})
\to\bm(\bZ_2,G)\to0.
\end{equation}

We now return to the calculation of discrete fluxes on $Y$.
Using \eqref{eqn:Y1}, we obtain a commutative diagramme
\[ \xymatrix@=.75pc{&0\ar[d]&0\ar[d]&0\ar[d]&&                           \\
0\ar[r]&(\pG_{[2]})^{\oplus2}\ar[d]\ar[r]^{\;\;(i_{\bZ_2}^1)^{\oplus2}}&
(Z(\tG)_{[2]})^{\oplus2}\ar[d]\ar[r]^{\;\;(j_{\bZ_2}^1)^{\oplus2}}&
(\ZG_{[2]})^{\oplus2}\ar[d]\ar[r]^{(\del_{\bZ_2}^1)^{\oplus2}}&\pG/2\pG
\ar[r]&\cdots                                                            \\
0\ar[r]&H^1(Y,\pG)\ar[d]\ar[r]^{i_Y^1}&H^1(Y,Z(\tG))\ar[d]\ar[r]^{j_Y^1}&
H^1(Y,\ZG)\ar[d]\ar[r]^{\del_Y^1}&H^2(Y,\ZG)\ar[r]&\cdots                \\
0\ar[r]&\pi^*H^1(C',\pG)\ar[d]\ar[r]^{i_C^1}&\pi^*H^1(C',Z(\tG))\ar[d]
\ar[r]^{j_C^1}&\pi^*H^1(C',\ZG)\ar[d]\ar[r(0.72)]&\hspace{-3.2em}0&     \\
&0&0&0&&}                                                                 \]
with exact rows and columns.                                                               
A similar diagramme chasing yields a short exact sequence
\begin{equation}\label{eqn:symC}
0\to(\ker(\del_{\bZ_2}^1))^{\oplus2}\to\ker(\del_Y^1)\to\pi^*H^1(C',\ZG)\to0.
\end{equation}
Taking the Pontryagin dual, we obtain
\begin{equation}\label{eqn:beC}
0\to(\pi^*H^1(C',\ZG))^\vee\to\be(Y,G)\to\be(\bZ_2,G)^{\oplus2}\to0.
\end{equation}
Using instead \eqref{eqn:Y2}, we have another commutative diagramme
\[ \xymatrix@=.75pc{&&0\ar[d]&0\ar[d]&0\ar[d]&                          \\
&\qquad\qquad0\ar[r]&\sfrac{H^1(C,\pG)}{\pi^*H^1(C',\pG)}\ar[d]
\ar[r]^{i_C^1}&\sfrac{H^1(C,Z(\tG))}{\pi^*H^1(C',Z(\tG))}\ar[d]
\ar[r]^{j_C^1}&\sfrac{H^1(C,\ZG)}{\pi^*H^1(C',\ZG)}\ar[d]\ar[r]&0    \\
\cdots\ar[r]&H^1(Y,\ZG)\ar[r]^{\del_Y^1}&H^2(Y,\pG)\ar[d]\ar[r]^{i_Y^2}
&H^2(Y,Z(\tG))\ar[d]\ar[r]^{j_Y^2}&H^2(Y,\ZG)\ar[d]\ar[r]&0             \\
\cdots\ar[r]&(\ZG_{[2]})^{\oplus2}\ar[r]^{\!\!(\del_{\bZ_2}^1)^{\oplus2}}&
(\pG/2\pG)^{\oplus2}\ar[d]\ar[r]^{\;\quad(i_{\bZ_2}^2)^{\oplus2}}&
(Z(\tG)/2Z(\tG))^{\oplus2}\ar[d]\ar[r]^{\quad(j_{\bZ_2}^2)^{\oplus2}}&
(\ZG/2\ZG)^{\oplus2}\ar[d]\ar[r]&0                                      \\
&&0&0&0&}                                                               \]
with exact rows and columns.
Consequently, there is another short exact sequence
\begin{equation}\label{eqn:bmC}
0\to\sfrac{H^1(C,\pG)}{\pi^*H^1(C',\pG)}\to\bm(Y,G)\to\bm(\bZ_2,G)^{\oplus2}
\to0.
\end{equation}
Thus we have determined the sets of rectified discrete electric and magnetic
fluxes when $Y=S^1\times_\iota C$.

Now consider the spacetime $4$-manifold $X=\tSig\times_\iota C$, where $\tSig$
is a closed orientable surface with an orientation reversing involution
$\iota$ such that $\Sig=\tSig/\iota$ has boundary $\bSig=\tSig^\iota$.
We want to compute the set $\ker(j_X^2)\subset H^2(X,\pGad)$ of discrete fluxes
of $G_\ad$-bundles over $X$ that can be lifted to $G$-bundles
(cf.~\S\ref{sec:Gad}).
This can be done by calculating the effect of $j_X^2$ on exact sequence
\eqref{eqn:HX2} with various coefficient groups in \eqref{eqn:piG}.
Using \eqref{eqn:cokerdel} and the above exact sequence of $\pi^*H^1(C',A)$
with the groups $A$ in \eqref{eqn:piG}, we get an exact sequence
\begin{equation}\label{eqn:kerjX}
\scalebox{.9}[1]{$0\to H^1(\bSig,\pG_{[2]})\oplus H^1\Big((\Sig,\bSig),
\sfrac{H^1(C,\pG)}{\pi^*H^1(C',\pG)}\Big)\to\ker(j_X^2)\to
H^0(\bSig,\bm(\bZ_2,G))\oplus H^1(\bSig,\pi^*H^1(C',\pG))\to0$}
\end{equation}
that determines $\ker(j_X^2)$.
In particular, if $\Sig$ is a cylinder, then \eqref{eqn:kerjX} reduces to
\eqref{eqn:Y1} and \eqref{eqn:bmC} because
\[ \ker(j_X^2)\cong\ker(j_Y^1)\oplus\ker(j_Y^2)\cong H^1(Y,\pG)\oplus\bm(Y,G).
\]
If $\Sig$ is a disc, then $\ker(j_X^2)\cong(\pG)_{[2]}\oplus\bm(\bZ_2,G)$.
More generally, if $\bSig$ is a single circle, then just as \eqref{eqn:filt}
for the cohomology of $X$, there is a filtration on $\ker(j_X^2)$ whose
graded components are
\begin{equation}\label{eqn:filtjX}
H^0(\bSig,\bm(\bZ_2,G)),\quad H^1(\hat\Sig,H^1(C,\pG)),\quad
H^1(\bSig,\pG_{[2]}),
\end{equation}
where $\hat\Sig$ is the closed surface obtained from $\Sig$ by attaching a
disc to $\bSig$.

\subsection{Two-dimensional interpretation of the discrete fluxes}
\label{sec:em2dn}
We interpret the sets of rectified discrete electric and magnetic fluxes,
$\be(Y,G)$ in \eqref{eqn:beC} and $\bm(Y,G)$ in \eqref{eqn:bmC} from the
point of view of the $2$-dimensional sigma-model.
Recall that $X=T^1\times Y$, $Y=S^1\times_\iota C$.
The sigma-model contains a map $u$ from $\Sig=T^1\times I$ to $\MH(C,G)$ so
that $u(\bSig)\subset\cN(C,G)$ and, on the boundary $\bSig=T^1\sqcup T^1$,
$u|_\bSig$ can be lifted to $u'\colon\bSig\to\MH(C',G)$.
Here $\MH(C',G)$ is the Hitchin moduli space of the non-orientable surface
$C'$, $\MH(C,G)$ is that of the orientation double cover $C$, and there is a
$\ZG_{[2]}$-covering $p\colon\MH(C',G)\to\cN(C,G)$, where $\cN(C,G)$ is a
part of $\MH(C,G)^\iota$ \cite{HWW} (cf.~\S\ref{sec:MHn}).

There are a number of differences between this case and the dimensional
reduction in \cite{KW} along orientable surfaces when the spacetime is a
product of two surfaces.
First, in our case, there is an absence of $m_0\in H^2(C,\pG)$ that labels
the connected components of $\MH(C,G)$ because $m_0=0$ is the only non-empty
sector.
Since all $G$-bundles over $C'$ pull back to topologically trivial bundles
over $C$, the world-volume $\cN(C,G)$ is contained solely in
$\MH^{m_0=0}(C,G)$.
So $\MH^{m_0=0}(C,G)$ is the only component that supports the type of branes
from our dimensional reduction.
 
Now set $\cM:=\MH^{m_0=0}(C,G)$ and $\cM':=\MH(C',G)$.
At any given time, the sigma-model configuration contains a pair of (unbased)
maps $(u,u')\colon(I,\bI)\to(\cM,\cM')$ whose homotopy classes form the set
$[(I,\bI),(\cM,\cM')]$ (cf.~\S\ref{sec:rel}).
Upon choosing base points, the set becomes the relative fundamental group
$\pi_1(\cM,\cM')$ which is determined by \eqref{eqn:MM'1}.
Without base points, we should count $\pi_0(\cM')$ twice as $I$ has two end
points.
Thus we have an exact sequence
\begin{equation}\label{eqn:2Y2}
0\to\sfrac{H^1(C,\pG)}{\pi^*H^1(C',\pG)}\to[(I,\bI),(\cM,\cM')]\to
(\pG/2\pG)^{\oplus2}\to0
\end{equation}
which matches \eqref{eqn:Y2} when $A=\pG$ for topological types of $G$-bundles
over $Y$.
{}From the $2$-dimensional point of view,
$(m_2,m'_2)\in(\pG/2\pG)^{\oplus2}$ labels the two components of $\cM'$ on
which an open string ends while $\mm_1\in H^1(C,\pG)/\pi^*H^1(C',\pG)$
classifies the remaining relative winding of the string in $\cM$.

We must also determine the discrete symmetries in the $2$-dimensional theory.
The presence of branes breaks the full symmetry $H^1(C,\ZG)$ on $\cM$.
Although the subgroup $H^1(C,\ZG)^\iota$ acts on the $\iota$-invariant subset
$\cM^\iota$, it does not always preserve the world-volume $\cN:=\cN(C,G)$ of
the brane.
However, the subgroup $\pi^*H^1(C',\ZG)$, which is proper in
$H^1(C,\ZG)^\iota$ if $\chi(C')$ is even, does act on $\cN$ and hence on
the sigma-model maps (at a fixed time) $u\colon(I,\bI)\to(\cM,\cN)$.
By \eqref{eqn:C1}, the group $H^1(C',\ZG)$ is an extension of
$\pi^*H^1(C',\ZG)$ by $\ZG_{[2]}$.
There is an action of $H^1(C',\ZG)$ on the pair $(\cM,\cM')$ lifting the
action of $\pi^*H^1(C',\ZG)$ on $\cM$; the subgroup $\ZG_{[2]}$ acts on $\cM$
trivially and it acts on $\cM'$ by deck transformations of the covering
$p\colon\cM'\to\cN$.
Likewise, on the extra degrees of freedom $u'\colon\bI\to\cM'$ at the
boundary, there is an action of $(\ZG_{[2]})^{\oplus2}$
(one copy for each end point of $I$).
So the discrete symmetry group in two dimensions is an extension of
$\pi^*H^1(C',\ZG)$ by $(\ZG_{[2]})^{\oplus2}$, which by \eqref{eqn:Y1} is
the same symmetry group $H^1(Y,\ZG)$ in the $4$-dimensional theory.

We then study the breaking of the discrete symmetry when the $2$-dimensional
theory is restricted to a topological sector.
Recall from \S\ref{sec:MHn} that the $\ZG_{[2]}$-covering of $\cM'$ over $\cN$
decomposes into $\ker(\del_{\bZ_2}^1)$-coverings of
$\cM'{}^{m_2}:=\MH^{m_2}(C',G)$ over $\cN^{\mm_2}:=\cN^{\mm_2}(C,G)$, where
$m_2$ ranges over $\pG/2\pG$ and $\mm_2$ is its coset in
$\bm(\bZ_2,G)=\coker(\del_{\bZ_2}^1)$.
We can thus impose a more refined boundary condition on the sigma-model so that
$u'$ maps each end of a string to $\cM'{}^{m_2}$ for some $m_2\in\pG/2\pG$.
This breaks the $\ZG_{[2]}^{\oplus2}$ part of the symmetry to
$\ker(\del_{\bZ_2}^1)^{\oplus2}$.
But the theory depends only on the cosets of $m_2$ in $\bm(\bZ_2,G)$ because
if $m_2$ changes by $\del_{\bZ_2}^1(z)$ for some $z\in\ZG_{[2]}$, then the
covering of $\cM'{}^{m_2}$ over the same $\cN^{\mm_2}$ is isomorphic
(see \S\ref{sec:MHn}) and hence it defines the same brane.
Henceforth we let $\cB^{\mm_2}$ be the brane supported on $\cN^{\mm_2}$ given
by the covering $q\colon\cM'{}^{m_2}\to\cN^{\mm_2}$.
Therefore $\bm(Y,G)$ in \eqref{eqn:bmC} matches the discrete parameters
$(\mm_2,\mm'_2)$ of the two branes that an open string ends on and the
relative winding parameter $\mm_1$.

The $\pi^*H^1(C',\ZG)$ part of the symmetry does not change the topological
types of bundles or branes.
So the unbroken discrete symmetry, which combines
$\ker(\del_{\bZ_2}^1)^{\oplus2}$ and $\pi^*H^1(C',\ZG)$, is the same discrete
symmetry group $\ker(\del_Y^1)$ in gauge theory given by \eqref{eqn:symC}.
The Hilbert spaces of the two theories in two and four dimensions decompose
according to the same character group $\be(Y,G)$ in \eqref{eqn:beC}, and in
both theories, in the sectors labelled by $e\in\be(Y,G)$, the phases in path
integrals associated to each discrete group element are the same.
But from the $2$-dimensional point of view, the two parts of $\ker(\del_Y^1)$
play different roles.
The subgroup $\ker(\del_{\bZ_2}^1)^{\oplus2}$ operates on the boundary data
whereas the quotient group $\pi^*H^1(C',\ZG)$ is a subgroup of $H^1(C,\ZG)$
which acts on the target space, and on the bulk data of the sigma-model.
In \eqref{eqn:beC}, elements $\ee_1\in(\pi^*H^1(C',\ZG))^\vee$ are the
discrete momenta of the translations in $\pi^*H^1(C',\ZG)$.
The symmetry $\ker(\del_{\bZ_2}^1)^{\oplus2}$ and its characters will be
analysed in more details below.

Suppose one end of the open string is on the brane $\cB^{\mm_2}$ defined by the
covering map $p\colon\cM'{}^{m_2}\to\cN^{\mm_2}$ for some $m_2\in\pG/2\pG$.
A group element $g_2\in\ker(\del_{\bZ_2}^1)$ acts as a deck transformation of
$p$.
In canonical quantisation, if $g_2=0$, then the corresponding boundary
component $T^1$ of the cylinder $\Sig$ is mapped to a loop in $\cM'{}^{m_2}$
by $u'$ which lifts $u\colon T^1\to\cN^{\mm_2}$.
If $g_2\ne0$ however, then the two ends of the time interval are mapped to
two points on $\cM'{}^{m_2}$ that differ by the deck transformation $g_2$.
But the projection of the curve to $\cN^{\mm_2}$ remains a loop, which is the
restriction of the map $u\colon\Sig\to\cM$ defined on the bulk of the
worldsheet.
In fact, by the exact sequence
\[   0\to\pi_1(\cM'{}^{m_2})\map{p_*}\pi_1(\cN^{\mm_2})
     \to\ker(\del_{\bZ_2}^1)\to0,                          \]
a loop in $\cN^{\mm_2}$ lifts to a loop in $\cM'{}^{m_2}$ if and only if it
represents an element in $\pi_1(\cN^{\mm_2})$ that maps to zero in
$\ker(\del_{\bZ_2}^1)$.
In the sector labelled by $\ee_2\in\ker(\del_{\bZ_2}^1)^\vee$, the phase
associated to $g_2$ is $\ee_2(g_2)^{-1}\in\mathrm U(1)$.
Summing over $\ee_2$ enforces the condition $g_2=0$ by \eqref{eqn:dis}, and
thus $u|_{T^1}$ can be lifted to $u'\colon T^1\to\cM'{}^{m_2}$.
Thus allowing all $\ee_2$ recovers the original extra degrees of freedom that
the sigma-model has on the boundary of the worldsheet.

In the $\ee_2$ sector, the phase $\ee_2(g_2)^{-1}$ is precisely the holonomy
along the loop $u|_{T^1}$ of the flat line bundle
$\ell^{\ee_2,\mm_2}:=\cM'{}^{m_2}\times_{\ee_2}\bC$ over $\cN^{\mm_2}$.
Therefore for each $(\ee_2,\mm_2)\in\be(\bZ_2,G)\oplus\bm(\bZ_2,G)$, we can
construct a brane $\cB^{\ee_2,\mm_2}$ supported on $\cN^{\mm_2}$ with the
Chan-Paton bundle $\ell^{\ee_2,\mm_2}$.
As explained above, the extra degrees of freedom on the boundary given by
the brane $\cB^{\mm_2}$ turns into multiple sectors labeled by $\ee_2$.
In each sector $\ee_2$, the theory is equivalent to having the end of the
string on the brane $\cB^{\ee_2,\mm_2}$ with the rank one Chan-Paton bundle
$\ell^{\ee_2,\mm_2}$.
Similarly, the other end of the string is on another brane of the same type.
If $\Sig$ has several boundary components, then each component ends on a brane
of the above type.
So we have $\mm_2\in H^0(\bSig,\bm(\bZ_2,G))$ and
$\ee_2\in H^1(\bSig,\be(\bZ_2,G))$.
This picture is relativistic invariant on the worldsheet.
But the $4$-dimensional relativistic invariance requires summing over $\ee_2$.

When $\Sig$ is closed, the discrete electric fluxes include $e_0\in\ZG^\vee$
that defines a $B$-field $e_0(\bar\xi(\cU))$ which the sigma-model couples to
(see \cite{KW} or \S\ref{sec:e2d}).
When $\bSig\ne\emptyset$, the absence of $e_0$ in gauge theory is evident
from the calculation of $\be(Y,G)$ in \eqref{eqn:beC}, but it can also be
explained in a purely $2$-dimensional way.
For a non-zero $e_0\in\ZG^\vee$, not only is the $B$-field $e_0(\bar\xi(\cU))$
itself is non-zero on $\cM$ (see \S\ref{sec:e2d}), its restriction to $\cN$ is
also non-zero.
Thus coupling to such the $B$-field is anomalous according to the anomaly-free
condition at the end of \S\ref{sec:4dn2d}.
To complete the explanation, we must show that the map
$\ZG^\vee\to H^2(\cN,\mathrm U(1))$ that defines the $B$-fields on $\cN$ is
injective.
In fact, after composing it with
$p^*\colon H^2(\cN,\mathrm U(1))\to H^2(\cM',\mathrm U(1))$ and
$\vro_{\cM',\mathrm U(1)}^2\colon H^2(\cM',\mathrm U(1))\to\pi_2(\cM')^\vee$,
the map $\ZG^\vee\to\pi_2(\cM')^\vee$ is still injective.
This is because the latter is the dual of the surjective map
$j\circ\zeta'_*\colon\pi_2(\cM')\to\ZG$ (cf.~\S\ref{sec:MH}).

Now suppose the worldsheet $\Sig$ is a general compact orientable surface
with boundary $\bSig$.
We compare the discrete parameters that are summed over in the path integrals
of gauge theory and sigma-model.
In four dimensions, they are the topological types of $G_\ad$-bundles over
$X$ that can be lifted to $G$-bundles and are in $\ker(j_X^2)$ and
$H^4(X,\pi_3(G))$.
In two dimensions, they are the homotopy types of pairs of maps
$(u,u')\colon(\Sig,\bSig)\to(\cM,\cM')$.
For simplicity, we assume that $\bSig$ is a single circle.
Then by \eqref{eqn:ZZ'}, there is a short exact sequence
\[  0\to\pi_2(\cM,\cM')\to[(\Sig,\bSig),(\cM,\cM')]_0\to
    H^1(\hat\Sig,H^1(C,\pG))\to0   \]
for the homotopy types of the based maps.
Here $\pi_2(\cM,\cM')$ is calculated by the exact sequence \eqref{eqn:MM'2}.
If we consider maps without base points, there is an additional set of
parameters from $\pi_0(\cM')=\pG/2\pG$.
Combining the above information, the set $[(\Sig,\bSig),(\cM,\cM')]$ has
a filtration whose graded components are
\begin{equation}\label{eqn:gr}
\pG/2\pG,\quad H^1(\hat\Sig,H^1(C,\pG)),\quad \pG_{[2]},\quad\pi_3(G).
\end{equation}
But the symmetry $\ZG_{[2]}$ acts on the maps $u'\colon\bSig\to\cM'$ and on
$m_2$ that labels the connected components of $\cM'$ by $\del_{\bZ_2}^1$.
So the actual parameters labelling non-isomorphic theories are in
$\bm(\bZ_2,G)$ instead of $\pG/2\pG$.
Meanwhile, $\pi_3(G)\cong H^4(X,\pi_3(G))$ contains the instanton number of
the gauge bundles.
If we adjust $\pG/2\pG$ to $\bm(\bZ_2,G)$ and exclude the non-torsion
$\pi_3(G)$ from \eqref{eqn:gr}, the graded components match those in
\eqref{eqn:filtjX} for $\ker(j_X^2)$.

As in \S\ref{sec:e2d}, the sigma-model map
$(u,u')\colon(\Sig,\bSig)\to(\cM,\cM')$ does not uniquely determine the
$G$-bundle $P$ over $X$ in gauge theory, and there is an
$|H^1(\Sig,\ZG)|$-fold ambiguity.
Therefore the partition functions $Z_{X,G}$ of the gauge theory and
$Z_{(\Sig,\bSig),(\cM,\cM')}$ of the sigma-model are related by
\begin{equation}\label{eqn:ZZ-rel}
Z_{X,G}=\mfrac{|H^1(\Sig,\ZG)|}{|\ZG|}Z_{(\Sig,\bSig),(\cM,\cM')}.
\end{equation}
By changing $G$ to $G_\ad$, a map pair $(u,u')\colon(\Sig,\bSig)\to(\cM,\cM')$
descends to $u_\ad\colon(\Sig,\bSig)\to(\cM_\ad,\cM'_\ad)$, where
$\cM_\ad:=\MH^{m_0=0}(C,G_\ad)$ and $\cM'_\ad:=\MH(C',G_\ad)$.
The multiplicity of $u$ over $u_\ad$ can be obtained by (\ref{eqn:diam}a).
Thus
\begin{equation}\label{eqn:ZZMad-rel}
Z_{(\Sig,\bSig),(\cM,\cM')}=|H^0(\bSig,\ZG_{[2]})|\,|\pi^*H^1(C',\ZG))|\,
Z_{(\Sig,\bSig),(\cM_\ad,\cM'_\ad)}^\dcirc,
\end{equation}
where $Z_{(\Sig,\bSig),(\cM_\ad,\cM'_\ad)}^\dcirc$ sums over the homotopy
classes $[u_\ad]$ that can be lifted to $[(u,u')]$.
By using the explicit formula \eqref{eqn:HX1} of $H^1(X,\ZG)$, the procedure
of passing from $Z_{X,G}$ to $Z_{(\Sig,\bSig),(\cM_\ad,\cM'_\ad)}$ through
\eqref{eqn:ZZ-rel} and \eqref{eqn:ZZMad-rel} is equivalent to first changing
the $G$ to $G_\ad$ by \eqref{eqn:ZGGad} and then taking the reduction to
sigma-model.

If we fix $\mm_2\in H^0(\bSig,\bm(\bZ_2,G))$ and
$\ee_2\in H^1(\bSig,\be(\bZ_2,G))$, then the $G$-bundles over $X$
are allowed to be twisted along the embedded surfaces $C'$.
Note that \eqref{eqn:ZZ-rel} remains valid for $Z_{X,G}^{\ee_2,\mm_2}$ and
$Z_{(\Sig,\bSig),(\cM,\cM'{}^{m_2})}^{\ee_2}$, where
$\cM'{}^{m_2}:=\MH^{m_2}(C',G)$.
The degeneracy of the map pairs $(u,u')$ over
$u_\ad\colon(\Sig,\bSig)\to(\cM_\ad,\cM_\ad^{\prime\mm_2})$, where
$\cM_\ad^{\prime\mm_2}:=\MH^{\mm_2}(C',G_\ad)$, can be found by
(\ref{eqn:diam}b).
So the sigma-models of different targets are related by
\[  Z_{(\Sig,\bSig),(\cM,\cM'{}^{m_2})}^{\ee_2}
    =|H^0(\bSig,\ker(\del_{\bZ_2}^1))|\,|\pi^*H^1(C',\ZG)|\,
    Z_{(\Sig,\bSig),(\cM_\ad,\cM_\ad^{\prime\mm_2})}^{\ee_2}.     \]
On the other hand, the multiplicity of passing such possibly twisted
$G$-bundles on $X$ to $G_\ad$-bundles is $|H^1(X,\ZG)|$ but the latter is
reduced by a factor
$|H^0(\bSig,\ker(i_{C'}^2))|=|H^0(\bSig,\im(\del_{\bZ_2}^1))|$ for allowing
twists.
Thus \eqref{eqn:ZGGad} becomes
\[ Z_{X,G}^{\ee_2,\mm_2}=\mfrac{|H^1(X,\ZG)|}
   {|\ZG|\,|H^0(\bSig,\im(\del_{\bZ_2}^1))|}Z_{X,G_\ad}^{\ee_2,\mm_2},   \]
and we get the same result by first changing the gauge group to $G_\ad$ and
then reducing to the sigma-model.

\subsection{Mirror symmetry of branes}\label{sec:mir-n}
Mirror symmetry in $2$-dimensional sigma-models is a consequence of the
electric-magnetic duality, or $S$-duality in four dimensions
\cite{BJSV,HMS,KW}.
For any orientable closed $3$-dimensional time-slice $Y$, $S$-duality
exchanges the rectified discrete electric and magnetic fluxes introduced in
\S\ref{sec:em4d} by \eqref{eqn:Srec}.
In the present case, $Y=S^1\times_\iota C$, where $C$ is the orientation
double cover of a non-orientable surface $C'$ and $\iota$ acts as orientation
reversing involutions on $C$ and $S^1$.
The sets $\be(Y,G)$ and $\bm(Y,G)$ of discrete fluxes are given by the exact
sequences \eqref{eqn:beC} and \eqref{eqn:bmC}, respectively.
We recall the notations $\bm(\bZ_2,G)$, $\be(\bZ_2,G)$ from \eqref{eqn:Z2G}.
When the gauge group $G$ is exchanged with its dual $\LG$, we have
$\pi_1(\LG)_{[2]}\cong(\ZG/2\ZG)^\vee$ and
$Z(\LG)/2Z(\LG)\cong(\pG_{[2]})^\vee$ by \eqref{eqn:LZG} and \eqref{eqn:two}.
By \eqref{eqn:coker}, there are isomorphisms
\begin{equation}\label{eqn:Z2dual}
\bm(\bZ_2,\LG)\cong\be(\bZ_2,G),\qquad\be(\bZ_2,\LG)\cong\bm(\bZ_2,G)
\end{equation}
since the Pontryagin dual of $\del_{\bZ_2}^1$ is the same map
${}^L\!\del_{\bZ_2}^1$ but for $\LG$.
Using \eqref{eqn:Z2dual} and \eqref{eqn:Ctau}, the duality \eqref{eqn:Srec}
between $\be(Y,G)$ and $\bm(Y,G)$ can be verified explicitly from
\eqref{eqn:beC} and \eqref{eqn:bmC}.
In this way, $S$-duality exchanges $\mm_1$ with $\ee_1$ and $\mm_2$ with
$\ee_2$.
For the spacetime $X=\tSig\times_\iota C$ where $\Sig=\tSig/\iota$ is not
necessarily a cylinder, the set $\ker(j_X^2)$ of discrete fluxes of
$G_\ad$-bundles over $X$ that lift to $G$-bundles should satisfy the duality
relation \eqref{eqn:SdualX} when the gauge group $G$ is exchanged with $\LG$.
This can be verified using \eqref{eqn:kerjX}, with the help of the
isomorphisms \eqref{eqn:LZG} and \eqref{eqn:Ctau}.

To see how $S$-duality is realised concretely from the $2$-dimensional point
of view, we introduce the Hitchin fibration of the moduli space $\MH(C',G)$
where $C'$ is a closed non-orientable surface.
Let $O'$ be the orientation line bundle of $C'$.
Given a conformal structure on $C'$, the Hodge star $*'$ on $C'$ maps
$\Om^1(C',\ad P')$ to $\Om^1(C',O'\otimes\ad P')$, and vice versa.
Therefore $*'$ acts on $\Om^1(C',\ad P')\oplus\Om^1(C',O'\otimes\ad P')$,
satisfying $(*')^2=-1$.
We can proceed with $C'$ alone, but the above direct sum can be identified
naturally with $\Om^1(C,\ad P)$, where $C$ is the orientation double cover
and $P=\pi^*P'$.
Furthermore, the conformal structure on $C'$ defines a complex structure on
$C$ so that the $\ii$-eigenspace of $*'$ matches $\Om^{1,0}(C,\ad P^\bC)$.
A Hitchin pair $(A',\phi')$ on $C'$ pulls back to $(A,\phi)$ on $C$.
The Hitchin fibration $h'\colon\MH(C',G)\to\B^\iota$ for $C'$ is defined by
$h':=h\circ p$ or $h'([(A',\phi')]):=h([(A,\phi)])$, where
$h\colon\MH(C,G)\to\B$ is the Hitchin map for $C$ (cf.~\S\ref{sec:S2d-o}).
So $h'$ factors through the space $\cN(C,G)\subset\MH(C,G)^\iota$.
The image of $h'$ is contained in the invariant subspace $\B^\iota$ because
the actions of $\iota$ on $\MH(C,G)$ and on $\B$ (as a real structure)
commutes with $h$.

Like the Hitchin map $h$ for an orientable surface \cite{Hi87}, $h'$ is also
proper and surjective.
For a regular $b\in\B^\iota$ (outside the discriminant divisor), the fibre
$\MH(C',G)_b:=(h')^{-1}(b)$ is smooth and is a disjoint union of affine real
tori $\MH^{m_2}(C',G)_b$ labelled by $m_2\in\pG/2\pG$.
Each $\MH^{m_2}(C',G)_b$ is Lagrangian in $\MH^{m_2}(C',G)$ with respect to
the symplectic structure $\om'_J$ in \S\ref{sec:MHn}.
Therefore $h'$ determines a real integrable system.
Under the regular $\ker(\del_{\bZ_2}^1)$-covering map
$p\colon\MH^{m_2}(C',G)\to\cN^{\mm_2}(C,G)$, the torus $\MH^{m_2}(C',G)_b$
maps to $\cN^{\mm_2}(C,G)_b:=p(\MH^{m_2}(C',G)_b)$ which depends only on the
coset element $\mm_2\in\bm(\bZ_2,G)$, and there is a disjoint union of
$\cN(C,G)_b:=p(\MH(C',G)_b)$,
\begin{equation}\label{eqn:realT}
\cN(C,G)_b=\msqcup{\mm_2\in\bm(\bZ_2,G)}{}\cN^{\mm_2}(C,G)_b.
\end{equation}
All real tori $\cN^{\mm_2}(C,G)_b$ are contained in the same torus component
$\MH^{m_0=0}(C,G)_b$.
The involution $\iota$ acts on $\MH(C,G)_b$ sending $\MH^{m_0}(C,G)_b$ to
$\MH^{-m_0}(C,G)_b$, and the fixed-point set $\MH(C,G)_b^\iota$ is precisely
$\cN(C,G)_b$ described in \eqref{eqn:realT}.
In fact, it is already known that $\MH(C,G)_b^\iota$, if smooth, is a union
of some real affine tori if $\iota$ is any anti-holomorphic involution on a
compact Riemann surface $C$ \cite{BS}.
The information on the precise real tori $\cN^{\mm_2}(C,G)_b$ appearing in
the decomposition \eqref{eqn:realT} will be crucial in verifying the
consequences of $S$-duality in two dimensions.

Recall that given $\mm_2\in\bm(\bZ_2,G)$ and $\ee_2\in\be(\bZ_2,G)$, the flat
line bundle $\ell^{\ee_2,\mm_2}=\MH^{m_2}(C',G)\times_{\ee_2}\bC$ defines a
brane $\cB^{\ee_2,\mm_2}$ of type $(A,B,A)$ on $\MH(C,G)$ whose world-volume
is $\cN^{\mm_2}(C,G)$.
If $b\in\B^\iota$ is regular, then on each fibre $\MH(C,G)_b$ of $h$, the
brane $\cB^{\ee_2,\mm_2}$ is supported on the real torus $\cN^{\mm_2}(C,G)_b$.
In four dimensions, the roles of $\mm_2$ and $\ee_2$ are interchanged by
$S$-duality.
In two dimensions, the mirror of the brane $\cB^{\ee_2,\mm_2}$ should be
${}^L\!\cB^{\mm_2,\ee_2}$ in the dual theory given by the line bundle
${}^L\!\ell^{\mm_2,\ee_2}:=\MH^{e_2}(C,\LG)\times_{\mm_2}\bC$ over
$\cN^{\ee_2}(C,\LG)$.
That is,
\begin{equation}\label{eqn:LBem}
(\cB^{\ee_2,\mm_2})^\vee={}^L\!\cB^{\mm_2,\ee_2}.
\end{equation}
For a regular $b\in\B^\iota$, the fibre $\MH(C',\LG)_b:=(\Lh')^{-1}(b)$ of
the dual fibration $\Lh'\colon\MH(C',\LG)\to\B^\iota$ is a union of real tori
$\MH^{e_2}(C',\LG)_b$ which are labelled by
$e_2\in\pi_1(\LG)/2\pi_1(\LG)\cong\ZG_{[2]}^\vee$, while
${}^L\!p(\MH(C',\LG)_b)=\cN(C,\LG)_b$ is a union of real tori
$\cN^{\ee_2}(C,\LG)_b:={}^L\!p(\MH^{e_2}(C',\LG)_b)$ labelled by the cosets
$\ee_2\in\bm(\bZ_2,\LG)\cong\be(\bZ_2,G)$.
The $\iota$-invariant tori in $\MH(C,G)_b$ and $\MH(C,\LG)_b$ are annihilators
of one another \cite{BS}.
In addition, we have \eqref{eqn:Nm}, \eqref{eqn:realT} and their duals for
the world-volumes of the branes.
By the fibrewise Fourier-Mukai transform \cite{Muk,BBH}, we see further that
the Chan-Paton line bundle $\ell^{\ee_2,\mm_2}$ on $\cN^{\mm_2}(C,G)_b$ shifts
the dual torus $\cN^{\ee_2=\bar 0}(C,\LG)_b$ in $\MH^{e_0=0}(C,\LG)_b$ to
$\cN^{\ee_2}(C,\LG)_b$ whereas the shift of $\cN^{\mm_2}(C,G)_b$ from
$\cN^{\mm_2=\bar 0}(C,G)_b$ in the original torus $\MH^{m_0=0}(C,G)_b$ gives
rise to the Chan-Paton line bundle ${}^L\!\ell^{\mm_2,\ee_2}$ on
$\cN^{\ee_2}(C,\LG)_b$.
This establishes the full identification \eqref{eqn:LBem} of
${}^L\!\cB^{\mm_2,\ee_2}$ as the mirror of $\cB^{\ee_2,\mm_2}$.

If the the gauge group $G$ is changed to the adjoint group $G_\ad$, the
target space of the $G_\ad$-theory is $\MH^{m_0=0}(C,G_\ad)$ and the brane
$\cB^{\ee_2,\mm_2}$ becomes $q_*\cB^{\ee_2,\mm_2}$ supported on
$\cN^{\mm_2}(C,G_\ad)$.
By \eqref{eqn:kerdel^}, $(\pi^*H^1(C',\ZG))^\vee$ is a subgroup of
$\ker(\del^1_{C'})^\vee$ and each $\ee_2\in\be(\bZ_2,G)$ can be regarded
as a $(\pi^*H^1(C',\ZG))^\vee$-coset in $\ker(\del^1_{C'})^\vee$.
Recall from \S\ref{sec:MHn} that $\cN^{\mm_2}(C,G)$ is a regular
$\pi^*H^1(C,\pG)$-cover of $\cN^{\mm_2}(C,G_\ad)$ while $\MH^{m_2}(C',G)$ is
a regular $\ker(\del_{\bZ_2}^1)$-cover of $\cN^{\mm_2}(C,G)$ and a regular
$\ker(\del_{C'}^1)$-cover of $\MH^{\mm_2}(C',G_\ad)=\cN^{\mm_2}(C,G_\ad)$.
So we have
$q_*\ell^{\ee_2,\mm_2}=\bigoplus_{\ee\in\ee_2}\ell_\ad^{\ee,\mm_2}$, where
$\ell_\ad^{\ee,\mm_2}:=\MH^{m_2}(C',G)\times_{\ee}\bC$ is a flat line bundle
over $\cN^{\mm_2}(C,G_\ad)$ determined by $\ee\in\ker(\del_{C'}^1)^\vee$.
If $\ee_2=0$, then the above simplifies to $q_*\ell^{\ee_2,\mm_2}=
\bigoplus_{\ee_1\in(\pi^*H^1(C',\ZG))^\vee}\ell_\ad^{\ee_1,\mm_2}$,
where $\ell_\ad^{\ee_1,\mm_2}:=\cN(C,G)\times_{\ee_1}\bC$.
But in general, the sum is over the coset $\ee_2$ which is a torsor over
$(\pi^*H^1(C',\ZG))^\vee$.
Let $\cB_\ad^{\ee,\mm_2}$ be the brane on $\MH^{m_0=0}(C,G_\ad)$ whose
world-volume is $\cN^{\mm_2}(C,G_\ad)$ and whose Chan-Paton bundle is
$\ell_\ad^{\ee,\mm_2}$.
Then the above means
\begin{equation}\label{eqn:qBem}
q_*\cB^{\ee_2,\mm_2}=\moplus{\ee\in\ee_2}{}\,\cB_\ad^{\ee,\mm_2}.
\end{equation}

Now suppose the gauge group is changed to the universal cover $\tG$.
A $G$-bundle over $C'$ of topological type $m_2$ lifts to a $\tG$-bundle
twisted by the discrete $B$-field $m_2$, and the latter pulls back to an
honest $\tG$-bundle over $C$.
Let $\MH^{m_2}(C',\tG)$ be the Hitchin moduli space from such twisted
$\tG$-bundles and let $\cN^{m_2}(C,\tG)\subset\MH(C,\tG)$ consist of
their pull-backs to $C$.
Then the regular $H^1(C,\pG)$-cover $\tq\colon\MH(C,\tG)\to\MH^{m_0=0}(C,G)$
restricts to a regular $\pi^*H^1(C',\pG)$-cover of $\cN^{m_2}(C,\tG)$ over
$\cN^{\mm_2}(C,G)$ (cf.~\S\ref{sec:MHn}).
Therefore the inverse image $\tq^{-1}(\cN^{\mm_2}(C,G))$ in $\MH(C,\tG)$ has
$\big|\frac{H^1(C,\pG)}{\pi^*H^1(C',\pG)}\big|$ connected components
including $\cN^{m_2}(C,\tG)$.
Since $H^1(C,\pG)$ acts on $\MH(C,\tG)$ as deck transformations and its
subgroup $\pi^*H^1(C',\pG)$ acts on $\cN^{m_2}(C,\tG)$, the set of connected
components form an $\frac{H^1(C,\pG)}{\pi^*H^1(C',\pG)}$-orbit, which is in
general a coset in $\coker(\udel{}^0_{C'})$ represented by $\mm_2$, due to
\eqref{eqn:cokerdel0}.
Thus we have
\begin{equation}\label{eqn:bqBem}
\tq^*\cB^{\ee_2,\mm_2}=\moplus{\mm\in\mm_2}{}\tB^{\ee_2,\mm},
\end{equation}
where $\tB^{\ee_2,\mm}$ is supported on the connected component
$\cN^\mm(C,\tG)\subset\tq^{-1}(\cN^{\mm_2}(C,G))$.
That the set of all components $\cN^\mm(C,\tG)$ has an Abelian group structure
will become more evident from the duality argument below.

We verify that the duality \eqref{eqn:LBem} of the branes is consistent with
the change of the gauge group.
The mirror of the $G$-theory has gauge group $\LG$ and the mirror of the brane
$\cB^{\ee_2,\mm_2}$ is precisely ${}^L\!\cB^{\mm_2,\ee_2}$ by \eqref{eqn:LBem}.
If $G$ is changed to $G_\ad$ in the original theory, then the brane
$\cB^{\ee_2,\mm_2}$ becomes $q_*\cB^{\ee_2,\mm_2}$ in \eqref{eqn:qBem}.
The mirror of the $G_\ad$-theory has gauge group $\tLG$ and the brane
${}^L\!\cB^{\mm_2,\ee_2}$ becomes ${}^L\!\tq^*({}^L\!\cB^{\mm_2,\ee_2})
=\bigoplus_{\ee\in\ee_2}{}^L\!\widetilde\cB^{\mm_2,\ee}$ by \eqref{eqn:bqBem}.
That $(q_*\cB^{\ee_2,\mm_2})^\vee={}^L\!\tq^*({}^L\!\cB^{\mm_2,\ee_2})$,
which is a special case of \eqref{eqn:qLB}, follows from
$(\cB_\ad^{\ee,\mm_2})^\vee={}^L\!\widetilde\cB^{\mm_2,\ee}$ which can be
verified directly.
In fact, consider the restrictions of two branes, $\cB_\ad^{\ee,\mm_2}$ and
$\cB_\ad^{\ee',\mm_2}$ ($\ee,\ee'\in\ee_2$), in the $G_\ad$-theory to a
regular torus component $\MH^{m_0=0}(C,G_\ad)_b$.
The two line bundles $\ell_\ad^{\ee,\mm_2}$ and $\ell_\ad^{\ee',\mm_2}$ over
the real torus $\cN^{\mm_2}(C,G_\ad)_b$ differ by the tensor product of the
line bundle $\ell_\ad^{\ee_1,\mm_2}$, where
$\ee_1:=\ee'-\ee\in(\pi^*H^1(C',\ZG))^\vee$.
In the mirror $\tLG$-theory, the branes ${}^L\!\widetilde\cB^{\mm_2,\ee}$ and
${}^L\!\widetilde\cB^{\mm_2,\ee'}$ are supported on $\cN^\ee(C,\tLG)$ and
$\cN^{\ee'}(C,\tLG)$.
Intersecting with the dual torus $\MH^{e_0=0}(C,\tLG)_b$, the real tori
$\cN^\ee(C,\tLG)_b$ and $\cN^{\ee'}(C,\tLG)_b$ are related by a translation
of $\ee_1\in\frac{H^1(C,\pi_1(\LG))}{\pi^*H^1(C',\pi_1(\LG))}$ by the property
of fibrewise Fourier-Mukai transform \cite{Muk}.
Moreover, there is an Abelian group structure on the set of connected
components $\cN^\ee(C,\tLG)$ just as the line bundles $\ell^{\ee,\mm_2}$
form an Abelian group in the original theory.
The argument also shows that the $\ee_1$-sector of the $G$-theory is mirror
to the sector of the $\LG$-theory with a relative winding $\ee_1$.
Similarly, if we change the gauge group to $\tG$ in the original theory and
$(\LG)_\ad$ in the mirror theory, then we have
$(\tB^{\ee_2,\mm})^\vee={}^L\!\cB_\ad^{\mm,\ee_2}$ and hence
$(\tq^*\cB^{\ee_2,\mm_2})^\vee={}^L\!q_*({}^L\!\cB^{\mm_2,\ee_2})$, which
verifies another part of \eqref{eqn:qLB}.
The relative winding $\mm_1$ in the $G$-theory is a character in the dual
theory.

\section{An application: quantisation of the moduli space via branes}
\label{sec:qt}
Finally, we apply the results to the quantisation of the Hitchin moduli space
$\MH(C',G)$ for a non-orientable surface $C'$.
To quantise a symplectic manifold $(M,\om)$, one finds a complexification
$M^\bC$ with an anti-holomorphic involution $\iota$ such that $M$ is a
connected component of the $\iota$-invariant part of $M^\bC$.
Assume that there is a holomorphic symplectic form $\om^\bC$ such that
$\om=\Re(\om^\bC)$ on $M$.
Assume also that there is a line bundle $\ell$ over $M^\bC$ with a lifted
action of $\iota$ preserving a unitary connection on $\ell$ whose curvature
is $\Re(\om^\bC)/\ii$.
Clearly, $\ell$ restricts to a pre-quantum line bundle over $M$.
Consider the $A$-model on $M^\bC$ with symplectic form $\Im(\om^\bC)$.
Then $\ell$ defines a space-filling canonical coisotropic $A$-brane $\cB_\cc$
on $M^\bC$.
On the other hand, a flat line bundle $\ell_0$ on $M$ defines a Lagrangian
$A$-brane $\cB_0$ on $M^\bC$.
The quantum Hilbert space $\cH$ of $(M,\om)$ with the pre-quantum line bundle
$\ell|_M\otimes\ell_0^{-1}$ is then $\Hom(\cB_\cc,\cB_0)$, the space of states
of open strings ending on $\cB_\cc$ and $\cB_0$ \cite{GW09}.
If $M^\bC$ has a mirror manifold $(M^\bC)^\vee$ with $B$-branes
$\cB_\cc^\vee,\cB_0^\vee$ that are dual to $\cB_\cc,\cB_0$, then the
Hilbert space $\cH$ can be written as $\Ext(\cB_\cc^\vee,\cB_0^\vee)$ in the
$B$-model on $(M^\bC)^\vee$ \cite{G}.

We need a slight generalisation with an $A$-brane $\cB'_0$ that comes from
a symplectic covering $p\colon(M',\om')\to(M,\om)$.
The flat line bundle $\ell_0$ over $M$ pulls back to $M'$.
Consider the same $A$-model on $M^\bC$ except the map $u$ from the worldsheet
$\Sig$ to $M^\bC$, when restricted to $u\colon\bSig\to M$, has a lifting
$u'\colon\bSig\to M'$.
Then the sigma-model has extra degrees of freedom on the boundary and has
a symmetry group $\Gam$, the group of deck transformations acting on $M'$,
which we assume to be Abelian for simplicity.
The sector labelled by the character $\al\in\Gam^\vee$ is equivalent to the
original $A$-model with a brane which we now construct (cf.~\S\ref{sec:em2dn}).
Let $\ell^\al:=M'\times_\al\bC$ and let $\cB_0^\al$ be the $A$-brane supported
on $M$ with the Chan-Paton line bundle $\ell_0\otimes\ell^\al$.
Since $p_*(p^*\ell_0)=\bigoplus_{\al\in\Gam^\vee}\ell_0\otimes\ell^\al$, we
have
\[ \Hom(\cB_\cc,\cB'_0)=\moplus{\al\in\Gam^\vee}{}\Hom(\cB_\cc,\cB^\al_0). \]
On the right hand side, $\Hom(\cB_\cc,\cB^\al_0)$, defined in the $A$-model
without augmenting $M'$, is the Hilbert space $\cH^\al$ from quantising
$(M,\om)$ with the pre-quantum line bundle
$\ell|_M\otimes\ell_0^{-1}\otimes(\ell^\al)^{-1}$.
The left hand side, defined in the extended $A$-model, is then the Hilbert
space $\cH'$ from quantising $(M',\om')$ with the pre-quantum line bundle
$p^*(\ell|_M\otimes\ell_0^{-1})$.
Thus $\cH'=\bigoplus_{\al\in\Gam^\vee}\cH^\al$ is simply the decomposition of
$\cH'$ according to the representation types of $\Gam$.
Such a relation of quantum mechanics on $M'$ and $M$ can be derived directly
in geometric quantisation or by path integrals.

In our case, given any $m_2\in\pG/2\pG$ or $\mm_2\in\bm(\bZ_2,G)$, the
symplectic manifold $(\MH^{m_2}(C',G),\om'_J)$ is a finite cover of
$\cN^{\mm_2}(C,G)$ whose complexification $\MH^{m_0=0}(C,G)$ has the
anti-holomorphic involution $\iota$ and the holomorphic symplectic form
$\om_J+\ii\om_K$ (both with respect to $I$).
Moreover, there is an Hermitian line bundle $\ell$ over $\MH^{m_0=0}(C,G)$
whose curvature is $\om_J/\ii$ with a lifted $\iota$-action \cite{Wu16}.
This defines the canonical coisotropic brane $\cB_\cc$ of type $(A,B,A)$.
It is clearly gradable in the sense of \cite{KL03,Li06} and thus the
sigma-model with this boundary condition is anomaly-free.
For simplicity, we pick the trivial line bundle over $\MH^{m_2}(C',G)$ which
pushes down via the projection $p$ to
$\bigoplus_{\ee_2\in\be(\bZ_2,G)}\ell^{\ee_2,\mm_2}$ on $\cN^{\mm_2}(C,G)$.
The quantum Hilbert space of $(\MH^{m_2}(C',G),\om'_J)$ is
(see \cite{Wu16} if $\del_{\bZ_2}^1=0$)
\[  \cH^{\mm_2}=\moplus{\ee_2\in\be(\bZ_2,G)}{}\cH^{\ee_2,\mm_2}, \]
where $\cH^{\ee_2,\mm_2}=\Hom(\cB_\cc,\cB^{\ee_2,\mm_2})$ in the $A$-model
with symplectic form $\om_K$.
The mirror theory is the $B$-model on $\MH^{e_0=0}(C,\LG)$ in the complex
structure $J$ \cite{KW}.
The mirror $\cB_\cc^\vee$ of $\cB_\cc$ is supported on a section of the dual
Hitchin fibration ${}^L\!h$ over the regular part of $\B$.
If $b\in\B$ is regular, the world-volume of $\cB_\cc^\vee$ intersects the
dual torus $\MH^{e_0=0}(C,\LG)_b$ at the point representing the line bundle
$\ell$ which is flat on the original torus $\MH^{m_0=0}(C,G)_b$.
Thus by \eqref{eqn:LBem}, we can write
$\cH^{\ee_2,\mm_2}=\Ext(\cB_\cc^\vee,{}^L\!\cB^{\mm_2,\ee_2})$ in the
$B$-model \cite{Wu16}.

\bigskip\medskip
\renewcommand{\thesection}{A}
\setcounter{subsection}{0}

\section*{Appendix. Some results in topology}

\subsection{Pontryagin duals and Poincar\'e duality}\label{sec:Poin}
If $A$ is a finitely generated Abelian group, the Pontryagin dual of $A$ is
$A^\vee:=\Hom(A,\mathrm U(1))$.
If $f\colon A\to B$ is a homomorphism of Abelian groups, the dual map
$f^\vee\colon B^\vee\to A^\vee$ satisfies
\begin{equation}\label{eqn:coker}
\ker(f)^\vee\cong\coker(f^\vee),\qquad\coker(f)^\vee\cong\ker(f^\vee).
\end{equation}
For example, if $B=A$ and $f$ is the multiplication by $2$, then
$\ker(f)=A_{[2]}$ is the $2$-torsion subgroup of $A$ while $\coker(f)=A/2A$
is another group of $2$-torsion elements, and \eqref{eqn:coker} becomes
\begin{equation}\label{eqn:two}
(A_{[2]})^\vee\cong A^\vee/2A^\vee,\qquad(A/2A)^\vee\cong(A^\vee)_{[2]}.
\end{equation}
If $A$ is finite, then $A\cong A^\vee$, though not naturally.
If $A$ and $B$ are finite, then
$|\ker(f)|/|\coker(f)|=|A|/|B|$ and $|\im(f)|=|\im(f^\vee)|$.
If in addition $f$ is surjective, then $f^\vee$ is injective, i.e.,
$B^\vee$ is a subgroup of $A^\vee$, and
\begin{equation}\label{eqn:dis}
\msum{e\in B^\vee}{}e(f(a))=\left\{\!\begin{array}{ll}
      |B| & \mbox{if } a\in\ker(f),\\
      0 & \mbox{if otherwise}.
      \end{array}\right.
\end{equation}

If $Y$ is a closed (i.e., compact and without boundary) orientable manifold
of dimension $n$, the cap product with the fundamental class
$[Y]\in H_n(Y,\bZ)$ defines the Poincar\'e duality isomorphism
$H^k(Y,A)\cong H_{n-k}(Y,A)$ ($0\le k\le n$).
There is also a natural isomorphism $H^k(Y,A^\vee)\cong H_k(Y,A)^\vee$.
Combining the two, we obtain an isomorphism
\begin{equation}\label{eqn:Poin}
H^k(Y,A^\vee)\cong H^{n-k}(Y,A)^\vee.
\end{equation}
for any integer $k$ ($0\le k\le n$).
Equivalently, there is a non-degenerate pairing
$H^k(Y,A^\vee)\times H^{n-k}(Y,A)\to\mathrm U(1)$.

Given a short exact sequence $0\to A'\map{i}A\map{j}A''\to0$ of Abelian groups,
there is a long exact sequence
\begin{equation}\label{eqn:longA}
\cdots\to H^k(Y,A')\map{i_Y^k}H^k(Y,A)\map{j_Y^k}H^k(Y,A'')\map{\del_Y^k}
H^{k+1}(Y,A')\to\cdots
\end{equation}
of cohomology groups of $Y$, where $i_Y^k,j_Y^k$ are induced by $i,j$ and
$\del_Y^k$ is the connecting homomorphism.
{}From the dual exact sequence
$0\to(A'')^\vee\map{\check i}A^\vee\map{\check j}(A')^\vee\to0$, where
$\check i=j^\vee$ and $\check j=i^\vee$, there is another long exact sequence
\begin{equation}\label{eqn:longA^}
\cdots\to H^k(Y,(A'')^\vee)\map{\check i_Y^k}H^k(Y,A^\vee)\map{\check j_Y^k}
H^k(Y,(A')^\vee)\map{\check\del_Y^k}H^{k+1}(Y,(A'')^\vee)\to\cdots
\end{equation}
with the induced maps $\check i_Y^k,\check j_Y^k$ and the connecting
homomorphism $\check\del_Y^k$.

We claim that $(\del_Y^{k-1})^\vee=\check\del_Y^{n-k}$,
$(i_Y^k)^\vee=\check j_Y^{n-k}$, $(j_Y^k)^\vee=\check i_Y^{n-k}$ under the
identification \eqref{eqn:Poin}.
Hence the long exact sequence \eqref{eqn:longA^} is the dual of
\eqref{eqn:longA}.
In fact, under Poincar\'e duality, $\del_Y^k$ intertwines with $\del^Y_{n-k}$,
where $\del^Y_k\colon H_k(Y,A'')\to H_{k-1}(Y,A')$ is the connecting
homomorphism in the homology long exact sequence.
On the other hand, the isomorphism $H^k(Y,A^\vee)\cong H_k(Y,A)^\vee$
intertwines $\check\del_Y^k$ and $(\del^Y_{k+1})^\vee$.
Therefore $(\del_Y^{k-1})^\vee=\check\del_Y^{n-k}$ and hence
$\ker(\check\del_Y^{n-k})=\coker(\del_Y^{k-1})^\vee$,
$\coker(\check\del_Y^{n-k})=\ker(\del_Y^{k-1})^\vee$.
The other two identifications of maps are shown similarly.
Thus the spaces $\ker(j_Y^k)\cong\coker(i_Y^k)$ and
$\ker(\check j_Y^{n-k})\cong\coker(\check i_Y^{n-k})$ are Pontryagin duals
to each other.

Any compact manifold $Y$ has a finite CW structure.
Let $S_\bullet(Y)$ be the cellular chain complex and let $c_k(Y)$ be the rank
of the free Abelian group $S_k(X)$ for $0\le k\le n$, where $n=\dim Y$.
Then the Euler characteristic of $Y$ is $\chi(Y)=\sum_{k=0}^n(-1)^kc_k(Y)$.
If $A$ is a finite Abelian group, then $H^\bullet(Y,A)$ are the cohomology
groups of the cochain complex $S^\bullet(Y,A):=\Hom(S_\bullet(Y),A)$.
Let $d^k\colon S^k(Y,A)\to S^{k+1}(Y,A)$ be the coboundary maps.
Then for each $k$, $|H^k(Y,A)|=|\ker(d^k)|/|\im(d^{k-1})|$ and
$|\ker(d^k)|\,|\im(d^k)|=|\Hom(S_k(Y),A)|=|A|^{c_k(Y)}$.
Therefore
\[ \mprod{k=1}n|H^k(Y,A)|^{(-1)^k}=\mprod{k=1}n|\Hom(C_k(Y),A)|^{(-1)^k}
   =\mprod{k=1}n|A|^{(-1)^kc_k(Y)}=|A|^{\chi(Y)}.          \]
If $Y$ is orientable, then $|H^k(Y,A)|=|H^{n-k}(Y,A^\vee)|=|H^{n-k}(Y,A)|$
by Poincar\'e duality.
So the above identity is trivial if $n$ is odd.
If $n=2l$ is even, it can be rewritten as
\begin{equation}\label{eqn:Euler}
\Big(\,\mprod{k=1}{l-1}|H^k(Y,A)|^{(-1)^k}\Big)\,
|H^l(Y,A)|^{(-1)^l/2}=|A|^{\chi(Y)/2}.
\end{equation}

Finally, let $Y'$ be a closed but non-orientable manifold of dimension $n$.
Then there is an exact sequence $1\to\pi_1(Y)\to\pi_1(Y')\to\bZ_2\to1$, where
$Y$ is the orientation double cover of $Y'$.
Given an Abelian group $A$, the action of $\bZ_2$ on $A$ as $\{\pm\id_A\}$
defines an action of $\pi_1(Y')$ on $A$ and a local system $\underline A$
on $Y'$.
Then Poincar\'e duality for $Y'$ is
\begin{equation}\label{eqn:Poin'}
H^k(Y',A^\vee)\cong H_k(Y',A)^\vee\cong H^{n-k}(Y',\underline A)^\vee.
\end{equation}
Given an exact sequence $0\to A'\to A\to A''\to0$ of Abelian groups, there is
a long exact sequence
\begin{equation}\label{eqn:longA'}
\cdots\to H^k(Y',\uA')\map{\ui_{Y'}^k}H^k(Y',\uA)\map{\uj_{Y'}^k}
H^k(Y',\uA'')\map{\udel_{Y'}^k}H^{k+1}(Y',\uA')\to\cdots,
\end{equation}
which is Poincar\'e dual to the homology long exact sequence of $Y'$.

\subsection{Principal bundles and the group of gauge transformations}
\label{sec:pi01}
Let $Y$ be a compact manifold of dimension $n$ and let $P$ be a principal
$G$-bundle over $Y$, where $G$ is a connected Lie group.
Obstructions to the triviality of $P$ or to constructing a global section of
$P$ are in $H^{k+1}(Y,\pi_k(G))$, $1\le k\le n-1$ \cite{St}.
In particular, the primary obstruction $\xi(P)\in H^2(Y,\pG)$ is the
obstruction to trivialising $P$ on a $2$-skeleton of $Y$ or to lifting
the structure group $G$ of $P$ to its universal cover $\tG$.
For any $k\ge1$ and Abelian group $A$, there is a natural homomorphism
\begin{equation}\label{eqn:vrho}
\vro^k_{Y,A}\colon H^k(Y,A)\to\Hom(\pi_k(Y),A)
\end{equation}
For example, $\vro^2_{Y,\pG}(\xi(P))\in\Hom(\pi_2(Y),\pG)$ is the map that
appears in the long exact sequence
\vspace{-2ex}

\[  \cdots\to\pi_2(P)\to\pi_2(Y)
\stackrel{\scalebox{.6}[.6]{$\vro^2_{Y,\pG}(\xi(P))$}}
{\xrightarrow{\hspace*{1cm}}}\pG\to\pi_1(P)\to\cdots   \]
of homotopy groups associated to the fibration $P\to Y$.
Principal $G$-bundles over a compact surface $C$ are classified by
$H^2(C,\pG)$.
If $G$ is compact, $G$-bundles over a compact $3$-manifold $Y$ are also
classified by $H^2(Y,\pG)$.
Over a compact $4$-manifold $X$, the set of topological types of $G$-bundles
fits in the exact sequence \cite{KR}
\begin{equation}\label{eqn:types}
0\to H^4(X,\pi_3(G))\to\{[P]:P\mbox{ is a }G\mbox{-bundle over }X\}
\map{\xi_*}H^2(X,\pG)\to0.
\end{equation}
If $\xi(P)=0$, then $k(P)\in H^4(X,\pi_3(G))$ is the obstruction to extending
the trivialisation of $P$ from the $2$-skeleton of $X$ to the whole space.
For a general $w\in H^2(X,\pG)$, the set $\xi_*^{-1}(w)$ is a torsor over
$H^4(X,\pi_3(G))$.

The group $\cG(P)$ of gauge transformations on a $G$-bundle $P$ over $Y$
is naturally the space of sections of $\Ad\,P:=P\times_\Ad G$, where $\Ad$
stands for the adjoint action of $G$ on $G$.
Two elements in $\cG(P)$ are in the same connected component if there is a
homotopy between them.
Obstructions to constructing such a homotopy are in $H^k(Y,\pi_k(G))$,
$0\le k\le n$ \cite{St}.
The map $\eta\colon\cG(P)\to H^1(Y,\pG)$ to the primary obstruction is a
homomorphism, and its kernel $\cG_1(P):=\ker(\eta)\subset\cG(P)$ consists of
sections of $\Ad\,P$ that can be lifted to $P\times_\Ad\tG$.
There is a commutative diagramme
\begin{equation}\begin{split}\label{eqn:pi0G}
\xymatrix@=1pc{1\ar[r]&\cG_1(P)\ar[r]\ar[d]&\cG(P)\ar[r]^{\hspace*{-5ex}\eta}
   \ar[d]&H^1(Y,\pG)\ar[r]\ar[d]^\id&1    \\
1\ar[r]&\pi_0(\cG_1(P))\ar[r]&\pi_0(\cG(P))\ar[r]^{\hspace*{-2ex}\eta_*}
   &H^1(Y,\pG)\ar[r]&1}
\end{split}\end{equation}
with exact rows and surjective vertical maps.
If $Y$ is a compact surface $C$, then $\pi_0(\cG(P))\cong H^1(C,\pG)$ and
$\cG_1(P)$ is connected.
If $G$ is compact, semisimple and $Y$ is a closed $3$-manifold, the second
row of \eqref{eqn:pi0G} becomes
\begin{equation}\label{eqn:pi0GY3}
1\to H^3(Y,\pi_3(G))\to\pi_0(\cG(P))\map{\eta_*}H^1(Y,\pG)\to1.
\end{equation}

A different but related approach is based on the observation \cite{Do02} that 
$\pi_0(\cG(P))$ coincides with the set of topological types of $G$-bundles
over $X:=T^1\times Y$ whose restrictions to a slice $\{t_0\}\times Y$ are $P$.
Given $g\in\cG(P)$, we can construct a bundle $P^g$ on $X$ by gluing the two
ends of the bundle $I\times P\to I\times Y$ using $g$.
Obstructions to the triviality of $P^g$ are in 
$H^{k+1}(X,\pi_k(G))\cong H^k(Y,\pi_k(G))\oplus H^{k+1}(Y,\pi_k(G))$.
The classes in $H^{k+1}(Y,\pi_k(G))$ are determined by the topology of $P$.
The remaining obstructions in $H^k(Y,\pi_k(G))$, $1\le k\le n$, detect the
homotopy class of $g\in\cG(P)$.
We have $\xi(P^g)=\eta(g)+\xi(P)$ according to the above decomposition when
$k=1$.
If $Y$ is a $3$-manifold and $G$ is semisimple, then the class in
$H^3(Y,\pi_3(G))$ is identified with $k(P^g)\in H^4(X,\pi_3(G))$.

Similarly, $\pi_1(\cG(P))$ can be identified with the set
of topological types of $G$-bundles over $S^2\times Y$ whose restrictions to
$D_\pm\times Y$, where $D_\pm$ are respectively the upper/lower hemispheres
of $S^2$, are the products $D_\pm\times P$.
A loop $\gam$ in $\cG(P)$ can be used to glue the two bundles $D_\pm\times P$
along the equator to form a bundle $P^\gam\to S^2\times Y$.
Topologically, $G$-bundles over $S^2\times Y$ are classified by $H^{k+1}
(S^2\times Y,\pi_k(G))\cong H^{k-1}(Y,\pi_k(G))\oplus H^{k+1}(Y,\pi_k(G))$.
Whereas the groups $H^{k+1}(Y,\pi_k(G))$ contain the characteristic classes of
$P$, the obstructions to contracting a loop in $\cG(P)$ to a point are in
$H^{k-1}(Y,\pi_k(G))$.
We have a map $\lam\colon L\cG(P)\to H^0(Y,\pG)\cong\pG$ to the set of primary
obstructions, and we denote its induced homomorphism by
$\lam_*\colon\pi_1(\cG(P))\to H^0(Y,\pG)$.
Then $\xi(P^\gam)=\lam_*([\gam])+\xi(P)$ according to the above decomposition
when $k=1$.
If $Y=C$ is a closed orientable surface, there is an exact sequence
\[ 0\to H^2(C,\pi_3(G))\to\pi_1(\cG(P))\map{\lam_*}H^0(C,\pG)\to0,   \]
where the class in $H^2(C,\pi_3(G))$ is identified with
$k(P^\gam)\in H^4(S^2\times C,\pi_3(G))$.
There is also an exact sequence
\begin{equation}\label{eqn:pi1GC2}
0\to H^2(C,\pi_3(G))\to\pi_1(\cG(P)/\ZG)\to H^0(C,\pGad)\to0.
\end{equation}

Now consider the space $\cB(P):=\cA(P)/\cG(P)$ of gauge equivalence
classes of connections on a $G$-bundle $P$ over $Y$.
The group $\cG(P)$ of gauge transformations acts on the space $\cA(P)$ of
connections while the centre $\ZG$ of $G$, regarded as a subgroup of $\cG(P)$,
acts on $\cA(P)$ trivially.
We can restrict to connections whose stabiliser is precisely $\ZG$ and still
denote the space by $\cA(P)$, which remains at least $2$-connected and on
which the group $\cG(P)/\ZG$ acts freely.
In this way, $\cB(P)$ is smooth (though infinite dimensional) and we have
\begin{equation}\label{eqn:isompi}
\pi_1(\cB(P))\cong\pi_0(\cG(P)/\ZG)\cong\pi_0(\cG(P)),\qquad
\pi_2(\cB(P))\cong\pi_1(\cG(P)/\ZG).
\end{equation}
Concretely, for any $g\in\cG(P)$, a connection on the bundle
$P^g\to T^1\times Y$ defines a loop in $\cB(P)$ representing the class $[g]$.
For any loop $\gam$ in $\cG(P)/\ZG$, there is a loop $\gam\cdot A_0$ in
$\cA(P)$ by choosing a base connection $A_0$.
Then a disc in $\cA(P)$ bounded by $\gam\cdot A_0$ descends to a sphere in
$\cB(P)$ representing an element of $\pi_2(\cB(P))$.

Finally, the action of $\pi_0(\cG(P))$ on $\pi_1(\cG(P)/\ZG)$ is given by
the conjugation of $\cG(P)$ on the identity component of $\cG(P)/\ZG$.
This action is trivial because for any $g\in\cG(P)$ and any loop $\gam$ in
the identity component of $\cG(P)/\ZG$, the bundles on $S^2\times Y$ defined
by $\gam$ and $g\gam g^{-1}$ can be shown to be topologically equivalent.
By the isomorphisms in \eqref{eqn:isompi}, the action of $\pi_1(\cB(P))$ on
$\pi_2(\cB(P))$ is also trivial.

\subsection{Gauge transformations on $G$-, $G_\ad$- and twisted bundles}
\label{sec:GGad}
Let $G$ be a connected compact semisimple Lie group.
A principal $G$-bundle $P$ over $Y$ defines a principal $G_\ad$-bundle
$P_\ad:=P/\ZG$ over the same space.
There is a commutative diagramme for $P_\ad$ similar to \eqref{eqn:pi0G}.
Let $q\colon\cG(P)\to\cG(P_\ad)$ be the map induced by the projections
$G\to G_\ad$, $P\to P_\ad$.
Since $P_\ad\times_\Ad G=P\times_\Ad G$ and
$P_\ad\times_\Ad\tG=P\times_\Ad\tG$, the restriction of $q$ to the subgroup
$\cG_1(P)=\ker(\eta)$ is a surjective map $q_1\colon\cG_1(P)\to\cG_1(P_\ad)$
with $\ker(q_1)\cong\ZG$.
There is a commutative diagramme
\[ \xymatrix@=.9pc{&1\ar[d]&&&  \\  &\ZG\ar[d]&&0\ar[d]& \\
1\ar[r]&\cG_1(P)\ar[r]\ar[d]^{q_1}&\cG(P)\ar[r]^{\hspace*{-5ex}\eta}\ar[d]^q&
H^1(Y,\pG)\ar[r]\ar[d]^{i_Y^1}&1   \\
1\ar[r]&\cG_1(P_\ad)\ar[r]\ar[d]&\cG(P_\ad)\ar[r]^{\hspace*{-5ex}\eta_\ad}
\ar@{..>}[dr]_{\hspace*{-2em}\zeta} &
H^1(Y,\pGad)\ar[r]\ar[d]^{j_Y^1}&1  \\
&1&&H^1(Y,\ZG)\ar[d]^{\del_Y^1}& \\ &&&H^2(Y,\pG)&}         \]
with exact rows and columns, the right column being \eqref{eqn:long}.
It follows from a straightforward diagramme chasing that
$\ker(q)\cong\ker(q_1)\cong\ZG$ and
$\coker(q)\cong\coker(i_Y^1)\cong\ker(\del_Y^1)$.
Therefore there is an exact sequence
\begin{equation}\label{eqn:GGad}
1\to\ZG\to\cG(P)\map{q}\cG(P_\ad)\map\zeta\ker(\del_Y^1)\to1,
\end{equation}
where $\zeta:=j_Y^1\circ\eta_\ad$.
For $g_\ad\in\cG(P_\ad)$ which is a section of $P_\ad\times_\Ad G_\ad$,
$\zeta(g_\ad)$ is the obstruction to lifting it to a section of
$P_\ad\times_\Ad G=P\times_\Ad G$ or to $\cG(P)$.
Taking the homotopy classes of gauge transformations, we obtain another exact
sequence
\begin{equation}\label{eqn:pi0GGad}
1\to\pi_0(\cG(P))\map{q_*}\pi_0(\cG(P_\ad))\map{\zeta_*}\ker(\del_Y^1)\to1.
\end{equation}
If $Y$ is a compact $3$-manifold, then $\pi_0(\cG(P))$ is given by
\eqref{eqn:pi0GY3} and thus \eqref{eqn:pi0GGad} becomes obvious by using
\eqref{eqn:long}.

There is another interpretation of \eqref{eqn:pi0GGad} in terms of homotopy
classes of paths in $\cB(P)$ using \eqref{eqn:isompi}.
Since $\cA(P)=\cA(P_\ad)$ and since $\ZG$ acts trivially on $\cA(P)$, we
deduce from \eqref{eqn:GGad} that $\cB(P)$ is a regular
$\ker(\del_Y^1)$-cover of $\cB(P_\ad)$.
Thus \eqref{eqn:pi0GGad} can be regarded as the short exact sequence of
fundamental groups of covering spaces
\begin{equation}\label{eqn:pi1BBad}
1\to\pi_1(\cB(P))\to\pi_1(\cB(P_\ad))\map{\zeta_*}\ker(\del_Y^1)\to1.
\end{equation}
A loop on $\cB(P_\ad)$ lifts to a loop on $\cB(P)$ if and only if its class
in $\pi_1(\cB(P_\ad))$ is in $\ker(\zeta_*)\cong\pi_0(\cG(P))$.
For a general $g\in\ker(\del_Y^1)$, $\zeta_*^{-1}(g)$ is the set
$\pi_1(\cB(P),g)$ of homotopy classes of paths $[A(t)]$ in $\cB(P)$ such that
$[A(1)]=g\cdot[A(0)]$.
The set $\zeta_*^{-1}(g)$ is a torsor over $\pi_1(\cB(P))$.

In \S\ref{sec:em4d}, we used the notion of principal bundles twisted by a
discrete $B$-field \cite{Mac,Wu15}.
Suppose $\{U_\al\}$ is an open cover of $Y$.
Then a principal $G$-bundle over $Y$ is a collection of transition functions
$g_{\al\beta}\colon U_\al\cap U_\beta\to G$ satisfying the cocycle condition
$g_{\al\beta}g_{\beta\gam}g_{\gam\al}=1$ on triple intersections
$U_\al\cap U_\beta\cap U_\gam$.
A discrete $B$-field in our setting (or a $\ZG$-gerbe) is an element of
$H^2(Y,\ZG)$ so that there is a `holonomy' in $\ZG$ for each closed surface
in $Y$.
In the \v Cech language, a cohomology class $[h]\in H^2(Y,\ZG)$ is described
by a collection of $h_{\al\beta\gam}\in\ZG$ that satisfy a cocycle condition
on quadruple intersections of the open sets.
A principal $G$-bundle twisted by the discrete $B$-field can be described
as a collection of transition functions
$g_{\al\beta}\colon U_\al\cap U_\beta\to G$ satisfying
$g_{\al\beta}g_{\beta\gam}g_{\gam\al}=h_{\al\beta\gam}$ on all triple
intersections $U_\al\cap U_\beta\cap U_\gam$ \cite{Mac}
(see \cite{Wu15} for a definition using bundle gerbes).
If $g'_{\al\beta}$ is related to $g_{\al\beta}$ by
$g'_{\al\beta}=f_\al^{-1}g_{\al\beta}f_\beta$ for some
$f_\al\colon U_\al\to G$, then $g'_{\al\beta}$ satisfy the same condition
and it defines an equivalent twisted $G$-bundle over $Y$.
A twisted $G$-bundle $P$ induces an honest $G_\ad$-bundle $P_\ad$ with
$\xi(P_\ad)\in(j_Y^2)^{-1}([h])\subset H^2(Y,\pGad)$.

We can define gauge transformations and connections on a twisted $G$-bundle
$P$.
Since the failure of the cocycle condition $h_{\al\beta\gam}$ is in the centre
$\ZG$, the definitions in the untwisted case work equally well here.
A gauge transformation on $P$ is a collection of maps from $U_\al$ to $G$
that are related by the conjugation of $g_{\al\beta}$ on $U_\al\cap U_\beta$.
It can be identified as a section of $\Ad P:=P_\ad\times_\Ad G$ which is a
well-defined (though not principal) bundle over $Y$ with fibre $G$.
They form a group $\cG(P)$ whose Lie algebra is $\Om^0(Y,\ad P)$, where the
adjoint bundle $\ad P:=P_\ad\times_\ad\fg$ ($\fg$ is the Lie algebra of $G$)
is also well defined.
A connection on $P$ is a collection of $A_\al\in\Om^1(U_\al,\fg)$ that are
related by the gauge transformation of $g_{\al\beta}$ on $U_\al\cap U_\beta$.
They form an affine space $\cA(P)$ modelled on $\Om^1(Y,\ad P)$.
The curvature is a collection of $2$-forms $F_\al\in\Om^2(U_\al,\fg)$ which
form $\ad P$-valued $2$-form $F\in\Om^2(Y,\ad P)$.
We still have an action of $\cG(P)$ on $\cA(P)$ such that $\ZG$ as a subgroup
of $\cG(P)$ acts trivially, with the quotient space $\cB(P):=\cA(P)/\cG(P)$.
The exact sequences \eqref{eqn:GGad}, \eqref{eqn:pi0GGad} and
\eqref{eqn:pi1BBad} remain valid when $P$ is a twisted $G$-bundle.

We will be mostly using twisted bundles in the following setting.
Suppose $P$ is a $G$-bundle on $Y$ which is lifted to a $\tG$-bundle $\tP$
twisted by the discrete $B$-field $i_Y^2(\xi(P))\in H^2(Y,Z(\tG))$.
Then as in \eqref{eqn:GGad} and \eqref{eqn:pi0GGad}, the group $\cG(\tP)$
of gauge transformations on $\tP$ fits in the exact sequences
\[ 1\to\pG\to\cG(\tP)\to\cG(P)\to H^1(C,\pG)\to1,  \]
\[ 1\to\pi_0(\cG(\tP))\to\pi_0(\cG(P))\to H^1(C,\pG)\to1. \]

\subsection{Homotopy groups of the Hitchin moduli spaces}\label{sec:MH}
Let $C$ be a closed orientable surface.
Recall from \S\ref{sec:4d2d} that the Hitchin moduli space $\MH(C,G)$ is the
space of pairs $(A,\phi)$ satisfying Hitchin's equations \eqref{eqn:H} modulo
gauge transformations.
Here $A$ is a connection on a $G$-bundle $P$ over $C$ and
$\phi\in\Om^1(C,\ad P)$.
We shall work with the smooth part of $\MH(C,G)$ for which we use the same
notation.

There is a decomposition according to the topological type of $G$-bundles
over $C$,
\begin{equation}\label{eqn:union}
\MH(C,G)=\msqcup{m_0\in H^2(C,\pG)}{}\MH^{m_0}(C,G).
\end{equation}
Each $\MH^{m_0}(C,G)$ is connected because $\MH(C,G)$ can be identified with
the moduli space $\Mf(C,G^\bC)$ of flat $G^\bC$-connections on $C$
\cite{Do87,Co} which has the same decomposition, and each $\Mf^{m_0}(C,G^\bC)$
is connected \cite{Li}.
This argument uses the complex structure $J$ on $\MH(C,G)$ because in
constructing $\Mf(C,G^\bC)$, instead of imposing the constraint $d_A*\phi=0$
(vanishing of the moment map for $\om_J$), we take the quotient by the group
$\cG(P^\bC)$ which acts holomorphically with respect to $J$.
The same conclusion can be reached using the complex structure $I$ as
$\MH(C,G)$ is also identified with the moduli space of Higgs pairs
\cite{Hi87,Si}.
The latter contains the cotangent bundle of the moduli space of stable
$G^\bC$-bundles \cite{NS,Rtn} or the moduli space $\Mf(C,G)$ of flat
$G$-connections.
The complement of $T^*\Mf(C,G)$ in $\MH(C,G)$, containing stable Higgs pairs
whose underlying holomorphic bundle is unstable, is of codimension at least
$g(C)$ when $G^\bC=\mathrm{SL}(2,\bC)$ \cite{Hi87}; there should be a lower
bound on the codimension for other types of semisimple Lie groups.
Thus $\pi_0(\MH(C,G))=\pi_0(T^*\Mf(C,G))=\pi_0(\Mf(C,G))$, which is
$H^2(C,\pG)\cong\pG$ by \cite{AB}.

Next, we study the fundamental group of $\MH(C,G)$.
If $m_0=0$, we have $\pi_1(\MH^0(C,G))\cong H^1(C,\pG)$ \cite{Hi87,KW,LR}.
In general, by Morse-theoretic arguments, we have \cite{Da,DU,BGG}
\begin{equation}\label{eqn:piMG}
\pi_k(\MH^{m_0}(C,G))\cong\pi_k(\Mf^{m_0}(C,G))\cong\pi_{k-1}(\cG(P)/\ZG)
\end{equation}
for small $k$.
(Strictly speaking, this was proved for $G^\bC=\mathrm{GL}(n,\bC)$ or
$\mathrm{SL}(n,\bC)$, but we expect their method applies to other types
of semisimple Lie groups.)
Since $\pi_0(\cG(P)/\ZG)=\pi_0(\cG(P))=H^1(C,\pG)$ (cf.~\S\ref{sec:pi01}),
we obtain
\begin{equation}\label{eqn:pi1M}
\pi_1(\MH^{m_0}(C,G))\cong H^1(C,\pG).
\end{equation}
Furthermore, since $\pi_1(\cG(P)/\ZG)$ is given by \eqref{eqn:pi1GC2}, we
obtain an exact sequence
\begin{equation}\label{eqn:pi2M}
0\to H^2(C,\pi_3(G))\to\pi_2(\MH^{m_0}(C,G))\map{\zeta_*}H^0(C,\pGad)\to0.
\end{equation}
Just like the homotopy groups of $\cB(P)$ in \S\ref{sec:pi01},
$\pi_1(\MH^{m_0}(C,G))$ is Abelian, and it acts trivially on
$\pi_2(\MH^{m_0}(C,G))$.

There is a regular $H^1(C,\ZG)$-covering
$q\colon\MH(C,G)\to\MH^\dcirc(C,G_\ad)$ \cite{HWW}, where 
\begin{equation}\label{eqn:union-ad}
\MH^\dcirc(C,G_\ad):=\msqcup{m_0\in H^2(C,\pG)}{}\MH^{m_0}(C,G_\ad)
\end{equation}
is the subset of $\MH(C,G_\ad)$ from the Hitchin pairs of the $G_\ad$-bundles
that lift to $G$-bundles.
Changing the gauge group to $G_\ad$, \eqref{eqn:pi1M} becomes
$\pi_1(\MH^{m_0}(C,G_\ad))\cong H^1(C,\pGad)$, where $m_0\in H^2(C,\pGad)$.
Thus \eqref{eqn:pi1M} is consistent with the behaviour of $\pi_1$ under the
covering map $q\colon\MH^{m_0}(C,G)\to\MH^{m_0}(C,G_\ad)$.
On the other hand, $\pi_2(\MH^{m_0}(C,G))$ depends only on $G_\ad$ or on the
Lie algebra of $G$, and it does not change when the gauge group becomes
$G_\ad$.
Indeed, covering maps do not change the second (and higher) homotopy groups.

If the gauge group is $\tG$, $\MH(C,\tG)$ has a single connected component
from topologically trivial bundles.
There is a regular $H^1(C,\pG)$-covering
$\tq\colon\MH(C,\tG)\to\MH^{m_0=0}(C,G)$.
More generally, we have the moduli space $\MH^{m_0}(C,\tG)$ of the gauge
equivalence classes of Hitchin pairs from a $\tG$-bundle twisted by a
discrete $B$-field $m_0\in H^2(C,\pG)$.
Such a twisted $\tG$-bundle descends to an honest $G$-bundle, and we have
a regular $H^1(C,\pG)$-covering
$\tq\colon\MH^{m_0}(C,\tG)\to\MH^{m_0}(C,G)$ \cite{Wu15}.
Each $\MH^{m_0}(C,\tG)$ is simply connected, which also explains the
fundamental group \eqref{eqn:pi1M}.
Collectively, we have a $H^1(C,\pG)$-covering map
$\tq\colon\MH^\dcirc(C,\tG)\to\MH(C,G)$, where
\begin{equation}\label{eqn:union-cov}
\MH^\dcirc(C,\tG):=\msqcup{m_0\in H^2(C,\pG)}{}\MH^{m_0}(C,\tG)
\end{equation}
is the Hitchin moduli space of possibly twisted $\tG$-bundles that descend
to honest $G$-bundles.

Now suppose $C'$ is a closed non-orientable surface.
The moduli space $\MH(C',G)$ can be defined (cf.~\S\ref{sec:non-or}) and it
can be identified with the moduli space $\Mf(C',G^\bC)$ of flat
$G^\bC$-connections on $C'$ \cite{HWW}.
Its smooth part (denoted by the same notation) has a K\"ahler form $\om'_J$.
If $C'$ is the connected sum of $g(C')>2$ copies of $\bR P^2$, the real
dimension of $\MH(C',G)$ or $\Mf(C',G^\bC)$ is $2(g(C')-2)\dim G$.

Similar to \eqref{eqn:union}, we have a decomposition
\begin{equation}\label{eqn:union'}
\MH(C',G)=\msqcup{m_2\in\pG/2\pG}{}\MH^{m_2}(C',G)
\end{equation}
according to the topological types of $G$-bundles over $C'$; the moduli
space $\Mf(C',G^\bC)$ has the same decomposition.
We expect that each $\MH^{m_2}(C',G)$ or equivalently each
$\Mf^{m_2}(C',G^\bC)$ is connected if $C'$ is the connected sum of $g(C')>2$
copies of $\bR P^2$ and if $G$ is compact, semisimple.
For the moduli space $\Mf(C',G)$ of flat $G$-connections, we do have the
connectedness of each $\Mf^{m_2}(C',G))$
(see \cite{HL03} for $g(C')=3$ and $g(C')\ge5$ and \cite{HL08} for $g(C')=4$).
The connectedness of $\MH^{m_2}(C',G)$ or $\Mf^{m_2}(C',G^\bC)$ for each
$m_2$ remains open, but it follows if $\MH^{m_2}(C',G)$ contains
$T^*\Mf^{m_2}(C',G))$ with a complement of high codimension.
See \cite{BC} for an affirmative result in the case $G=\mathrm{SU}(n)$,
$G^\bC=\mathrm{SL}(n,\bC)$.

For higher homotopy groups, we expect the validity of the isomorphisms
\eqref{eqn:piMG} for small $k$ when $C$ is replaced by a non-orientable
surface $C'$.
This is consistent with the known result \cite{Rs} on flat vector bundles
over $C'$.
If so, then $\pi_1(\MH^{m_2}(C',G))\cong H^1(C',\pG)$ as for orientable
surfaces.
Since $H^2(C',\pi_3(G))=0$, \eqref{eqn:pi2M} reduces to the statement that
the map $\zeta'_*\colon\pi_2(\MH^{m_2}(C',G))\to H^0(C',\pGad)\cong\pGad$
is an isomorphism.
These results, together with their consequence in \S\ref{sec:MHn}, are
compatible with the physics requirements (see \S\ref{sec:em2dn}).

\subsection{The universal bundle, its connection and curvature}
\label{sec:univ}
Let $G$ be a compact semisimple Lie group and let $P$ be a principal
$G$-bundle over a smooth manifold $M$.
The group $\cG(P)$ of gauge transformations acts on $P$ on the left and the
action commutes with the right action of $G$ on $P$.
It also acts on the space $\cA(P)$ of connections on $P$ on the right by
pulling back, i.e., $A\mapsto g^*A$ by $g\in\cG(P)$.
Therefore $\cG(P)$ acts on $\cA(P)\times P$ on the left by
$g\colon(A,p)\mapsto((g^{-1})^*A,g\cdot p)$, where
$g\in\cG(P),A\in\cA(P),p\in P$.
The Lie algebra of $\cG(P)$ can be identified with $\Om^0(M,\ad P)$ while the
tangent space of $\cA(P)$ at any $A$ is $T_A\cA(P)\cong\Om^1(M,\ad P)$.
For $v\in\Om^0(M,\ad P)$, let $V_v$ be the induced vector field on $P$.
Then the induced vector field on $\cA(P)\times P$ is $(-d_Av,V_v(p))$ at
$(A,p)\in\cA(P)\times P$.

The centre $\ZG$ of $G$ is a subgroup of $\cG(P)$ as constant gauge
transformations and it acts trivially on $\cA(P)$.
We consider a smooth part of the orbit space $\cB(P):=\cA(P)/\cG(P)$ that
comes from the connections whose stabiliser subgroup is precisely $\ZG$.
For these connections $A$, the operator
$d_A\colon\Om^0(M,\ad P)\to\Om^1(M,\ad P)$ is injective and, by picking
a Riemannian metric on $M$, $d_A^*d_A$ is an invertible operator on
$\Om^0(M,\ad P)$.
Then $\cA(P)\to\cB(P)$ is a principal bundle of structure group $\cG(P)/\ZG$.
We define a connection on this bundle so that the horizontal subspaces are
the orthogonal compliments of the orbits of $\cG(P)$.
The connection $1$-form at $A\in\cA(P)$ is $(d_A^*d_A)^{-1}d_A^*$ and the
curvature at $A$ on two horizontal vectors
$\al,\beta\in\ker(d_A^*)\subset\Om^1(M,\ad P)$ is
$(d_A^*d_A)^{-1}(b_\al^*\beta-b_\beta^*\al)$.
Here $b_\al\colon\Om^0(M,\ad P)\to\Om^1(M,\ad P)$ is the operator
$v\mapsto[\al,v]$, and $b_\al^*$ is its adjoint operator.

Consider the pull-back of the $G$-bundle $P$ under the projection
$\cA(P)\times M\to M$.
Restricting to the connections whose stabiliser is $\ZG$, there is a free
action of $\cG(P)$ on the total space $\cA(P)\times P$ covering its action
on the base $\cA(P)\times M$.
Taking the quotients by $\cG(P)$, we get the universal bundle
$\cU(P):=\cA(P)\times_{\cG(P)}P\to\cB(P)\times M$ whose structure group is
$G_\ad$ \cite{AS}.
We define a connection of the $G$-bundle $\cA(P)\times P\to\cA(P)\times M$
whose horizontal subspace at $(A,p)\in\cA(P)\times P$ consists of tangent
vectors $(\al,S)\in\Om^1(M,\ad P)\oplus T_pP$ such that
$V_{(d_A^*d_A)^{-1}d_A^*\al}(p)+S\in T_pP$ is horizontal with respect to the
connection $A$.
Then for all $v\in\Om^0(M,\ad P)$, the vector $(-d_Av,V_v(p))$ is horizontal
at $(A,p)$ and the connection is invariant under the action of $\cG(P)$.
The curvature vanishes when contracted with the vector $(-d_Av,V_v)$.
Therefore the connection descends to the universal (or tautological)
connection $A^\cU$ on $\cU(P)$, and for $\al,\beta\in\ker(d_A^*)$ and tangent
vectors $S,T\in T_pM$, the curvature is \cite{AS}
\[ F^\cU_{[(A,p)]}((\al,S),(\beta,T))=(d_A^*d_A)^{-1}(b_\al^*\beta-
      b_\beta^*\al)(p)+\al_p(T)-\beta_p(S)+(F_A)_p(S,T).               \]
The above construction of $A^\cU$ follows the general procedure of passing
from connections on $P$ (or $P_\ad$) to one on the Borel construction
$\cA(P)\times_{\cG(P)}P=\cA(P)\times_{\cG(P)/\ZG}P_\ad$ of the $\cG(P)$-action
on $P$, and the universal curvature $F^\cU$ can also be obtained by applying
the Chern-Weil map on the $\cG(P)$-equivariant extension of $F_A$.

In the context of Hitchin's moduli space, the universal bundle and its
connection can be constructed similarly.
Let $P$ be a principal $G$-bundle over a closed orientable surface $C$
(though the restriction to two dimensions is not essential) and let
$\AH(P)\subset\cA(P)\times\Om^1(C,\ad P)$ be the subset of Hitchin pairs,
i.e., $(A,\phi)$ satisfying Hitchin's equations \eqref{eqn:H}.
As above, we only consider a smooth subset of the Hitchin moduli space
$\MH(P):=\AH(P)/\cG(P)$ from Hitching pairs with minimal stabiliser.
The tangent space of $\AH(P)$ at $(A,\phi)$ consists of vectors $(\al,\xi)$
satisfying the linearised Hitchin equations $d_A\al=[\phi,\xi]$,
$d_A^*\xi+b_\phi^*\al=0$, $d_A\xi+[\phi,\al]=0$.
There is a connection on the principal bundle $\AH(P)\to\MH(P)$ by choosing
the horizontal spaces as the orthogonal compliments of the $\cG(P)$-orbits.
A vector $(\al,\xi)$ is orthogonal to the orbits of $\cG(P)$ if and only if
$(\al,\xi)\in\ker(d_A,b_\phi)^*$ or equivalently, $d_A^*\al+b_\phi^*\xi=0$.
The curvature on horizontal vectors
$(\al,\xi),(\beta,\eta)\in\ker(d_A,b_\phi)^*$ is $(d_A^*d_A
+b_\phi^*b_\phi)^{-1}(b_\al^*\beta-b_\beta^*\al+b_\xi^*\eta-b_\eta^*\xi)$.

The universal bundle $\UH(P):=\AH(P)\times_{\cG(P)}P$ is a principal
$G_\ad$-bundle over $\MH(P)\times C$.
We define a connection on the $G$-bundle $\AH(P)\times P\to\AH(P)\times C$
by declaring that a vector $(\al,\xi,S)$ at $(A,\phi,p)$ is horizontal if
$V_{(d_A^*d_A+b_\phi^* b_\phi)^{-1}(d_A^*\al+b_\phi^*\xi)}(p)+S$ is so at
$p\in P$ with respect to $A$.
This connection is invariant under $\cG(P)$ and for any $v\in\Om^0(C,\ad P)$,
the vector $(-d_Av,-b_\phi v,V_v)$ from the infinitesimal $\cG(P)$-action is
horizontal at $(A,\phi,p)$.
The curvature vanishes when contracted with the vector
$(-d_Av,-b_\phi v,V_v)$.
So the connection descends to the universal (or tautological) connection
$A^\cU$ on $\UH(P)\to\MH(P)\times C$ whose curvature we denote by $F^\cU$.
For $(\al,\xi),(\beta,\eta)\in\ker(d_A,b_\phi)^*$ and tangent vectors
$S,T\in T_pC$, the curvature is
\begin{equation}\label{eqn:univ-F}
F^\cU_{[(A,\phi,p)]}((\al,\xi,S),(\beta,\eta,T))
=(d_A^*d_A+b_\phi^*b_\phi)^{-1}(b_\al^*\beta-b_\beta^*\al
+b_\xi^*\eta-b_\eta^*\xi)(p)+\al_p(T)-\beta_p(S)+(F_A)_p(S,T).
\end{equation}
The universal connection $A^\cU$ can be understood from a similar Borel
construction and the universal curvature \eqref{eqn:univ-F} is obtained
by a Chern-Weil map on the $\cG(P)$-equivariant extension of $F_A$.

In addition to the universal connection $A^\cU$, there is a universal `Higgs'
field $\phi^\cU$ constructed as follows.
First, there is a tautological $1$-form on $\AH(P)\times C$ with values in
$\AH(P)\times\ad P$ given by $(A,\phi,p)\mapsto\phi_p$.
Since it is basic with respect to the action of $\cG(P)$, it descends to
$\phi^\cU\in\Om^1(\MH(P)\times C,\AH(P)\times_{\cG(P)}\ad P)$.
We have
\[ (d_{A^\cU}\phi^\cU)_{[(A,\phi,p)]}((\al,\xi,S),(\beta,\eta,T))
   =\xi_p(T)-\eta_p(S)+(d_A\phi)_p(S,T),   \]
where $(\al,\xi,S),(\beta,\eta,T)$ are as above.
We call $(A^\cU,\phi^\cU)$ the universal Hitchin pair; it satisfies the
Hitchin equations along $C$ for each point in $\MH(P)$.
See \cite{H} for another construction using the language of algebraic
geometry.

\subsection{Characteristic classes of the universal bundle}\label{sec:univ-c}
Let $G$ be a compact semisimple Lie group.
Suppose $P$ is a $G$-bundle over a closed orientable surface $C$.
Let $\AH(P)$ be the set of Hitchin pairs and $\MH(P):=\AH(P)/\cG(P)$ be the
Hitchin moduli space.
The quotient of $\AH(P)\times P$ by the group $\cG(P)$ of gauge
transformations is the total space of the universal bundle
$\UH(P)\to\MH(P)\times C$.
Since $\cG(P)$ contains $\ZG$ which acts trivially on $\AH(P)$, $\UH(P)$ is
a principal $G_\ad$-bundle (\S\ref{sec:univ}).
We will write $\cM=\MH(P)$ and $\cU=\UH(P)$ for short.

The discrete flux $\xi(\cU)\in H^2(\cM\times C,\pGad)$ is the obstruction
to lifting $\cU$ to a $\tG$-bundle.
Using the induced maps in \eqref{eqn:longA} but for the space $\cM\times C$,
the obstruction to lifting $\cU$ to a $G$-bundle is
$j_{\cM\times C}^2(\xi(\cU))\in H^2(\cM\times C,\ZG)$.
We write $\xi(\cU)=\xi^{2,0}(\cU)+\xi^{1,1}(\cU)+\xi^{0,2}(\cU)$ according to
the K\"unneth decomposition
\[ H^2(\cM\times C,\pGad)\cong H^2(\cM,\pGad)\oplus H^1(\cM,H^1(C,\pGad))
   \oplus H^2(C,\pGad). \]
Then $j_{\cM\times C}^2(\xi(\cU))=j_\cM^2(\xi^{2,0}(\cU))
+j_{\cM\times C}^2(\xi^{1,1}(\cU))+j_C^2(\xi^{0,2}(\cU))$ according to the
same decomposition but with the coefficient group $\ZG$.
Clearly, $j_C^2(\xi^{0,2}(\cU))$ vanishes because the restriction of $\cU$ to
$C$ comes from a $G$-bundle.
In fact, since $\xi^{0,2}(\cU)=\xi(P_\ad)=i_C^2(\xi(P))$, we have
$j_C^2(\xi^{0,2}(\cU))=(j_C^2\circ i_C^2)(\xi(P))=0$.
We claim that $j_{\cM\times C}^2(\xi^{1,1}(\cU))=0$ as well.
Applying \eqref{eqn:vrho} on $\xi^{1,1}(\cU)$, we obtain a homomorphism
$\vro^1_{\cM,H^1(C,\pGad)}(\xi^{1,1}(\cU))$ from $\pi_1(\cM)=H^1(C,\pG)$ to
$H^1(C,\pGad)$.
By the definition of the universal bundle, this map is precisely $i_C^1$.
So
\[ \vro^1_{\cM,H^1(C,\ZG)}(j_{\cM\times C}^2(\xi^{1,1}(\cU)))
   =j_C^1\circ\vro^1_{\cM,H^1(C,\pGad)}(\xi^{1,1}(P))=j_C^1\circ i_C^1=0.  \]
The vanishing of $j_{\cM\times C}^2(\xi^{1,1}(\cU))$ then follows since
$\vro^1_{\cM,H^1(C,\pGad)}$ is an isomorphism.

As for the $(2,0)$-component, we apply \eqref{eqn:vrho} on $\xi^{2,0}(\cU)$
to obtain $\vro^2_{\cM,\pGad}(\xi^{2,0}(\cU))\colon\pi_2(\cM)\to\pGad$.
Again by the definition of $\cU$, this is the map $\zeta_*$ in
\eqref{eqn:pi2M}.
The vanishing of $j_{\cM\times C}^2(\xi^{1,1}(\cU))$ and
$j_C^2(\xi^{0,2}(\cU))$ means that the only obstruction to lifting $\cU$
to a $G$-bundle is
\[  \bar\xi(\cU):=j_{\cM\times C}^2(\xi(\cU))=j_\cM^2(\xi^{2,0}(\cU))
    \in H^2(\cM,\ZG),      \]
and $\vro^2_{\cM,\ZG}(\bar\xi(\cU))=j\circ(\vro^2_{\cM,\pGad}(\xi^{2,0}(\cU)))
=j\circ\zeta_*\in\Hom(\pi_2(\cM),\ZG)$.
By \eqref{eqn:pi2M}, this map is surjective, and hence is non-zero unless
$\ZG=0$.
Thus the structure group $G_\ad$ of $\cU$ can not be lifted to $G$ unless
$G=G_\ad$ \cite{Rn}.

The universal bundle also has an instanton `number' or the first Pontryagin
class $k(\cU)\in H^4(\cM\times C,\pi_3(G_\ad))$.
The component $k^{2,2}(\cU)\in H^2(\cM,H^2(C,\pi_3(G)))$ defines a map
$\vro^2_{\cM,\pi_3(G)}(k^{2,2}(\cU))$ from $\pi_2(\cM)$ to
$H^2(C,\pi_3(G))\cong\pi_3(G)$.
Suppose $u\colon S^2\to\cM$ represents a class $[u]\in\pi_2(\cM)$.
Then the universal bundle pulls back to $P^u:=(u\times\id_C)^*\cU$ over
the $4$-manifold $S^2\times C$.
The bundle $P^u$ can be obtained by gluing two bundles $D_\pm\times P$ over
$D_\pm\times C$ along $S^1\times C$ using a loop in $\cG(P)$.
Here $D_\pm$ are the upper/lower hemispheres in $S^2$.
The loop in $\cG(P)$ represents an element of $\pi_1(\cG(P))$ which maps to
$[u]$ in $\pi_2(\cM)$ by the isomorphism \eqref{eqn:piMG}.
If $[u]$ belongs to the subgroup $H^2(C,\pi_3(G))$, then the image of $[u]$
in $H^0(C,\pGad)\cong H^2(S^2,\pGad)$ is zero by \eqref{eqn:pi2M}.
In this case, the restriction of $P^u$ to the $S^2$ factor is trivial.
The element of $\pi_1(\cG(P))$ in the gluing construction is also identified
with the (appropriately normalised) instanton number $k(P^u)=(u\times\id_C)^*
k^{2,2}(\cU)\in H^2(S^2,H^2(C,\pi_3(G)))\cong H^2(C,\pi_3(G))$.

Recall the $G_\ad$-bundle $P_\ad:=P/\ZG$.
Let $\cM_\ad:=\MH(P_\ad)$ and let $\cU_\ad\to\cM_\ad\times C$ be the
universal bundle.
Then there is a regular $H^1(C,\ZG)$-covering $q\colon\cM\to\cM_\ad$ on the
smooth parts \cite{HWW,Wu15} and $(q\times\id_C)^*\cU_\ad\cong\cU$ as
$G_\ad$-bundles (by the definition of universal bundles).
Consequently, $q^*\bar\xi(\cU_\ad)=\bar\xi(\cU)$ and
$(q\times\id_C)^*k(\cU_\ad)=k(\cU)$.
On the other hand, if $\widetilde P$ is a twisted $\tG$-bundle that descends
to an honest $G$-bundle, let $\tM:=\MH(\widetilde P)$ be the Hitchin moduli
space.
The universal bundle $\tU:=\UH(\widetilde P)\to\tM\times C$ is an
honest $G_\ad$-bundle and, following the above arguments, there is a
characteristic class $\bar\xi(\tU)\in H^2(\tM,\ZG)$.
Under the $H^1(C,\pG)$-covering $\tq\colon\tM\to\cM$, we have
$(\tq\times\id_C)^*\cU=\tU$ and hence $\tq^*\bar\xi(\cU)=\bar\xi(\tU)$,
$(\tq\times\id_C)^*k(\cU)=k(\tU)$.

Much of the above discussion is valid for a closed non-orientable surface
$C'$.
Let $\cM':=\MH(P')$ be the Hitchin moduli space associated to a $G$-bundle
$P'$ on $C'$ and let $\cU':=\UH(P')$ be the universal bundle over
$\cM'\times C'$.
There is a similar decomposition of its discrete flux $\xi(\cU')=
\xi^{2,0}(\cU')+\xi^{1,1}(\cU')+\xi^{0,2}(\cU')\in H^2(\cM'\times C',\pGad)$,
which is the obstruction to lifting $\cU'$ to a $\tG$-bundle.
As above, $\xi^{0,2}(\cU')\in H^2(C',\pGad)$ is in the subgroup $H^2(C',\pG)$,
$\vro^1_{\cM',H^1(C',\pGad)}(\xi^{1,1}(\cU'))\colon\pi_1(\cM')\to
H^1(C',\pGad)$ is the map $i_{C'}^1$, and
$\vro^2_{\cM',\pGad}(\xi^{2,1}(\cU'))\colon\pi_2(\cM')\to\pGad$ is the
isomorphism $\zeta'_*$ in \S\ref{sec:MH}.
So the obstruction to lifting $\cU'$ to a $G$-bundle is
$\bar\xi(\cU'):=j_{\cM'}^2(\xi^{2,0}(\cU'))\in H^2(\cM',\ZG)$, and
$\vro^2_{\cM',\ZG}(\bar\xi(\cU'))\in\Hom(\pi_2(\cM'),\ZG)$ is the surjective
map $j\circ\zeta'_*$.
It follows that $\cU'$ can not be lifted to a $G$-bundle unless $G=G_\ad$.
Finally, let $\cM'_\ad:=\MH(P'_\ad)$ and let $\cU'_\ad\to\cM'_\ad\times C'$
be the universal bundle.
There is a regular $H^1(C',\ZG)$-covering $q'\colon\cM'\to\cM'_\ad$ and
$(q')^*\cU'_\ad\cong\cU'$, $(q')^*\bar\xi(\cU'_\ad)=\bar\xi(\cU')$.

Let $\pi\colon C\to C'$ be the orientation double cover.
We set $\cM:=\MH^{m_0=0}(C,G)$, $\cM':=\MH(C',G)$ and consider the universal
bundles $\cU:=\UH(C,G)\to\cM\times C$ and $\cU':=\UH(C',G)\to\cM'\times C'$.
Then there is a bundle isomorphism
\begin{equation}\label{eqn:ppull}
(\id_\cM\times\pi)^*\cU'\cong(p\times\id_C)^*\cU.
\end{equation}
Consequently, we have $p^*(\bar\xi(\cU))=\bar\xi(\cU)$.
To show \eqref{eqn:ppull}, we choose a $G$-bundle $P'\to C'$ and let
$P=\pi^*P'$.
Consider the universal bundles $\cU(P)=\cA(P)\times_{\cG(P)}P$ and
$\cU(P')=\cA(P')\times_{\cG(P')}P'$ over $\cB(P)\times C$ and
$\cB(P')\times C'$, respectively.
Let $p\colon\cB(P')\to\cB(P)$ be defined by pulling back connections.
Then we have the isomorphism
$(\id_{\cB(P')}\times\pi)^*\cU(P')\cong(p\times\id_C)^*\cU(P)$ because
they are the same $G_\ad$-bundle $\cA(P')\times_{\cG(P')}P$ over
$\cB(P')\times C$, where $\cG(P')$ acts on $P$ via the (injective)
homomorphism $\cG(P')\to\cG(P)$.
The argument for the Hitchin moduli spaces is identical upon replacing 
$\cA(P)$, $\cA(P')$ by the spaces $\AH(P)$, $\AH(P')$ of Hitchin pairs.

\subsection{Puppe sequence and relative homotopy groups}\label{sec:rel}
Let $(X,x_0)$ and $(Z,z_0)$ be topological spaces with base points.
Consider the set of homotopy classes $[(X,x_0),(Z,z_0)]$, or $[X,Z]_0$ if
there is no confusion, of continuous maps from $(X,x_0)$ to $(Z,z_0)$.
If $f\colon(X,x_0)\to(Y,y_0)$ is a continuous map, the mapping cone
$C_f:=(I\times X)\sqcup Y/\sim$ of $f$, where the identification is
$(0,x)\sim(0,x')$ and $(1,x)\sim f(x)$ for $x,x'\in X$, contains $Y$
and $f(X)$ as subspaces, and has an obvious base point $c_0$ given by
$(1,x_0)\sim y_0$.
The Puppe sequence is a long exact sequence of pointed sets
\begin{equation}\label{eqn:puppe0}
\cdots\to[\sus(Y),Z]_0\map{\sus(f)^*}[\sus(X),Z]_0\map{j^*}[C_f,Z]_0\map{i^*}
[Y,Z]_0\map{f^*}[X,Z]_0
\end{equation}
(see for example \cite{Sp}, \S7.1).
Here $\sus$ stands for taking a (reduced) suspension, $i\colon Y\to C_f$ is
the inclusion and $j\colon C_f\to\sus(X)$ collapses $i(Y)$ to a point.
For simplicity, we have omitted the base points in \eqref{eqn:puppe0}.

Let $\Sig$ be a closed orientable surface of genus $g(\Sig)$.
It has a cellular structure whose $0$-skeleton $\Sig^{(0)}$ is a single point
and whose $1$-skeleton $\Sig^{(1)}=\bigvee_{i=1}^{2g(\Sig)}S^1$ is a bouquet
of $2g(\Sig)$ circles; $\Sig$ itself is obtained by attaching a
$2$-cell via a map $f\colon(S^1,s_0)\to(\Sig^{(1)},\Sig^{(0)})$.
In this way, $\Sig$ is the mapping cone $C_f$ and the Puppe sequence
\eqref{eqn:puppe0} is
\[ \cdots\to\textstyle\prod_{i=1}^{2g(\Sig)}\pi_2(Z)\map{\sus(f)^*}\pi_2(Z)
   \map{j^*}[\Sig,Z]_0\map{i^*}
   \textstyle\prod_{i=1}^{2g(\Sig)}\pi_1(Z)\map{f^*}\pi_1(Z).   \]
The map $f^*$ is $(a_1,b_1,\dots,a_{g(\Sig)},b_{g(\Sig)})\mapsto
\prod_{i=1}^{g(\Sig)}a_ib_ia_i^{-1}b_i^{-1}$;
when $g(\Sig)=1$, this is known as the Whitehead product.
Therefore $\ker(f^*)=\Hom(\pi_1(\Sig),\pi_1(Z))$ and
$i^*\colon[\Sig,Z]_0\to\ker(f^*)$ sends the homotopy class $[u]$ of a map
$u\colon(\Sig,\Sig^{(0)})\to(Z,z_0)$ to its induced homomorphism
$u_*\colon\pi_1(\Sig)\to\pi_1(Z)$.
The map $\sus(f)^*$ has a similar expression, but it is zero because
$\pi_2(Z)$ is Abelian.
Thus the Puppe sequence reduces to a short exact sequence
\begin{equation}\label{eqn:[S,Z]}
0\to\pi_2(Z)\to[\Sig,Z]_0\to\Hom(\pi_1(\Sig),\pi_1(Z))\to0.
\end{equation}
The exactness at $[\Sig,Z]_0$ can be checked directly:
if $u_*\colon\pi_1(\Sig)\to\pi_1(Z)$ is zero, the restriction of $u$ to
$\Sig^{(1)}$ is null homotopic, and hence there is a (based) map $\bar u$ from
$\Sig/\Sig^{(1)}=S^2$ to $Z$, whose homotopy class $[\bar u]$ is in $\pi_2(Z)$.
If $\pi_1(Z)$ is Abelian and it acts on $\pi_2(Z)$ trivially, then in
\eqref{eqn:[S,Z]}, $[\Sig,Z]_0$ can be replaced by the set $[\Sig,Z]$ of
homotopy classes of unbased maps and $\Hom(\pi_1(\Sig),\pi_1(Z))$ simplifies
to $H^1(\Sig,\pi_1(Z))$.

Now suppose $(X,X',x_0)$ and $(Z,Z',z_0)$ are pairs of based topological
spaces.
Consider the set of homotopy classes $[(X,X',x_0),(Z,Z',z_0)]$, or
$[(X,X'),(Z,Z')]_0$ if there is no confusion, of continuous maps from
$(X,X',x_0)$ to $(Z,Z',z_0)$; we reserve $[(X,X'),(Z,Z')]$ for the set
of homotopy classes of unbased maps.
Given $f\colon(X,X',x_0)\to(Y,Y',y_0)$, let $f':=f|_X$.
Then $(C_f,C_{f'},c_0)$ is a pair of based spaces.
We have the inclusion map $i\colon(Y,Y',y_0)\to(C_f,C_{f'},c_0)$ and the map
$j\colon(C_f,C_{f'},y_0)\to(\sus(X),\sus(X'),x_0)$ that collapses
$(i(Y),i(Y'))$  to a point.
The relative Puppe sequence
\begin{equation}\label{eqn:puppe}
\scalebox{.85}[1]{$\cdots\to[(\sus(Y),\sus(Y')),(Z,Z')]_0\map{\sus(f)^*}
[(\sus(X),\sus(X')),(Z,Z')]_0\map{j^*}[(C_f,C_{f'}),(Z,Z')]_0\map{i^*}
[(Y,Y'),(Z,Z')]_0\map{f^*}[(X,X'),(Z,Z')]_0$}
\end{equation}
is a long exact sequence of pointed set (cf.~\cite{Sp}, \S7.1), which
simplifies to \eqref{eqn:puppe0} if $X',Y',Z'$ are points.

The set $[(D^n,S^{n-1},s_0),(Z,Z',z_0)]$ is the $n$th relative homotopy group
$\pi_n(Z,Z',z_0)$, or $\pi_n(Z,Z')$ for short, of the pair $(Z,Z')$.
It is indeed a group when $n\ge2$ and the group is Abelian when $n\ge3$.
If $Z'$ is a point, then $\pi_n(Z,z_0,z_0)=\pi_n(Z,z_0)$ is the usual $n$th
homotopy group.
If $f\colon(I,\{0\})\to(I,\bI)$ is the inclusion, then \eqref{eqn:puppe}
reduces to a long exact sequence of relative homotopy groups
(cf.~\cite{Sp}, \S7.2)
\begin{equation}\label{eqn:rel}
\cdots\to\pi_n(Z',z_0)\map{i_*}\pi_n(Z,z_0)\map{j_*}\pi_n(Z,Z',z_0)
\map{\partial_*}\pi_{n-1}(Z',z_0)\to\cdots.
\end{equation}
Here $i\colon(Z',z_0)\to(Z,z_0)$ and $j\colon(Z,z_0,z_0)\to(Z,Z',z_0)$ are
inclusion maps and $\partial_*$ is the restriction of maps
$(D^n,S^{n-1},s_0)\to(Z,Z',z_0)$ to $(S^{n-1},s_0)\to(Z',z_0)$.
For example, $\pi_1(Z,Z',z_0)$ is the set of homotopy classes $[\gam]$ of
paths $\gam\colon I=[0,1]\to Z$ such that $\gam(0)=z_0$ and $\gam(1)\in Z'$.
Though $\gam(1)$ is path connected to $z_0$ in $Z$, it may not be so in $Z'$,
and $\partial_*([\gam])\in\pi_0(Z',z_0)$ labels the connected component of
$Z'$ containing $\gam(1)$.

In \S\ref{sec:non-or}, we need a slightly generalised setup, replacing the
pair $(Z,Z')$ with base point $z_0$ by a map $p\colon(Z',z'_0)\to(Z,z_0)$.
(If $Z'$ is a subset of $Z$, then $p$ is the inclusion map.)
Consider the set $[(X,X',x_0),((Z,z_0),(Z',z'_0))]$, or $[(X,X'),(Z,Z')]_0$
for short, of homotopy classes of pairs of maps $(f,f')\colon(X,X')\to(Z,Z')$
with $f\colon(X,x_0)\to(Z,z_0)$ and $f'\colon(X',x_0)\to(Z',z'_0)$ satisfying
$f|_{X'}=p\circ f'$.
Consider the mapping cylinder $M_p:=(I\times Z')\sqcup Z/\sim$ of $p$, where
the identification is $(1,z')\sim p(z')$ for $z'\in Z'$.
Then $M_p$ contains both $Z$ and $Z'$ as subspaces, $Z$ is a deformation
retract of $M_p$, and $p$ is the composition of the inclusion $Z'\to M_p$
and the retraction $M_p\to Z$.
Thus $[(X,X'),(Z,Z')]_0=[(X,X'),(M_f,Z')]_0$ and the Puppe sequence
\eqref{eqn:puppe} remains valid in this case.

Similarly, $\pi_n((Z,z_0),(Z',z'_0))$, or $\pi_n(Z,Z')$ for short, of the map
$p\colon Z'\to Z$ is the set of homotopy classes of pairs of maps $(u,u')$,
where $u\colon(D^n,s_0)\to(Z,z_0)$ and $u'\colon(S^{n-1},s_0)\to(Z',z'_0)$
satisfy $u|_{S^{n-1}}=p\circ u'$.
Since $\pi_n(M_p,Z',z_0)\cong\pi_n((Z,z_0),(Z',z'_0))$, the exact sequence
\eqref{eqn:rel} for the pair $(M_p,Z')$ becomes
\begin{equation}\label{eqn:rel-p}
\cdots\to\pi_n(Z',z'_0)\map{p_*}\pi_n(Z,z_0)\map{j_*}\pi_n((Z,z_0),(Z',z'_0))
\map{\partial_*}\pi_{n-1}(Z',z'_0)\to\cdots,
\end{equation}
where the connecting homomorphism $\partial_*$ is simply given by
$\partial_*\colon[(u,u')]\mapsto[u']$.

We want to calculate $[(\Sig,\bSig),(Z,Z')]_0$ for
$p\colon(Z',z'_0)\to(Z,z_0)$ as above, where $\Sig$ is a compact orientable
surface with boundary $\bSig$.
Let $\hat\Sig$ be the closed surface obtained by attaching a disc to each
boundary circle of $\bSig$; it has a cellular structure whose
$\hat\Sig^{(0)}$ is a point, $\hat\Sig^{(1)}$ is bouquet of $2g(\hat\Sig)$
circles, and $\hat\Sig$ is obtained by attaching a $2$-cell to $\hat\Sig^{(1)}$
via a map $\hat f\colon(S^1,s_0)\to(\hat\Sig^{(1)},\hat\Sig^{(0)})$.
For simplicity, we assume that $\bSig$ is a circle.
Then $(\Sig,\bSig)$ is the mapping cone $(C_f,C_{f'})$ of
$(f,f')\colon(I,\bI)\to(Z,Z')$, where $f'$ maps $\bI$ to $\Sig^{(0)}$ while
$f$ maps $(I,\bI)$ to $(S^1,s_0)$ and then to
$(\hat\Sig^{(1)},\hat\Sig^{(0)})$ via $\hat f$.
The relative Puppe sequence \eqref{eqn:puppe} is
\[ \cdots\to\textstyle\prod_{i=1}^{2g(\hat\Sig)}\pi_2(Z)\map{\sus(f,f')^*}
   \pi_2(Z,Z')\to[(\Sig,\bSig),(Z,Z')]_0\to
   \textstyle\prod_{i=1}^{2g(\hat\Sig)}\pi_1(Z)\map{(f,f')^*}\pi_1(Z,Z'). \]
The maps $(f,f')^*$ and $\sus(f,f')^*$ factor through $\pi_1(Z)$ and
$\pi_2(Z)$ via $\hat f^*$ and $\sus(\hat f)^*$, respectively.
If $\pi_1(Z)$ is Abelian, then both maps are zero.
In this case, the above Puppe sequence simplifies to a short exact sequence
\begin{equation}\label{eqn:ZZ'}
0\to\pi_2(Z,Z')\to[(\Sig,\bSig),(Z,Z')]_0\to H^1(\hat\Sig,\pi_1(Z))\to0.
\end{equation}

\subsection{Cohomology of orientable and non-orientable surfaces}
\label{sec:HC}
As in \S\ref{sec:non-or}, let $C'$ be a closed non-orientable surface and
let $\pi\colon C\to C'$ be its orientation double cover.
The non-trivial deck transformation $\iota$ acts on $C$ as an
orientation-reversing involution satisfying $\pi\circ\iota=\pi$ and
$\iota^2=\id_C$.
We wish to study the pull-back $\pi^*\colon H^k(C',A)\to H^k(C,A)$ of
cohomology groups with coefficients in any finitely generated Abelian group
$A$.
Clearly, $\pi^*$ maps into the $\iota$-invariant part $H^k(C,A)^\iota$.
For $k=2$, by the naturality of the universal coefficient formula, the
pull-back map $\pi^*$ from $H^2(C',A)=\Ext(H_1(C'),A)\cong A/2A$ to
$H^2(C,A)=\Hom(H_2(C),A)\cong A$ is zero whereas
$H^2(C,A)^\iota=A_{[2]}$ can be non-zero.
The case $k=1$ is surprisingly more involved and will occupy the rest of the
discussion below.

Consider the fibration $\veps$ over $B\bZ_2$ whose fibre is $C$ and whose
total space $E\bZ_2\times_\iota C$ is homotopic to $C'$.
The Leray-Serre spectral sequence has $E_2^{pq}=H^p(\bZ_2,H^q(C,A))$, where
$\bZ_2$ acts on $H^q(C,A)$ via the pull-back of $\iota$, and converges to
$H^\bullet(C',A)$.
If $g(C)\ge1$, it coincides with the Lyndon-Hochschild-Serre spectral
sequence of group cohomology associated to
$1\to\pi_1(C)\to\pi_1(C')\to\bZ_2\to1$.
For example, we have $E_2^{10}=H^1(\bZ_2,A)=A_{[2]}$,
$E_2^{01}=H^1(C,A)^\iota$ and $E_2^{20}=H^2(\bZ_2,A)=A/2A$.
The map $\pi^*\colon H^1(C',A)\to H^1(C,A)^\iota$ is contained in the exact
sequence
\[ 0\to E_2^{10}\map{\veps^*}H^1(C',A)\map{\pi^*}E_2^{01}\map{d^{01}_2}
   E_2^{20}\map{\veps^*}H^2(C',A).             \]
So we get $\ker(\pi^*)\cong A_{[2]}$ and the exact sequence
\begin{equation}\label{eqn:C1}
0\to A_{[2]}\to H^1(C',A)\to\pi^*H^1(C',A)\to0.
\end{equation}

The homology spectral sequence with $E^2_{pq}=H_p(\bZ_2,H_q(C,A))$ converges
to $H_\bullet(C',A)$.
We have an exact sequence
\[ H_2(C'A)\map{\veps_*}E^2_{20}\map{d^2_{20}}E^2_{01}\map{\pi_*}H_1(C',A)
   \to E^2_{10}\map{\veps_*}0,       \]
where $E^2_{10}=A/2A$, $E^2_{20}=A_{[2]}\cong H_2(C',A)$ and
$E^2_{01}=H_1(C,A)/(1-\iota_*)H_1(C,A)$.
The map $E^2_{01}\to H_1(C',A)$ induced by
$\pi_*\colon H_1(C,A)\to H_1(C',A)$ is also denoted by $\pi_*$.
This leads to an exact sequence
\begin{equation}\label{eqn:C_1}
0\to\sfrac{H_1(C,A)}{\ker(\pi_*\!\colon\!H_1(C,A)\to H_1(C',A))}\to H_1(C',A)
\to A/2A\to 0.
\end{equation}

Generally, we have
$(1-\iota_*)H_1(C,A)\subset\ker(\pi_*)\subset H_1(C,A)^{-\iota}$ and
$(1+\iota^*)H^1(C,A)\subset\pi^*H^1(C',A)\subset H^1(C,A)^\iota$.
To study these inclusions, we use explicit cellular structures of $C$ and $C'$.
The generators $[C]$ of $S_2(C)$ and $[C']$ of $S_2(C')$ satisfy
$\iota_*[C]=-[C]$, $\pi_*[C]=0$.
If $\chi(C')$ is odd, $C'$ is the connected sum of an orientable surface
(of genus $g$) and an $\bR P^2$.
Then $S_1(C)$ is generated by $a_i,b_i,\ta_i,\tb_i$ and $S_1(C')$ is
generated by $a'_0,a'_i,b'_i$ with $\iota_*a_i=\ta_i$, $\iota_*b_i=\tb_i$,
$\pi_*a_i=a'_i$, $\pi_*b_i=b'_i$ ($1\le i\le g)$;
the only non-zero boundary is $\partial[C']=2a'_0$.
A cohomology class $H^1(C,A)$ is determined by its values on the
generators of $S_1(C)$.
All three spaces in $H_1(C,A)$ are the same and are freely generated by
$a_i-\ta_i$, $b_i-\tb_i$.
The conditions for $\gam\in H^1(C,A)$ being in the three spaces are also
the same: $\gam(a_i)=\gam(\ta_i)$, $\gam(b_i)=\gam(\tb_i)$.
If $\chi(C')$ is even, $C'$ is the connected sum of an orientable surface
(of genus $g$) and a Klein bottle.
There are additional generators $a_0,b_0$ of $S_1(C)$ and $a'_0,b'_0$ of
$S_1(C')$ satisfying $\iota_*a_0=-a_0$, $\iota_*b_0=b_0$, $\pi_*a_0=a'_0$,
$\pi_*b_0=2b'_0$.
The three spaces $(1-\iota_*)H_1(C,A)$, $\ker(\pi_*)$, $H_1(C,A)^{-\iota}$
in $H_1(C,A)$ have extra parts $2Aa_0$, $2Aa_0\oplus A_{[2]}b_0$,
$Aa_0\oplus A_{[2]}b_0$, respectively.
The extra conditions for $\gam$ being in the three spaces
are $\gam(a_0)\in A_{[2]}$ for $\gam\in H^1(C,A)^\iota$,
$\gam(a_0)\in A_{[2]}$ and $\gam(b_0)\in 2A$ for $\gam\in\pi^*H^1(C',A)$,
$\gam(a_0)=0$ and $\gam(b_0)\in 2A$ for $\gam\in(1+\iota^*)H^1(C,A)$.
Thus the inclusions can be proper and
\begin{equation}\label{eqn:quot}
\sfrac{\pi^*H^1(C',A)}{(1+\iota^*)H^1(C,A)}\cong
\sfrac{\ker(\pi_*\!\colon\!H_1(C,A)\to H_1(C',A))}{(1-\iota_*)H_1(C,A)}
\cong A_{[2]},\qquad\sfrac{H^1(C,A)^\iota}{\pi^*H^1(C',A)}\cong
\sfrac{H_1(C,A)^{-\iota}}{\ker(\pi_*\!\colon\!H_1(C,A)\to H_1(C',A))}
\cong A/2A.
\end{equation}
To summarise, both maps $\veps^*\colon A/2A\to A/2A$ and
$\veps_*\colon A_{[2]}\to A_{[2]}$ are the multiplication by $\chi(C')$.

To study these subspaces under Poincar\'e duality, recall the isomorphism
$\al\in H^1(C,A)\mapsto\al\cap[C]$, where $[C]\in H_2(C,\bZ)$ denotes the
fundamental class of $C$.
This defines an inclusion $\pi^*H^1(C',A)\to\ker(\pi_*)\subset H_1(C,A)$
because $\pi_*(\pi^*(\al')\cap[C])=\al'\cap\pi_*([C])=0$ for all
$\al'\in H^1(C'A)$, and $\pi_*([C])=0$.
It is actually an isomorphism, i.e.,
\begin{equation}\label{eqn:Cpi}
\pi^*H^1(C',A)\cong\ker(\pi_*\colon H_1(C,A)\to H_1(C',A)).
\end{equation}
Consequently, by the Poincar\'e duality \eqref{eqn:Poin'} for $C'$, we can
rewrite the exact sequence \eqref{eqn:C_1} as
\begin{equation}\label{eqn:C1'}
0\to\sfrac{H^1(C,A)}{\pi^*H^1(C',A)}\to H^1(C',\uA)\to A/2A\to 0.
\end{equation}
Next, we consider the non-degenerate pairing
\begin{equation}\label{eqn:pair}
H^1(C,A^\vee)\times H^1(C,A)\map\cup
H^2(C,A^\vee\times A)\to H^2(C,\mathrm U(1))\cong\mathrm U(1)
\end{equation}
that induces another Poincar\'e duality map
$\al\in H^1(C,A^\vee)\mapsto\al^\flat\in H^1(C,A)^\vee$.
Since $H^2(C',\mathrm U(1))=0$, by the naturality of cup product, the pairing
\eqref{eqn:pair} is zero when restricted to
$\pi^*H^1(C',A^\vee)\times\pi^*H^1(C',A)$.
So if $\al\in\pi^*H^1(C',A^\vee)$, then $\al^\flat(\pi^*H^1(C',A))=1$.
(We write the Abelian group structure on $\mathrm U(1)$ multiplicatively,
but write the cohomology groups additively).
The converse is also true, that is, Poincar\'e duality restricts to
an isomorphism
\begin{equation}\label{eqn:Ctau}
\pi^*H^1(C',A^\vee)\cong\{\al\in H^1(C,A)^\vee:\al^\flat(\pi^*H^1(C',A))=1\}
\cong\Big(\!\sfrac{H^1(C,A)}{\pi^*H^1(C',A)}\!\Big)^{\!\!\vee}.
\end{equation}
The isomorphisms \eqref{eqn:Cpi}, \eqref{eqn:Ctau} can be verified by the cup
and cap products in the above cellular decompositions of $C$ and $C'$ or by
the intersection form, $a_i\cdot b_j=-\ta_i\cdot\tb_j=\del_{ij}$
($1\le i\le g$), $a_0\cdot b_0=1$ if $\chi(C')$ is even, on $C$.

Curiously, the subspaces $(1+\iota^*)H^1(C,A^\vee)$ and $H^1(C,A)^\iota$ are
complementary to each other under \eqref{eqn:pair}.
In fact, since $\iota$ reverses the orientation of $C$, we have
$(\iota^*\al)^\flat=(\al^\flat\circ\iota^*)^{-1}$ for any $\al\in H^1(C',A^\vee)$
and thus
\[ (\al-\iota^*\al)^\flat=\al^\flat\cdot((\iota^*\al)^\flat)^{-1}
   =\al^\flat\cdot(\al^\flat\circ\iota^*)=\al^\flat\circ(1+\iota^*).  \]
Relating cohomology with homology, we have
$((1\pm\iota^*)\al)\cap[C]=(1\mp\iota_*)(\al\cap[C])$ as $\iota_*([C])=-[C]$.
So we get
\[ (1+\iota^*)H^1(C,A^\vee)\cong(1-\iota_*)H_1(C,A)\cong
   \Big(\!\sfrac{H^1(C,A)}{H^1(C,A)^\iota}\!\Big)^{\!\!\vee},
   \qquad H^1(C,A^\vee)^\iota\cong H_1(C,A)^{-\iota}\cong
   \Big(\!\sfrac{H^1(C,A)}{(1+\iota^*)H^1(C,A)}\!\Big)^{\!\!\vee}. \]
These isomorphisms are the same as \eqref{eqn:Ctau}, \eqref{eqn:Cpi} if
$\chi(C')$ is odd but are different from them if $\chi(C')$ is even.

\subsection{Cohomology of the $3$- and $4$-manifolds containing a
non-orientable surface}\label{sec:HYX}
As in \S\ref{sec:HC}, $C'$ is a closed non-orientable surface with an
orientation double cover $\pi\colon C\to C'$, $\iota$ is the non-trivial
deck transformation on $C$.

Let $\iota$ also act on the circle $S^1$ by reflection along an axis;
then the quotient of $S^1$ by $\iota$ is an interval, say $I=[0,1]$.
Let $Y:=S^1\times_\iota C$, where $\iota$ acts diagonally.
Since $\iota$ acts freely on $C$ and reverses the orientations on both $S^1$
and $C$, $Y$ is a smooth closed orientable $3$-manifold
(cf.~\S\ref{sec:em4dn}).
We want to find the cohomology of $Y$ with coefficients in any finitely
generated Abelian group $A$.
There is a surjective map $\pi_Y\colon Y\to I$.
The inverse image $\pi_Y^{-1}(s)$ is $C$ if $0<s<1$ and is $C'$ if $s=0,1$.
Let $Y_0:=\pi_Y^{-1}([0,2/3))$ and $Y_1:=\pi_Y^{-1}((1/3,1])$.
Then $Y_0,Y_1$ are both homotopic to $C'$, $Y_0\cap Y_1$ is homotopic to $C$,
and $Y_0\cup Y_1=Y$.
The Mayer-Vietoris long exact sequence is
\[ \cdots\to H^{k-1}(C',A)^{\oplus2}\map{\;\mu^{k-1}_Y}H^{k-1}(C,A)\to
   H^k(Y,A)\to H^k(C',A)^{\oplus2}\map{\mu^k_Y}H^k(C,A)\to\cdots    \]
where $\mu^k_Y\colon(a_0,a_1)\in H^k(C',A)^{\oplus2}\mapsto
\pi^*(a_1-a_0)\in H^k(C,A)$.
For $k=2$, since the map $\pi^*\colon H^2(C',A)\cong A/2A\to H^2(C,A)\cong A$
is zero (see \S\ref{sec:HC}), we have a short exact sequence
\begin{equation}\label{eqn:Y2}
0\to\sfrac{H^1(C,A)}{\pi^*H^1(C',A)}\to H^2(Y,A)\to(A/2A)^{\oplus2}\to0
\end{equation}
that calculates $H^2(Y,A)$.
For $k=1$, we have
$H^1(Y,A)=\ker(\mu^1_Y\colon H^1(C',A)^{\oplus2}\to H^1(C,A))$.
By \eqref{eqn:C1}, the group $H^1(Y,A)$ contains $A_{[2]}^{\oplus2}$ as a
subgroup and the quotient group is isomorphic to $\pi^*H^1(C',A)$.
Consequently, $H^1(Y,A)$ fits in the short exact sequence
\begin{equation}\label{eqn:Y1}
0\to A_{[2]}^{\oplus2}\to H^1(Y,A)\to\pi^*H^1(C',A)\to0.
\end{equation}
The exact sequences \eqref{eqn:Y2}, \eqref{eqn:Y1} are compatible with
Poincar\'e duality \eqref{eqn:Poin} by the isomorphisms \eqref{eqn:two},
\eqref{eqn:Ctau}.

Now let $\tSig$ be a closed orientable surface with an orientation reversing
involution, also denoted by $\iota$, with a non-empty fixed-point set
$\tSig^\iota$.
The quotient $\Sig$ is a compact orientable surface with boundary
$\bSig=\tSig^\iota$.
Let $X=\tSig\times_\iota C$ with $\iota$ acting diagonally.
Then $X$ is a smooth closed orientable $4$-manifold (cf.~\S\ref{sec:4dn2d}).
There is a surjective map $\pi_X\colon X\to\Sig$ so that the inverse image
$\pi_X^{-1}(\sig)$ is $C$ if $\sig$ is in the interior of $\Sig$ and is
$C'$ if $\sig\in\bSig$.
Recall from \S\ref{sec:4dn2d} that $X=X_0\cup X_1$, where $X_0$, $X_1$ and
$X_0\cap X_1$ are homotopic to $\Sig\times C$, $\bSig\times C'$ and
$\bSig\times C$, respectively.
Instead of the Mayer-Vietoris sequence, we use a commutative diagramme
\[ \scalebox{.9}[1]{$\xymatrix@=1pc{\cdots\ar[r]&
H^{k-1}(\bSig\times C',A)\ar[r]^{\mu_X^{k-1}}\ar[d]_{(\id_\bSig\times\pi)^*}&
H^k((X,\bSig\times C'),A)\ar[r]\ar[d]_{\tht^k}&H^k(X,A)\ar[r]\ar[d]&
H^k(\bSig\times C',A)\ar[r]^{\!\!\!\!\!\!\!\!\!\mu_X^k}
\ar[d]_{(\id_\bSig\times\pi)^*}&
H^{k+1}((X,\bSig\times C'),A)\ar[r]\ar[d]_{\tht^{k+1}}&\cdots              \\
\cdots\ar[r]&
H^{k-1}(\bSig\times C,A)\ar[r]^{\!\!\!\mu_{\Sig\times C}^{k-1}}&
H^k((\Sig,\bSig)\times C,A)\ar[r]&H^k(\Sig\times C,A)\ar[r]&
H^k(\bSig\times C,A)\ar[r]^{\!\!\!\!\!\!\!\!\mu_{\Sig\times C}^k}&
H^{k+1}((\Sig,\bSig)\times C,A)\ar[r]&\cdots}$}                           \]
whose rows are the long exact sequences of the pairs $(X,\bSig\times C')$ and
$(\Sig,\bSig)\times C$.
The vertical maps $\tht^k$ are excision isomorphisms.
The Mayer-Vietoris sequence for $X=X_0\cup X_1$ follows from the above
diagramme \cite{BaW}.

The cohomology group $H^k(X,A)$ thus fits in the short exact sequence
\[  0\to\coker(\mu_X^{k-1})\to H^k(X,A)\to\ker(\mu_X^k)\to0.   \]
The kernel and cokernel of $\mu_X^k$ are isomorphic to those of the
composite map $\mu_{\Sig\times C}^k\circ(\id_\bSig\times\pi)^*$.
In general, given two homomorphisms $f$ and $g$ of Abelian groups that can be
composed, there is an exact sequence
\[ 0\to\ker(f)\to\ker(g\circ f)\to\ker(g)\to\coker(f)\to\coker(g\circ f)\to
   \coker(g)\to0.   \]
Using this, a lengthy but routine calculation yields two exact sequences
\begin{equation}\label{eqn:HX1}
0\to H^0(\bSig,A_{[2]})\oplus H^1(\Sig,A)\to H^1(X,A)\to\pi^*H^1(C',A)\to0,
\end{equation}
\begin{equation}\label{eqn:HX2}
\scalebox{.95}[1]{$0\to H^1(\bSig,A_{[2]})\oplus H^1\Big((\Sig,\bSig),
\sfrac{H^1(C,A)}{\pi^*H^1(C',A)}\Big)\to H^2(X,A)\to H^0(\bSig,A/2A)\oplus
H^1(\Sig,\pi^*H^1(C',A))\to0.$}
\end{equation}
It can be checked that these results are consistent with the counting in
\eqref{eqn:Euler}.
The group $H^3(X,A)$ can be calculated by Poincar\'e duality \eqref{eqn:Poin}
from \eqref{eqn:HX1}.
For $H^2(X,A)$, the exact sequence \eqref{eqn:HX2} is consistent with
\eqref{eqn:Poin} by \eqref{eqn:two}, \eqref{eqn:Ctau} and a relative version
of Poincar\'e duality for $\Sig$.
If $\Sig$ is a cylinder, then \eqref{eqn:HX1} and \eqref{eqn:HX2} reduce to
the exact sequences \eqref{eqn:Y2} and \eqref{eqn:Y1}.
If $\Sig$ is a disk, then $H^1(X,A)\cong H^1(C',A)$ as in \eqref{eqn:C1} and
$H^2(X,A)\cong A_{[2]}\oplus A/2A$.
More generally, if $\bSig$ is a single circle, then both $H^1(\Sig,A)$ and
$H^1((\Sig,\bSig),A)$ are equal to $H^1(\hat\Sig,A)$, where $\hat\Sig$ is
the closed surface obtained from $\Sig$ by filling the boundary circle with
a disc (cf.~\S\ref{sec:rel}).
Therefore \eqref{eqn:HX2} is equivalent to a filtration of $H^2(X,A)$ whose
graded components are
\begin{equation}\label{eqn:filt}
H^0(\bSig,A/2A),\quad H^1(\hat\Sig,H^1(C,A)),\quad H^1(\bSig,A_{[2]}).
\end{equation}

By the above explicit constructions, a compact non-orientable surface $C'$ can
always be smoothly embedded in the orientable $3$-manifold $Y$ and in the
orientable $4$-manifold $X$.
For $A=\bZ_2$ (which is a field), the cohomology and homology groups are dual
to each other as vector spaces over $\bZ_2$.
For example, $H^2(C',\bZ_2)\cong\bZ_2$ is dual to $H_2(C',\bZ_2)\cong\bZ_2$.
In our case the maps from $H_2(C',\bZ_2)$ to $H_2(Y,\bZ_2)$ and $H_2(X,\bZ_2)$
induced by the embeddings are injective.
For constraints on embedding non-orientable surfaces in a fixed
$3$- or $4$-manifold, see \cite{BW,LRS}.

\subsection{More on moduli spaces associated to a non-orientable surface}
\label{sec:MHn}
Again, $C'$ is a closed non-orientable surface with an orientation double
covering $\pi\colon C\to C'$ and $G$ is a compact, semisimple Lie group.

We begin by recalling from \cite{HWW} some facts on the Hitchin's moduli space
$\MH(C',G)$ and its relation to $\MH(C,G)$;
some parallel results on the moduli spaces $\Mf(C',G)$ and $\Mf(C,G)$ of flat
$G$-connections can be found in \cite{Ho,HL03}.
There is a map $p\colon\MH(C',G)\to\MH(C,G)$ defined by pulling back
bundles, connections and sections from $C'$ to $C$.
Restricting to appropriate smooth parts of the moduli spaces, $p$ is a
regular $\ZG_{[2]}$-covering onto its image $\cN(C,G)$ in $\MH(C,G)$.
The action of $z\in\ZG_{[2]}$ on $\MH(C',G)$ modifies the holonomy of a
connection along a loop $\al$ in $C'$ by $z$ if $\al$ does not come from a
loop in $C$; the holonomy is unchanged if there is a loop in $C$ that
descends to $\al$.
The requirement $2z=0$ is because for any loop $\al$ in $C'$, $\al\cdot\al$
always comes from a loop in $C$.
The set $\cN(C,G)$ is contained in the connected component
$\MH^{m_0=0}(C,G)$ of $\MH(C,G)$.
It is also inside the fixed-point set  $\MH(C,G)^\iota$ of the involution
induced by the non-trivial deck transformation $\iota$ on $C$.
Since the involution is anti-holomorphic in $I$ and $K$ but holomorphic in
$J$, the subset $\MH(C,G)^\iota$ is Lagrangian with respect to $\om_I$ and
$\om_K$ but complex with respect to $J$ (see also \cite{BHH,Sc,BS}), and the
covering $p$ is K\"ahler with respect to $\om'_J$ on $\MH(C',G)$ and the
restriction of $\om_J$ to $\cN(C,G)$.
For coverings over other parts of $\MH(C,G)^\iota$, we need $G$-bundles on
$C'$ that are twisted by discrete $B$-fields \cite{Wu15}.
It is possible to show that all $\MH(C,G)^\iota$ is contained in
$\MH^{m_0=0}(C,G)$.

To construct the branes used in \S\ref{sec:em2dn} (see also \cite{Wu16}),
we need a more refined understanding of the covering map $p$.
Recall the map $\del_{\bZ_2}^1$ in \eqref{eqn:del2}.
The group $\ZG_{[2]}$ acts on $\MH(C',G)$ but only its subgroup
$\ker(\del_{\bZ_2}^1)$ preserves each component $\MH^{m_2}(C',G)$ labelled by 
$m_2\in H^2(C',\pG)\cong\pG/2\pG$.
There is a decomposition
\begin{equation}\label{eqn:Nm}
\cN(C,G)=\msqcup{\mm_2\in\bm(\bZ_2,G)}{}\cN^{\mm_2}(C,G),
\end{equation}
where $\cN^{\mm_2}(C,G):=p(\MH^{m_2}(C',G))$ and $\MH^{m_2}(C',G)$ is a
regular $\ker(\del_{\bZ_2}^1)$-cover over $\cN^{\mm_2}(C,G)$.
Each $\cN^{\mm_2}(C,G)$ is connected if we expect that $\MH^{m_2}(C',G)$ is so
(cf.~\S\ref{sec:MH}).
For any $z\in\ZG_{[2]}$, there is a commutative diagramme
\[ \xymatrix@=1pc{\MH^{m_2}(C',G)\ar[rr]^{\hspace*{-1.5em}z}\ar[rd]^p&&
   \MH^{m_2+\del_{\bZ_2}^1(z)}(C',G)\ar[ld]_p\\
   &\cN^{\mm_2}(C,G)&\hspace*{-4.7em}\subset\,\cN(C,G)} \]
showing that the isomorphism type of the covering depends only on the coset
$\mm_2$.
Since each fibre of the cover $\MH(C',G)$ over $\cN(C,G)$ is a single
$\ZG_{[2]}$-orbit, the union \eqref{eqn:Nm} must be disjoint.

The homotopy groups of $\cM:=\MH^{m_0=0}(C,G)$ and $\cM':=\MH(C',G)$ are
studied in \S\ref{sec:MH}.
First, $\cM$ is connected and $\pi_0(\cM')$ is expected to be $\pG/2\pG$.
For $k=1,2$, the maps $p_*\colon\pi_k(\cM')\to\pi_k(\cM)$ agree with the
pull-backs of cohomology groups from $C'$ to $C$ via $\pi$.
Concretely, we have commutative diagrammes
\[ \xymatrix@=1pc{\pi_1(\cM')\ar[d]_{p_*}\ar[r]^{\hspace*{-1.75em}\cong}
   &H^1(C',\pG)\ar[d]_{\pi^*}\\
   \pi_1(\cM)\ar[r]^{\hspace*{-1.75em}\cong}& H^1(C,\pG),}\qquad\qquad
\xymatrix@=1pc{&&\pi_2(\cM')\ar[d]_{p_*}\ar[r]^{\hspace*{-1.75em}\zeta'_*}
   _{\hspace*{-1.75em}\cong} & H^0(C',\pGad)\ar[d]_{\pi^*}^\cong & \\
   0\ar[r]&H^2(C,\pi_3(G))\ar[r]&\pi_2(\cM)\ar[r]^{\hspace*{-1.75em}\zeta_*}
   & H^0(C,\pGad)\ar[r]&0}   \]
from which we obtain $\ker(p_*\colon\pi_1(\cM')\to\pi_1(\cM))\cong\pG_{[2]}$
and $\coker(p_*\colon\pi_2(\cM')\to\pi_2(\cM))\cong\pi_3(G)$.
By the long exact sequence \eqref{eqn:rel-p} for the map $p\colon\cM'\to\cM$,
we obtain two short exact sequences
\begin{equation}\label{eqn:MM'1}
0\to\sfrac{H^1(C,\pG)}{\pi^*H^1(C',\pG)}\to\pi_1(\cM,\cM')\to\pG/2\pG\to0,
\end{equation}
\begin{equation}\label{eqn:MM'2}
0\to\pi_3(G)\to\pi_2(\cM,\cM')\to\pG_{[2]}\to0.
\end{equation}
Comparing with \eqref{eqn:C1'}, we find out that
$\pi_1(\cM,\cM')\cong H^1(C',\upG)$.
We also have an exact sequence
\begin{equation}\label{eqn:MN1}
0\to\sfrac{H^1(C,\pG)}{\pi^*H^1(C',\pG)}\to\pi_1(\cM,\cN)\to\bm(\bZ_2,G)\to0,
\end{equation}
where $\cN:=\cN(C,G)$.
Therefore $\pi_1(\cM,\cN)\cong\coker(\udel^0_{C'})$ by \eqref{eqn:cokerdel0}.

Each component $\MH^{m_2}(C',G)$ has a free action of
$\ker(\del_{C'}^1)\subset H^1(C',\ZG)$, and the quotient map is a covering
$q'\colon\MH^{m_2}(C',G)\to\MH^{\mm_2}(C',G_\ad)$, where we have identified
$i_{C'}^2(m_2)$ with $\mm_2\in\coker(\del_{\bZ_2}^1)$ by \eqref{eqn:cokerdel}.
There is a also a covering $p\colon\MH^{m_2}(C',G)\to\cN^{\mm_2}(C,G)$ by
taking the quotient of the subgroup $\ker(\del_{\bZ_2}^1)$ of
$\ker(\del_{C'}^1)$.
The group $\pi^*H^1(C',\ZG)$ acts freely on $\cN^{\mm_2}(C,G)$ and the quotient
$\cN^{\mm_2}(C,G_\ad)$ can be identified with $\MH^{\mm_2}(C',G_\ad)$.
(See \cite{Wu15} for these results.)
Summarising, if $\MH^\dcirc(C',G_\ad)$ is the part of $\MH(C',G_\ad)$ such
that the structure group can be lifted to $G$ and $\cN^\dcirc(C,G_\ad)$ is its
image in $\MH(C,G_\ad)$, we have two commutative diagrammes of covering maps
\begin{equation}\begin{split}\label{eqn:diam}
\scalebox{.95}[1]{$\xymatrix@=1pc{&\MH(C',G)\ar[ld]_{\scriptstyle\ZG_{[2]}}
   \ar[rd]^{\scriptstyle\;\;\;H^1(C',\ZG)}&                            \\
   \cN(C,G)\ar[rd]_{\scriptstyle\hspace*{-4em}\pi^*H^1(C',\ZG)}&&
   \MH^\dcirc(C',G_\ad)\ar[ld]^{\scriptstyle\;\;1}                   \\
   & \cN^\dcirc(C,G_\ad) & \\ & \raisebox{4ex}{(a)} &} \qquad
   \xymatrix@=1pc{&\MH^{m_2}(C',G)\ar[ld]_{\scriptstyle\ker(\del_{\bZ_2}^1)}
   \ar[rd]^{\scriptstyle\;\;\ker(\del_{C'}^1)}&                      \\
   \cN^{\mm_2}(C,G)\ar[rd]_{\scriptstyle\hspace*{-4em}\pi^*H^1(C',\ZG)}&&
   \MH^{\mm_2}(C',G_\ad)\ar[ld]^{\scriptstyle\;\;1}                     \\
   & \cN^{\mm_2}(C,G_\ad) & \\ & \raisebox{4ex}{(b)} &}$}
\end{split}\end{equation}
\vspace{-5ex}

\noindent
in which the label on each arrow is the group of deck transformations of the
cover.

Consider the moduli space $\MH^{m_2}(C',\tG)$ of Hitchin pairs from a
$\tG$-bundle on $C'$ twisted by a discrete $B$-field
$m_2\in H^2(C',\pG)\cong\pG/2\pG$; such a twisted $\tG$-bundle reduces to an
honest $G$-bundle on $C'$ of topological type $m_2$ and pulls back to an
honest $\tG$-bundle on $C$.
As in \S\ref{sec:MH}, $\MH^{m_2}(C',\tG)$ is expected to
be connected and simply connected.
There is an action of $H^1(C',Z(\tG))$ on $\MH^{m_2}(C',\tG)$.
The quotient by the subgroup $H^1(C',\pG)$ is a regular $H^1(C',\pG)$-covering
$\tq'\colon\MH^{m_2}(C',\tG)\to\MH^{m_2}(C',G)$, whereas the quotient by the
subgroup $Z(\tG)_{[2]}$ is a regular $Z(\tG)_{[2]}$-covering
$\tilde p\colon\MH^{m_2}(C',\tG)\to\cN^{m_2}(C,\tG)$ onto its image, which is
also connected, in $\MH^{m_0=0}(C,\tG)$.
Moreover, $\pi^*H^1(C',\pG)$ acts on $\cN^{m_2}(C,\tG)$ and the quotient is
$\cN^{\mm_2}(C,G)$.
Let $\MH^\dcirc(C',\tG)$ be the union of $\MH^{m_2}(C',\tG)$ for all
$m_2\in\pG/2\pG$ and let $\cN^\dcirc(C,\tG)$ be its image in
$\MH^{m_0=0}(C,\tG)$.
We have commutative diagrammes \cite{Wu15}
\[ \begin{split}
\scalebox{.95}[1]{$\xymatrix@=1pc{&\MH^\dcirc(C',\tG)
   \ar[ld]_{\scriptstyle Z(\tG)_{[2]}}\ar[rd]^{\scriptstyle\;\;H^1(C',\pG)}&\\
   \cN^\dcirc(C,\tG)\ar[rd]&&\MH(C',G)\ar[ld]^{\scriptstyle\;\;\ZG_{[2]}}   \\
   &\cN(C,G)&} \qquad
   \xymatrix@=1pc{&\MH^{m_2}(C',\tG)\ar[ld]_{\scriptstyle Z(\tG)_{[2]}}
   \ar[rd]^{\scriptstyle\;\;H^1(C',\pG)}&                                   \\
   \cN^{m_2}(C,\tG)
   \ar[rd]_{\scriptstyle\hspace*{-4em}\pi^*H^1(C',\pG)}&&\MH^{m_2}(C',G)
   \ar[ld]^{\scriptstyle\;\;\ker(\del_{\bZ_2}^1)} \\ &\cN^{\mm_2}(C,G)&} $}
\end{split} \]
of regular coverings in which the labels on the arrows are again the groups of
deck transformations.

\bigskip\medskip\noindent{\bf Acknowledgments.}
The work is supported in part by grants Nos.~105-2115-M-007-001-MY2 and
106-2115-M-007-005-MY2 from MOST (Taiwan).
The author thanks Haibao~Duan and Nan-Kuo~Ho for discussions and the referee
for comments.


\begin{thebibliography}{99}
\bibitem{KW}
A.~Kapustin and E.~Witten,
Electric-magnetic duality and the geometric Langlands program,
Commun.\ Number Theory Phys.\ 1 (2007) 1--236, {\tt arXiv:hep-th/0604151}

\bibitem{GW08}
S.~Gukov and E.~Witten,
Gauge theory, ramification, and the geometric Langlands program,
in: Current developments in mathematics, 2006, eds.\ D.~Jenison et al.,
pp.~35--180, Int.\ Press, Somerville, MA (2008), {\tt arXiv:hep-th/0612073}

\bibitem{W08}
E.~Witten,
Gauge theory and wild ramification,
Anal.\ Appl.\ (Singap.) 6 (2008) 429--501, {\tt arXiv:0710.0631\,[hep-th]}

\bibitem{FW08}
E.~Frenkel and E.~Witten,
Geometric endoscopy and mirror symmetry,
Commun.\ Number Theory Phys.\ 2 (2008) 113--283,
{\tt arXiv:0710.5939\,[math.AG]}

\bibitem{GW10}
S.~Gukov and E.~Witten,
Rigid surface operators,
Adv.\ Theor.\ Math.\ Phys.\ 14 (2010) 87--177,
{\tt arXiv:0804.1561\,[hep-th]}

\bibitem{W18}
E.~Witten,
More on gauge theory and geometric Langlands,
Adv.\ Math.\ 327 (2018) 624--707, {\tt arXiv:1506.04293\,[hep-th]}

\bibitem{tH78}
G.~'t~Hooft,
On the phase transition towards permanent quark confinement,
Nucl.\ Phys.\ B138 (1978) 1--25

\bibitem{tH79}
G.~'t~Hooft,
A property of electric and magnetic flux in non-abelian gauge theories,
Nucl.\ Phys.\ B153 (1979) 141--160

\bibitem{GW09}
S.~Gukov and E.~Witten,
Branes and quantization,
Adv.\ Theor.\ Math.\ Phys.\ 13 (2009) 1445--1518,
{\tt arXiv:0809.0305\,[hep-th]}

\bibitem{G}
S.~Gukov,
Quantization via mirror symmetry,
Japan.\ J.\ Math.\ 6 (2011) 65--119, {\tt arXiv:1011.2218\,[hep-th]}

\bibitem{St}
N.~Steenrod,
The topology of fibre bundles,
Princeton Univ.\ Press, Princeton (1951)

\bibitem{BPST}
A.A.~Belavin, A.M.~Polyakov, A.S.~Schwartz and Yu.S.~Tyupkin,
Pseudoparticle solutions of the Yang-Mills equations,
Phys.\ Lett.\ 59B (1975) 85--87

\bibitem{JR}
R.~Jackiw and C.~Rebbi,
Vacuum periodicity in a Yang-Mills quantum theory,
Phys.\ Rev.\ Lett.\ 37 (1976) 172--175

\bibitem{CDG}
C.G.~Callan, Jr., R.F.~Dashen and D.J.~Gross,
The structure of the gauge theory vacuum,
Phys.\ Lett.\ 63B (1976) 334--340

\bibitem{Wu15}
S.~Wu,
Twisted character varieties, covering spaces and gerbes,
Theor.\ Math.\ Phys.\ 185 (2015) 1769--1788

\bibitem{VW}
C.~Vafa and E.~Witten,
A Strong coupling test of $S$-duality,
Nucl.\ Phys.\ B431 (1994) 3--77, {\tt arXiv:hep-th/9408074}

\bibitem{Wu08}
S.~Wu,
$S$-duality in Vafa-Witten theory for non-simply laced gauge groups,
J.\ High Energy Phys.\ 0805 (2008) 009, {\tt arXiv:0802.2047\,[hep-th]}

\bibitem{AB}
M.F.~Atiyah and R.~Bott,
The Yang-Mills equations over Riemann surfaces,
Philos.\ Trans.\ Roy.\ Soc.\ London Ser.~A 308 (1983) 523--615

\bibitem{FL}
Th.~Friedrich and L.~Habermann,
Yang-Mills equations on the two-dimensional sphere,
Commun.\ Math.\ Phys.\ 100 (1985) 231--243

\bibitem{Ka06}
A.~Kapustin,
Wilson-'t~Hooft operators in four-dimensional gauge theories and $S$-duality,
Phys.\ Rev.\ D 74 (2006) 025005, {\tt arXiv:hep-th/0501015}

\bibitem{GNO}
P.~Goddard, J.~Nuyts and D.I.~Olive,
Gauge theories and magnetic charge,
Nucl.\ Phys.\ B125 (1977) 1--28

\bibitem{MO}
C.~Montonen and D.I.~Olive,
Magnetic monopoles as gauge particles?,
Phys.\ Lett.\ 72B (1977) 117--120

\bibitem{DFHK}
N.~Dorey, C.~Fraser, T.J.~Hollowood and M.A.C.~Kneipp,
$S$-duality in $N=4$ supersymmetric gauge theories,
Phys.\ Lett.\ 383B (1996) 422--428, {\tt arXiv:hep-th/9605069}

\bibitem{Sen}
A.~Sen,
Electric-magnetic duality in string theory,
Nucl.\ Phys.\ B404 (1993) 109--126, {\tt arXiv:hep-th/9207053}

\bibitem{AKS}
P.C.~Argyres, A.~Kapustin and N.~Seiberg,
On $S$-duality for non-simply-laced gauge groups,
J.\ High Energy Phys.\ 06 (2006) 043, {\tt arXiv:hep-th/0603048}

\bibitem{WO}
E.~Witten and D.I.~Olive,
Supersymmetry algebras that include topological charges,
Phys.\ Lett.\ 78B (1978) 97--101

\bibitem{Os}
H.~Osborn,
Topological charges for $N=4$ supersymmetric gauge theories and monopoles of
spin~1,
Phys.\ Lett.\ 83B (1979) 321--326

\bibitem{Ya}
J.~P.~Yamron,
Topological action from twisted supersymmetric theories,
Phys.\ Lett.\ 213B (1988) 325--330

\bibitem{Ma}
N.~Marcus,
The other topological twisting of $N=4$ Yang-Mills,
Nucl.\ Phys.\ B452 (1995) 331--345, {\tt arXiv:hep-th/9506002}

\bibitem{BJSV}
M.~Bershadsky, A.~Johansen, V.~Sadov and C.~Vafa,
Topological reduction of 4D SYM to 2D $\sigma$-models,
Nucl.\ Phys.\ B448 (1995) 166--186, {\tt arXiv:hep-th/9501096}

\bibitem{HMS}
J.A.~Harvey, G.~Moore and A.~Strominger,
Reducing $S$-duality to $T$-duality,
Phys.\ Rev.\ D52 (1995) 7161--7167, {\tt arXiv:hep-th/9501022}

\bibitem{Hi87}
N.~Hitchin,
The self-duality equations on a Riemann surface,
Proc.\ London Math.\ Soc.\ 55 (1987) 59--126

\bibitem{Mu}
M.K.~Murray,
Bundle gerbes,
J.\ London Math.\ Soc.\ 54 (1996) 403--416, {\tt arXiv:dg-ga/9407015}

\bibitem{GHR}
S.J.~Gates,~Jr., C.M.~Hull and M.~Ro\v cek,
Twisted multiplets and new supersymmetric non-linear-$\sig$-models,
Nucl.\ Phys.\ B248 (1984) 157--186

\bibitem{Gu}
M.~Gualtieri,
Generalized complex geometry,
D.\ Phil.\ thesis, Oxford University (2003),
{\tt arXiv:math/0401221\,[math.DG]}

\bibitem{W92}
E.~Witten,
Mirror manifolds and topological field theory,
in: Essays on mirror manifolds, ed.\ S.T.~Yau, International Press,
Hong Kong (1992), pp.~120--158, {\tt arXiv:hep-th/9112056}

\bibitem{KL07}
A.~Kapustin and Y.~Li,
Topological sigma-models with $H$-flux and twisted generalized complex
manifolds,
Adv.\ Theor.\ Math.\ Phys.\ 11 (2007) 261--290, {\tt arXiv:hep-th/0407249}

\bibitem{Pu}
V.~Pustun,
Topological strings in generalized complex space,
Adv.\ Theor.\ Math.\ Phys.\ 11 (2007) 399--450, {\tt arXiv:hep-th/0603145}

\bibitem{Hi03}
N.~Hitchin,
Generalized Calabi-Yau manifolds,
Quart.\ J.\ Math.\ 54 (2003) 281--308, {\tt arXiv:math/0209099\,[math.DG]}

\bibitem{HWW}
N.-K.~Ho, G.~Wilkin and S.~Wu,
Hitchin's equations on a nonorientable manifold,
Commun.\ Anal.\ Geom.\ 26 (2018) 857--886, {\tt arXiv:1211.0746\,[math.DG]}

\bibitem{DG}
R.~Donagi and D.~Gaitsgory,
The gerbe of Higgs bundles,
Transf.\ Groups 7 (2002) 109--153,
{\tt arXiv:math/0005132\,[math.AG]}

\bibitem{Hi92}
N.~Hitchin,
Lie groups and Teichm\"uller space,
Topology 31 (1992) 449--473

\bibitem{HT}
T.~Hausel and M.~Thaddeus,
Mirror symmetry, Langlands duality, and the Hitchin system,
Invent.\ Math.\ 153 (2003) 197--229, {\tt arXiv:math/0205236\,[math.AG]}

\bibitem{DP12}
R.~Donagi and T.~Pantev,
Langlands duality for Hitchin systems,
Invent.\ Math.\ 189 (2012) 653--735, {\tt arXiv:math/0604617\,[math.AG]}

\bibitem{SYZ}
A.~Strominger, S.-T.~Yau and E.~Zaslow,
Mirror symmetry is $T$-duality,
Nucl.\ Phys.\ B479 (1996) 243--259, {\tt arXiv:hep-th/9606040}

\bibitem{Hi01}
N.~Hitchin,
Lectures on special Lagrangian submanifolds,
in: Winter school on mirror symmetry, vector bundles and Lagrangian
submanifolds (Cambridge, MA, 1999), AMS/IP Stud.\ Adv.\ Math., 23,
pp.~151--182, Amer.\ Math.\ Soc., Providence, RI (2001)

\bibitem{DP08}
R.~Donagi and T.~Pantev,
Torus fibrations, gerbes, and duality,
Mem.\ Amer.\ Math.\ Soc.\ 193, no.~901 (2008) 90 pp.,
{\tt arXiv:math/0306213\,[math.AG]}

\bibitem{BEM}
P.~Bouwknegt, J.~Evslin and V.~Mathai,
Topology and $H$-flux of $T$-dual manifolds,
Phys.\ Rev.\ Lett.\ 92 (2004) 181601, {\tt arXiv:hep-th/0312052}

\bibitem{Muk}
S.~Mukai,
Duality between $D(X)$ and $D(\hat X)$ with its application to Picard sheaves,
Nagoya Math.\ J.\ 81 (1981) 153--175

\bibitem{BBH}
C.~Bartocci, U.~Bruzzo and D.~Hern\'andez Ruip\'erez,
Fourier-Mukai and Nahm transforms in geometry and physics,
Progr.\ Math., 276, Birkh\"auser, Boston (2009)

\bibitem{Me}
M.A.~Metlitski,
$S$-duality of $u(1)$ gauge theory with $\tht=\pi$ on non-orientable
manifolds: applications to topological insulators and superconductors,
{\tt arXiv:1510.05663\,[hep-th]}

\bibitem{Hor}
K.~Hori, S.~Katz, A.~Klemm, R.~Pandharipande, R.~Thomas, C.~Vafa, R.~Vakil
and E.~Zaslow,
Mirror symmetry, Clay Math.\ Monogr., vol.~1,
Amer.\ Math.\ Soc., Providence, RI (2003), \S39.3 

\bibitem{KL03}
A.~Kapustin and Y.~Li,
Stability conditions for topological $D$-branes: a worldsheet approach,
preprint, {\tt arXiv:hep-th/0311101}

\bibitem{Kon}
M.~Kontsevich,
Homological algebra of mirror symmetry,
Proc. Intern.\ Congr.\ Math., Vol.~1 (Z\"urich, 1994), ed.~S.D.~Chatterji,
pp.~120--139, Birkh\"auser, Basel (1995), {\tt arXiv:alg-geom/9411018}

\bibitem{dS}
V.~de~Silva,
Products in the symplectic Floer homology of Lagrangian intersections,
D.\ Phil.\ thesis, Oxford University (1998)

\bibitem{Liu}
C.-C.M.~Liu,
Moduli of $J$-holomorphic curves with Lagrangian boundary conditions and open
Gromov-Witten invariants for an $S^1$-equivariant pair,
Ph.D.\ thesis, Harvard (2002), {\tt arXiv:math/0210257\,[math.SG]}

\bibitem{So}
J.P.~Solomon,
Intersection theory on the moduli space of holomorphic curves with Lagrangian
boundary conditions,
Ph.D.\ thesis, MIT (2006), {\tt arXiv:math/0606429\,[math.SG]}

\bibitem{FO3}
K.~Fukaya, Y.-G.~Oh, H.~Ohta and K.~Ono,
Lagrangian intersection Floer theory: anomaly and obstruction, Part~II,
AMS/IP Studies Adv.\ Math., 46.2, Amer.\ Math.\ Soc., Providence, RI and
Intern.\ Press, Somerville, MA (2009), chap.~8

\bibitem{Ge}\scalebox{.975}[1]{$\!\!$
P.~Georgieva,
The orientability problem in open Gromov-Witten theory,
Geom.\ Topol.\ 17 (2013) 2485--2512, {\tt arXiv:1207.5471\,[math.SG]}}

\bibitem{Sei}
P.~Seidel,
Graded Lagrangian submanifolds,
Bull.\ Soc.\ Math.\ France 128 (2000) 103--149,
{\tt arXiv:math/9903049\,[math.SG]}

\bibitem{Ka00}
A.~Kapustin,
D-branes in a topologically nontrivial $B$-field,
Adv.\ Theor.\ Math.\ Phys.\ 4 (2000) 127--154, {\tt arXiv:hep-th/9909089}

\bibitem{Th}
R.~Thom,
Quelques propri\'et\'es globales des vari\'et\'es diff\'erentiables,
Comment.\ Math.\ Helv.\ 28 (1954) 17--86

\bibitem{FW}
D.S.~Freed and E.~Witten,
Anomalies in string theory with $D$-branes,
Asian J.\ Math.\ 3 (1999) 819--851, {\tt arXiv:hep-th/9907189}

\bibitem{BS}
D.~Baraglia and L.P.~Schaposnik,
Higgs bundles and $(A,B,A)$-branes,
Commun.\ Math.\ Phys.\ 331 (2014) 1271--1300, {\tt arXiv:1305.4638\,[math.DG]}

\bibitem{Wu16}
S.~Wu,
Quantization of Hitchin's moduli space of a non-orientable surface,
in: Geometric methods in physics, Trends Math., eds.\ P.~Kielanowski et al.,
pp.~343--363, Birkh\"auser/Springer, Cham (2016)

\bibitem{Li06}
Y.~Li,
Anomalies and graded coisotropic branes,
J.\ High Energy Phys.\ 0603 (2006) 100, {\tt arXiv:hep-th/0405280}

\bibitem{KR}
T.P.~Killingback and E.G.~Rees,
Topology of gauge theories on compact $4$-manifolds,
Class.\ Quant.\ Grav.\  4 (1987) 357--373

\bibitem{Do02}
S.K.~Donaldson,
Floer homology groups in Yang-Mills theory,
Cambridge Univ.\ Press, Cambridge (2002), \S2.5.2

\bibitem{Mac}
M.~Mackaay,
A note on the holonomy of connections in twisted bundles,
Cah.\ Topol.\ G\'eom.\ Diff\'er.\ Cat\'eg.\ 44 (2003) 47--71,
{\tt 	arXiv:math/0106019\,[math.DG]}

\bibitem{Do87}
S.K.~Donaldson,
Twisted harmonic maps and the self-duality equations,
Proc.\ London Math.\ Soc.\ 55 (1987) 127--131

\bibitem{Co}
K.~Corlette,
Flat $G$-bundles with canonical metrics,
J.\ Diff.\ Geom.\ 28 (1988) 361--382

\bibitem{Li}
J.~Li,
The space of surface group representations,
Manuscripta Math.\ 78 (1993) 223--243

\bibitem{Si}
C.T.~Simpson,
Higgs bundles and local systems,
Inst.\ Hautes \'Etudes Sci.\ Publ.\ Math.\ 75 (1992) 5--95

\bibitem{NS}
M.S.~Narasimhan and C.S.~Seshadri,
Stable and unitary vector bundles on a compact Riemann surface,
Ann.\ Math.\ 82 (1965) 540--567

\bibitem{Rtn}
A.~Ramanathan,
Stable principal bundles on a compact Riemann surface,
Math.\ Ann.\ 213 (1975) 129--152

\bibitem{LR}
S.~Lawton and D.~Ramras,
Covering spaces of character varieties,
New York J.\ Math.\ 21 (2015) 383--416, {\tt arXiv:1402.0781\,[math.AT]}

\bibitem{Da}
G.D.~Daskalopoulos,
The topology of the space of stable bundles on a compact Riemann surface,
J.\ Diff.\ Geom.\ 36 (1992) 699--746

\bibitem{DU}
G.D.~Daskalopoulos and K.K.~Uhlenbeck,
An application of transversality to the topology of the moduli space of
stable bundles, Topology 34 (1995) 203--215

\bibitem{BGG}
S.B.~Bradlow, O.~Garc\'ia-Prada and P.B.~Gothen,
Homotopy groups of moduli spaces of representations,
Topology 47 (2008) 203--224, {\tt arXiv:math/0407523\,[math.AG]}

\bibitem{HL03}
N.-K.~Ho and C.-C.M.~Liu,
Connected components of the space of surface group representations,
Inter.\ Math.\ Res.\ Notices 44 (2003) 2359--2371,
{\tt arXiv:math/0303255\,[math.SG]}

\bibitem{HL08}
N.-K.~Ho and C.-C.M.~Liu,
Yang-Mills connections on nonorientable surfaces,
Commun.\ Anal.\ Geom.\ 16 (2008) 617--679, {\tt arXiv:math/0605587\,[math.SG]}

\bibitem{BC}
V.V.~Benyash-Krivets and V.I.~Chernousov,
Representation varieties of the fundamental groups of compact non-orientable
surfaces, Sbornik: Math.\ 188 (1997) 997--1039

\bibitem{Rs}
D.A.~Ramras,
On the Yang-Mills stractification for surfaces,
Proc.\ Amer.\ Math.\ Soc.\ 139 (2011) 1851--1863;
The Yang-Mills stratification for surfaces revisited,
{\tt arXiv:0805.2587\,[math.DG]}

\bibitem{AS}
M.F.~Atiyah and I.M.~Singer, 
Dirac operators coupled to vector potentials,
Proc.\ Nat.\ Acad.\ Sci.\ USA\ 81 (1984) 2597--2600

\bibitem{H}
T.~Hausel,
Vanishing of intersection numbers on the moduli space of Higgs bundles,
Adv.\ Theor.\ Math.\ Phys.\ 2 (1998) 1011--1040, {\tt arXiv:math.AG/9805071}

\bibitem{Rn}
S.~Ramanan,
The moduli space of vector bundles over an algebraic curve,
Math.\ Ann.\ 200 (1973) 69--84

\bibitem{Sp}
E.H.~Spanier,
Algebraic topology,
Springer-Verlag, New York (1966)

\bibitem{BaW}
M.G.~Barratt and J.H.C.~Whitehead, 
The first nonvanishing group of an $(n+1)$-ad, 
Proc.\ London Math.\ Soc.\ 6 (1956) 417--439

\bibitem{BW}
G.E.~Bredon and J.W.~Wood,
Non-orientable surfaces in orientable $3$-manifolds,
Invent.\ Math.\ 7 (1969) 83--110

\bibitem{LRS}
A.S.~Levine, D.~Ruberman and S.~Strle,
Nonorientable surfaces in homology cobordisms (Appendix by I.M.~Gessel),
Geom.\ Topol.\ 19 (2015) 439--494, {\tt arXiv:1310.8516\,[math.GT]}

\bibitem{Ho}
N.-K.~Ho,
The real locus of an involution map on the moduli space of flat connections
on a Riemann surface,
Inter.\ Math.\ Res.\ Notices 61 (2004) 3263--3285,
{\tt arXiv:math/0312426\,[math.SG]}

\bibitem{BHH}
I.~Biswas, J.~Huisman and J.~Hurtubise,
The moduli space of stable vector bundles over a real algebraic curve,
Math.\ Ann.\ 347 (2010) 201--233, {\tt arXiv:0901.3071\,[math.AG]}

\bibitem{Sc}
F.~Schaffhauser,
Real points of coarse moduli schemes of vector bundles on a real algebraic curve, J.\ Symplectic Geom.\ 10 (2012) 503--534,
{\tt arXiv:1003.5285\,[math.AG]}

\end{thebibliography}
\end{document}